\documentclass[prc,aps,floatfix,showpacs,twocolumn]{revtex4-1}
\usepackage{dcolumn}
\usepackage{bm}
\usepackage{color}
\usepackage{graphicx}
\usepackage{pstricks}
\usepackage{amssymb,amsmath}

\def\het{{{}^3{\rm He}}}
\def\heq{{{}^4{\rm He}}}
\def\a{\alpha}

\def\bmr{{\bm r}}
\def\bmx{{\bm x}}
\def\bmp{{\bm p}}
\def\bmk{{\bm k}}
\def\bmq{{\bm q}}
\def\bmR{{\bm R}}
\def\bmK{{\bm K}}
\def\bmG{{\bm G}}
\def\bmP{{\bm P}}
\def\bmL{{\bm L}}

\def\be{\begin{equation}}
\def\ee{\end{equation}}
\def\bea{\begin{eqnarray*}}
\def\eea{\end{eqnarray*}}

\def\bi{\begin{itemize}}
\def\ei{\end{itemize}}
\def\n{\phantom{0}}
\def\m{\phantom{-}}

\newcommand{\bmsi}{{\bm \sigma}}
\newcommand{\bmna}{{\bm \nabla}}

\def\m{{\phantom{-}}}
\def\nnlo{{\rm N2LO}}

\begin{document}
\title{A chiral effective field theory study of hadronic parity violation
in few-nucleon systems}
\author{M.\ Viviani$^{\,{\rm a}}$, A. Baroni$^{\,{\rm b}}$,
L.\ Girlanda$^{\,{\rm c}}$, A.\ Kievsky$^{\,{\rm a}}$, L.E.\ Marcucci$^{\,{\rm  d,a}}$, and R.\ Schiavilla$^{\,{\rm b,e}}$}
\affiliation{
$^{\rm a}$\mbox{INFN-Pisa, 56127 Pisa, Italy}\\
$^{\rm b}$\mbox{Department of Physics, Old Dominion University, Norfolk, VA 23529, USA} \\
$^{\rm c}$\mbox{Department of Physics, University of Salento and INFN-Lecce, 73100 Lecce, Italy}\\
$^{\rm d}$\mbox{Department of Physics, University of Pisa, 56127 Pisa, Italy}\\
$^{\rm e}$\mbox{Jefferson Lab, Newport News, VA 23606}\\
}

\date{\today}

\begin{abstract}
We reconsider the derivation of the nucleon-nucleon parity-violating (PV)
potential within a chiral effective field theory framework.  We construct the potential
up to next-to-next-to-leading order by including one-pion-exchange,
two-pion-exchange, contact, and $1/M$ ($M$ being the nucleon mass)
terms, and use dimensional regularization
to renormalize the pion-loop corrections.  A detailed analysis of the number of
independent low-energy constants (LEC's) entering the potential is carried out.
We find that it depends on six LEC's: the pion-nucleon PV coupling
constant $h^1_\pi$ and five parameters multiplying contact interactions.
We investigate PV effects induced by this potential on several few-nucleon
observables, including the $\vec{p}\,$-$p$ longitudinal asymmetry, the
neutron spin rotation in $\vec{n}$-$p$ and $\vec{n}$-$d$ scattering, and the longitudinal
asymmetry in the $^3$He$(\vec{n},p)^3$H charge-exchange reaction.  An estimate for the
range of values of the various LEC's is provided by using available
experimental data.
\end{abstract}



\maketitle

\section{Introduction}
\label{sec:intro}

A number of experiments aimed at studying hadronic parity violation in
low-energy processes involving few nucleon systems are being completed,
or are in an advanced stage of planning, at cold neutron facilities,
such as the Los Alamos Neutron Science Center (LANSCE), the National
Institute of Standards and Technology (NIST) Center for
Neutron Research, and the Spallation Neutron Source (SNS) at Oak Ridge National Laboratory. The
primary objective of this program is to determine the fundamental
parameters of the parity-violating (PV) nucleon-nucleon ($NN$) potential
(for a review of the current status of experiment and theory, see, for
example, Refs.~\cite{MP06,HH13,SS13}).

Until a few years ago, the standard framework by which nuclear PV processes
were analyzed theoretically was based on meson-exchange potentials, in particular
the model proposed by Desplanques, Donoghue, and Holstein (DDH)~\cite{DDH}
which included pion and vector-meson exchanges with
seven unknown meson-nucleon PV coupling constants.

More recently, however, the emergence of chiral effective field theory ($\chi$EFT)~\cite{Weinberg90}
has provided renewed impetus to the development of nuclear forces in
a field-theoretic framework~\cite{ORVK96,Epelbaum09,ME11}.
The $\chi$EFT approach is based on the observation that
the chiral symmetry exhibited by quantum chromodynamics (QCD)
severely restricts the form of the interactions of pions among
themselves and with other particles~\cite{W66,CCWZ69}.  In particular,
the pion couples to the nucleon by powers of its momentum $Q$, and the
Lagrangian describing these interactions can be expanded in powers of
$Q/\Lambda_\chi$, where $\Lambda_\chi \simeq 1$ GeV specifies the
chiral symmetry breaking scale.  As a consequence, classes of
Lagrangians emerge, each characterized by a given power of
$Q/\Lambda_\chi$ and each involving a certain number of unknown
coefficients, so called low-energy constants (LEC's), which are then
determined by fits to experimental data (see, for example, the
review papers~\cite{Epelbaum09,Bernard95} and
references therein).

Chiral effective field theory has been used to study two- and many-nucleon
interactions~\cite{Epelbaum09,ME11} and the interaction of electroweak probes with
nuclei~\cite{Park96,Park03,Pastore09,Koelling09,Pastore11,Koelling11,Piarulli13}.
Its validity is restricted to processes occurring at low energies.
In this sense, it has a more limited range of applicability than meson-exchange
or more phenomenological models of these interactions, which in fact
quantitatively and successfully account for a wide variety of nuclear
properties and reactions up to energies, in some cases, well beyond
the pion production threshold (for a review, see Ref.~\cite{Carlson98}).
However, it is undeniable that $\chi$EFT has put nuclear physics on a
more fundamental basis by providing, on the one hand, a direct connection
between the QCD symmetries---in particular, chiral symmetry---and the
strong and electroweak interactions in nuclei, and, on the
other hand, a practical calculational scheme, which can, at least in principle,
be improved systematically.

The $\chi$EFT approach has also been used to study PV $NN$ potentials,
which are induced by hadronic weak interactions---these follow from
weak interactions between quarks inside hadrons~\cite{KS93}.  It is well
known that the weak interaction in the Standard Model contains both
parity-conserving (PC) and PV components.  The part of the weak
interactions contributing to the PC $NN$ potential is obviously totally ``hidden'' by
the strong and electromagnetic interactions, and is therefore not accessible
experimentally.  However, their PV part can be revealed in dedicated experiments.
Since the fundamental weak
Lagrangian of quarks is not invariant under chiral symmetry, one constructs the
most general PV Lagrangian of nucleons and pions by requiring that the pattern
of chiral symmetry breaking at the hadronic level be the same as
at the quark level.  Moreover, since the combination of charge conjugation
($C$) and parity ($P$)  is known to be violated to a
much lesser extent, it is customary to consider only
$P$-violating but $CP$-conserving terms~\cite{KS93}.

Following this scheme, Kaplan and Savage~\cite{KS93}
constructed an effective Lagrangian describing
the interactions of pions and nucleons up to one derivative
(i.e., at order $Q$).  This Lagrangian includes at leading order (LO), or $Q^0$,
a Yukawa-type pion-nucleon interaction with no derivatives.
The coupling constant multiplying this term is
denoted as $h^1_\pi$, the pion-nucleon
weak coupling constant.  It gives rise to a long-range,
one-pion-exhange (OPE) contribution to the PV $NN$ potential.
Many experiments have attempted to determine  this long-range component
and to obtain a determination of $h^1_\pi$, a task which
has proven so far to be elusive (for a review see Ref.~\cite{HH13}).
Very recently, an attempt has also been made to estimate the
value of $h^1_\pi$ in a lattice QCD calculation~\cite{Wasem2012}.

The Kaplan-Savage Lagrangian also includes
five next-to-leading order (NLO) pion-nucleon interaction, or $Q^1$,
terms with one derivative (and accompanying LEC's), which, however,
do not enter the PV $NN$ potential when considering processes at either
tree level or one loop~\cite{KS93,Zhu05}.

Since the pioneering study of Ref.~\cite{KS93}, there have been several
studies of the PV $NN$ potential in $\chi$EFT~\cite{Zhu00,Zhu01}.  The first 
derivation up to next-to-next-to-leading order ($\nnlo$) was carried out by
Zhu {\it et al.}~\cite{Zhu05}. This potential
includes the long-range OPE component, medium-range components originating
from two-pion-exhange (TPE) processes, and short-range components
deriving from ten four-nucleon contact terms involving one derivative of
the nucleon field.  In a subsequent analysis~\cite{Girlanda08}, it was shown
that there exist only five independent contact terms entering the
potential at $\nnlo$, corresponding to the five PV
S-P transition amplitudes at low energies~\cite{Danilov65}.  Zhu {\it et al.}~\cite{Zhu05} also included
three pion-nucleon PV interaction terms of order $Q^2$.   This potential was recently
used in a calculation of the longitudinal analyzing power in $\vec{p}$-$p$
scattering~\cite{Vries13}. 

In subsequent years, the $\nnlo$ contributions due to TPE and contact
terms were independently studied in a series of papers~\cite{Desp08,Liu07,Hyun07,Hyun08}
by a different collaboration.  In particular, this collaboration carried
out a calculation of the photon asymmetry in the radiative capture
$^1$H$(\vec{n},\gamma)^2$H.  

The objectives of the present work are twofold.  The first is to reconsider the problem
of how many independent PV Lagrangian terms are allowed at
order $Q^2$, and to construct the complete PV $NN$ potential at
$\nnlo$. A similar analysis for the parity- and time-reversal
violating Lagrangian terms (with the aim to study the electric dipole moment of
nucleons and light nuclei) has been recently reported in
Ref.~\cite{VMTK13}.
The second objective is to use this potential to investigate PV effects
in several processes involving few-nucleon systems, including the $\vec p\,$-$p$ longitudinal
asymmetry, the neutron spin rotation in $\vec n$-$p$ and $\vec n$-$ d$ scattering, and 
the longitudinal asymmetry in the $^3$He($\vec{n},p$)$^3$H reaction, and to provide estimates
for the values of the various LEC's by fitting available experimental data. 

To date, measurements are available for the following PV observables: the longitudinal analyzing
power in $\vec{p}\,$-$p$~\cite{Balzer80}--\cite{Berdoz03} and $\vec{p}$-$\alpha$~\cite{Lang85}
scattering, the photon asymmetry and photon circular polarization in, respectively, the
$^1$H($\vec{n},\gamma$)$^2$H~\cite{Cavaignac77}--\cite{Gericke09}
and $^1$H($n,\vec{\gamma}$)$^2$H~\cite{Knyazkov84} radiative captures, and the neutron
spin rotation in $\vec{n}$-$\alpha$ scattering~\cite{Snow09,Bass09}.  The planned 
experiments include measurements of the neutron spin
rotation in $\vec{n}$-$p$~\cite{Snow09} and $\vec{n}$-$d$~\cite{Markoff07} scattering,
and of the longitudinal asymmetry in the charge-exchange reaction $^3$He($\vec{n},p$)$^3$H
at cold neutron energies~\cite{Bowman07}.  Recent studies of these
observables in the framework of the DDH potential can be found in 
Refs.~\cite{Schiavilla04,Schiavilla08,Viviani10,Gudkov10,Song11,Song12}.

We conclude this overview by noting that there exists 
a different approach to the derivation of PV (and PC) nuclear forces, based on an effective
field theory in which pion degrees of freedom are integrated
out (so called pionless EFT) and only contact interaction terms
are considered.  Such a theory, which is only valid at energies much less
than the pion mass~\cite{SS13}, has been used to study
PV effects in nucleon-nucleon scattering~\cite{PRR09},
PV asymmetries in the $^1$H($\vec{n},\gamma$)$^2$H
capture~\cite{SS09}, spin rotations in $\vec n$-$p$ and $\vec
n$-$d$ scattering~\cite{GSS11}, as well as other observables~\cite{SS13}.

The present paper is organized as follows.  In Sec.~\ref{sec:pvl} we study the PV
chiral Lagrangian up to order $Q^2$, while in Sec.~\ref{sec:pvnn} we derive the
PV $NN$ potential at $\nnlo$.  In Sec.~\ref{sec:res}, we report results obtained
for the $\vec p\, $-$p$ longitudinal asymmetry, the neutron spin rotation in
$\vec n$-$p$ and $\vec n$-$ d$ scattering, and the longitudinal asymmetry in the
$^3$He($\vec{n},p$)$^3$H reaction, and provide estimates for the values
of the various LEC's.  Finally, in Sec.~\ref{sec:conc}
we present our conclusions and perspectives.  A number of technical details are
relegated in Appendices~\ref{app:note1}-\ref{app:pvr}.

\section{The PV Lagrangian}
\label{sec:pvl}

Weak interactions between quarks induce a PV $NN$ potential.  This
potential can be constructed starting from a pion-nucleon effective
Lagrangian including all terms for which the pattern of chiral symmetry
violation is the same as in the fundamental (quark-level) Lagrangian.

We begin with a brief summary of the building blocks used
to construct the chiral Lagrangian (for reviews, see
Refs.~\cite{Epelbaum09,Bernard95,SS13}).  The pion field $\vec\pi$
enters the chiral Lagrangian via the unitary $SU(2)$ matrix
\begin{equation}
 U=1+{i\over f_\pi} \vec\tau\cdot\vec\pi-{1\over 2f_\pi^2}\, \vec\pi^{\,2}
    -{i\alpha\over f_\pi^3}  \,\vec\pi^{\, 2}\,
    \vec\tau\cdot\vec\pi+{8\alpha-1\over 8f_\pi^4} \vec\pi^{\, 4}+\ldots
    \ , \label{eq:uumatrix}
\end{equation}
where $f_\pi\simeq 92$ MeV is the bare pion decay
constant and $\alpha$ is an arbitrary coefficient reflecting our freedom in the
choice of the pion field.  Observables must be independent of $\alpha$.  Standard choices
are $\alpha=0$ (non-linear sigma model) or $\alpha=1/6$, corresponding
to the exponential parametrization $U=\exp(i\,\vec\tau\cdot\vec\pi/f_\pi)$.
In this and in the next section as well as in the Appendices,
except for Appendix~\ref{app:pvr}, all the LEC's entering the Lagrangian
are to be considered as bare parameters.  Renormalized LEC's
will be denoted by an overline above each symbol.

The other building blocks are $u=\sqrt{U}$ and
\begin{eqnarray}
\nabla_\mu U  &=& \partial_\mu
    U -i r_\mu U + i U \ell_\mu \ ,\label{eq:nablau}\\
  u_\mu &=& i(u^\dag\partial_\mu u - u\, \partial_\mu u^\dag)
    +u^\dag r_\mu u - u\, \ell_\mu u^\dag\ , \label{eq:umu}\\
  \Gamma_\mu &=& {1\over 2} (u^\dag\partial_\mu u + u\, \partial_\mu
  u^\dag) - \frac{i}{2}(     u^\dag r_\mu u + u\,  \ell_\mu u^\dag) \ , \label{eq:gammau}\\
  \chi_\pm &=& u^\dag \chi \,u^\dag \pm u\, \chi^\dag u\ ,\label{eq:chipm}\\
  F_{\mu\nu}^\pm&=& u^\dag F_{\mu\nu}^R u \pm u\, F_{\mu\nu}^L u^\dag
  \ ,\label{eq:chief}\\
  X^a_{L} &=& u\, \tau_a u^\dag\ ,\qquad\qquad
  X^a_{R} = u^\dag \tau_a u \ ,\label{eq:xxu}
\end{eqnarray}
with
\begin{eqnarray}
  r_\mu &=& v_\mu + a_\mu, \quad \ell_\mu =v_\mu - a_\mu\ , \quad
  \chi = s + i p\ ,\label{eq:av}\\
   F_{\mu\nu}^R &=& \partial_\mu r_\nu - \partial_\nu r_\mu - i
   [r_\mu,r_\nu] \ , \label{eq:fmunuR}\\
    F_{\mu\nu}^L &=& \partial_\mu \ell_\nu - \partial_\nu \ell_\mu - i
    [\ell_\mu,\ell_\nu]\ .\label{eq:fmunuL}
\end{eqnarray}
Here $v_\mu$, $a_\mu$, $p$, and $s$ are, respectively,
$SU(2)$ matrices of vector, axial-vector, pseudoscalar, and scalar
``external'' fields, which are assumed  to transform as
\begin{eqnarray}
 r_\mu&\to& R\, r_\mu R^\dag + i\, R\,\partial_\mu R^\dag\ ,\label{eq:lrc1}\\
 l_\mu&\to& L\, l_\mu L^\dag + i\, L\, \partial_\mu L^\dag\ ,\label{eq:lrc2}\\
 \chi&\to& R\,\chi\, L^\dag\ , \label{eq:lrc3}
\end{eqnarray}
where $L$ ($R$) represents a local rotation in the isospin space of
the left (right) components.  The Lagrangian constructed in terms of these
fields is invariant under local $SU(2)_L\times SU(2)_R$ chiral transformations.

In this paper we are interested in the $NN$ potential, and so ultimately we set
$v_\mu=a_\mu=p=0$ and
\begin{equation}
 s \to 2 B  M_q\ ,\qquad M_q= \left( \begin{array}{cc} m_u & 0 \\ 0 &
  m_d \end{array} \right)\ , \label{eq:qmass}
\end{equation}
$m_u$, $m_d$ being the masses of ``current'' up and down quarks,
respectively, and $B$ is a parameter related to the $\bar q q$ quark
condensate, $ B\, (m_u+m_d)\sim m^2_\pi$, $m_\pi$ being the pion mass.
Then, $\chi$ takes into account the explicit chiral
symmetry breaking due to the non-vanishing current-quark
masses. In the following, however, we construct all possible Lagrangian
terms in the presence of external fields, since this will be useful when
considering the coupling of nucleons and pions to electromagnetic
and/or weak probes. In studying PV Lagrangian terms, it is customary
to disregard isospin violation due to the $u$-$d$ quark
mass difference, so we will assume $m_u=m_d$.

The transformation properties under the (non-linear) chiral symmetry
$SU(2)_L\times SU(2)_R$ of the nucleon field $\psi$ and of the
quantities defined in Eqs.~(\ref{eq:uumatrix})--(\ref{eq:xxu})
are the following (see, for example, Ref.~\cite{Bernard95})
\begin{eqnarray}
  U&\rightarrow&R\, U\,L^\dag\qquad\qquad
   \nabla_\mu U \rightarrow R\, (\nabla_\mu U)\, L^\dag\ ,\nonumber\\
  \psi&\rightarrow&h\, \psi\ ,\qquad\qquad
  u\rightarrow R\, u\, h^\dag = h\, u\, L^\dag \ , \nonumber\\
  u_\mu &\rightarrow & h\, u_\mu h^\dag \ , \qquad\qquad
  \Gamma_\mu \rightarrow h \,\Gamma_\mu h^\dag +h\, \partial_\mu h^\dag\
  , \label{eq:bbt} \\
  \chi_\pm &\rightarrow& h \chi_\pm  h^\dag\ ,\qquad\qquad
   F_{\mu\nu}^\pm \to  h\,  F_{\mu\nu}^\pm h^\dag  \ , \nonumber\\
  X^a_{L} &\rightarrow& h\, u \, L^\dag \tau_a L\, u^\dag h^\dag \ ,\qquad
  X^a_{R} \rightarrow h\, u^\dag R^\dag \tau_a R \,u \,h^\dag \nonumber\ ,
\end{eqnarray}
where  $h$ is a $SU(2)$ matrix depending in a
complicate way on $L$, $R$, and $\vec\pi(x)$, and expressing the
non-linearity of the transformation.  Note that the operator
\begin{equation}
  D_\mu = \partial_\mu + \Gamma_\mu\ ,
\end{equation}
when acting on the field $\psi$, transforms covariantly, namely
\begin{equation}
  D_\mu \psi \rightarrow h \, D_\mu \psi\ .
\end{equation}
The terms $\overline{\psi} X_{L}^a\psi$ and $\overline{\psi}
X_{R}^a\psi$ transform like the operators
$\overline{q}_L\tau_a q_L$ and $\overline{q}_R\tau_a q_R$, respectively,
where $q$ represents the doublet of $u$ and $d$ quark fields,
and $q_L$ and $q_R$ are the left and right components.
Under $SU(2)_L\times SU(2)_R$ $q_L$ and $q_R$ transform
as $q_L\rightarrow L \, q_L$ and $q_R\rightarrow R \, q_R$, and therefore
\begin{equation}
  \overline{q}_L\tau_a q_L\rightarrow\overline{q}_L L^\dag \tau_a\, L \, q_L\ , \quad
  \overline{q}_R\tau_a q_R\rightarrow\overline{q}_R R^\dag \tau_a \, R\, q_R\ . 
  \label{eq:quarkt}
\end{equation}
Such terms enter the weak interaction Lagrangian at quark level.
Therefore, quantities like $\overline{\psi}\, X_{L}^a\psi$ and $\overline{\psi}\,
X_{R}^a\psi$ can be used to construct PV Lagrangian terms at hadronic
level, which mimic the corresponding terms entering the weak interaction
at quark level~\cite{KS93}.  For reasons mentioned earlier, the part
of weak interaction contributing to the PC $NN$ potential is of no
interest, and only $P$-violating but $CP$-conserving terms are considered below.

More precisely, at quark level the weak interaction includes terms which
under chiral symmetry transform as
isoscalar, isovector, and isotensor operators~\cite{KS93}.  At hadronic level, isoscalar
terms can be constructed without involving the $X_{L,R}$
matrices, isovector terms are linear in $X_{L}$ or $X_{R}$, and
isotensor terms involve combinations like ${\cal I}_{ab} (X^a_{L}
X^b_{L}\pm X^a_{R} X^b_{R})$, where
\begin{equation}
  {\cal I}_{ab}=
\left(
\begin{array}{ccc}
-1 & 0 & 0\\
0 & -1& 0 \\
0 & 0  &  +2 \\
\end{array}
\right)
\ .\label{eq:Iab}
\end{equation}
Note that our definition of ${\cal I}_{ab}$ above is
different from that commonly adopted in the literature.
In the following, we also consider the quantities
\begin{equation}
 X^a_{\pm}=X^a_{L}\pm X^a_{R}\ ,\qquad 
\end{equation}
which transform simply under $P$ and $C$, and
use the notation $\langle\dots \rangle$ to denote the
trace in flavor space, and define
\[
\hat A = A - \frac{1}{2} \langle A \rangle\ .
\]

The Lagrangian is ordered in classes of operators with increasing
chiral dimension, according to the number of derivatives and/or quark
mass insertions, e.g.,
\begin{eqnarray*}
u_\mu \sim Q\ , \qquad
 F^\pm_{\mu\nu} \sim Q^2\ , \qquad
\chi_\pm \sim Q^2\ .
\end{eqnarray*}
As per the covariant derivative $D_\mu$, it counts as $Q$, except
when it acts on the nucleon field, in which case it is $Q^0$
due to the presence of the nucleon mass scale.  The $\gamma^5$ matrix
mixes the small and large components of the Dirac spinors, so that it
should also be counted as $Q$.  By considering all possible $P$-odd but $CP$-even
interaction terms one can construct the most general Lagrangian.  In doing so,
use can be made of the equations of motion (EOM) for the nucleon and
pion fields at lowest order, namely
\begin{equation}
i D_\mu\gamma^\mu  \psi = \left( M + \frac{g_A}{2} \gamma_5 \gamma^\mu u_\mu
\right) \psi + {\cal O}(Q^2)\ ,\label{eq:eomn}
\end{equation}
\begin{equation}
[D_\mu, u^\mu ] = \frac{i}{2} \hat \chi_-  + {\cal
  O}(Q^4)\ ,\label{eq:eomp}
\end{equation}
$M$ being the nucleon mass, and of a number of other identities,
\begin{equation}
[D_\mu ,D_\nu] =\frac{1}{4} [ u_\mu,u_\nu ] - \frac{i}{2}
F^+_{\mu\nu}\ ,\label{eq:rel1} 
\end{equation}
\begin{equation}
[D_\mu,u_\nu] - [D_\nu,u_\mu]= F^-_{\mu\nu}\ .\label{eq:rel2} 
\end{equation}
Note that covariant derivatives of $u_\mu$ only appear in the
symmetrized form
\begin{equation}
h_{\mu\nu}=[D_\mu,u_\nu] + [D_\nu,u_\mu] \ ,\label{eq:hmunu}
\end{equation}
and that further simplifications follow from the
Cayley-Hamilton relations, valid for any 2$\times$2 matrices $A$ and
$B$,
\begin{equation}
A B + B A = A \langle B \rangle + B \langle A \rangle + \langle A B
\rangle - \langle A \rangle \langle B \rangle\ ,\label{eq:CH}
\end{equation}
and from the traceless property of $u_\mu$ and $X^a_{L/R}$.
As discussed before, we disregard isospin violation due to the $u$-$d$ quark
mass difference, therefore, we assume $\chi=m_\pi^2\, I$ and set
$\langle\chi_-\rangle=\langle F^{\mu\nu}_-\rangle=0$.

Care must be taken when constructing combinations of terms like
$D_\mu  X^a_{L/R}$, since they do not transform as given in Eq.~(\ref{eq:quarkt}),
see discussion in Appendix~\ref{app:note1}. There it is also shown that it is convenient 
to work instead with the following quantities
\begin{eqnarray}
  \left(X^a_R\right)_\mu &=& \left[D_\mu + i u^\dag r_\mu u\,  ,\, 
    X^a_R\right] \ ,\label{eq:xxRL1} \\
  \left(X^a_L\right)_\mu &=& \left[D_\mu + i u\, \ell_\mu u^\dag \, ,\, 
    X^a_L\right] \ .\label{eq:xxRL2}
\end{eqnarray}
These, in turn, reduce to
\begin{equation}
   \left(X^a_R\right)_\mu = {i\over 2} \left[ u_\mu\, ,\, X^a_R\right]\ ,\qquad
   \left(X^a_L\right)_\mu = -{i\over 2} \left[ u_\mu\, , \,X^a_L\right]\ .
   \label{eq:xxid}
\end{equation}
These identities are used in Appendix~\ref{app:lq2} in order to reduce the number of
terms entering the PV Lagrangian.

We now first list the possible $\pi$-$N$ independent Lagrangian terms up to order $Q^2$.
First, we need to recall the transformation properties of various
quantities under hermitean conjugation, parity ($P$), and charge conjugation ($C$), which
is done in Appendix~\ref{app:trasf}.  The detailed discussion of the
possible independent Lagrangian terms at order $Q^2$ is provided in
Appendix~\ref{app:lq2}.  We include in the definition of
the interaction terms one power of the inverse nucleon mass for each  covariant
derivative $D_\mu$ acting on a nucleon field.  All $Q^2$ terms have dimension
(MeV)$^5$.

\subsection{The $\Delta I=0$ sector}
\label{sec:delta0}
 In the $\Delta I=0$ sector, the Lagrangian starts at order $Q$ with one operator~\cite{KS93},
\begin{equation}
  O^{(0)}_{0V}=\bar \psi \gamma_\mu u^\mu \psi\ .\label{eq:o0_0v}
\end{equation}
At order $Q^2$, as shown in Appendix~\ref{app:lq2_0}, we find the following two
operators
\begin{eqnarray}
&& O^{(0)}_1={1\over M}\bar \psi \gamma^\mu \gamma^5 D^\nu \psi \langle u_\mu u_\nu
  \rangle+{\mathrm{h.c.}}\ ,\label{eq:o0_1}
\\
&&O^{(0)}_2=\bar \psi F^-_{\mu\nu} \sigma^{\mu\nu} \psi\ , \label{eq:o0_2}
\end{eqnarray}
where $ {\mathrm{h.c.}}$ stands for the hermitean conjugate.

\subsection{The $\Delta I=1$ sector}
\label{sec:delta1}
In the $\Delta I=1$ sector the Lagrangian, linear in the $X^a_{R/L}$
operators,  starts at order $Q^0$ with
\begin{equation}
   O^{(1)}_\pi=\bar \psi X^3_- \psi\ .\label{eq:o1_pi}
\end{equation}
At order $Q$ there are two additional operators~\cite{KS93},
\begin{eqnarray}
 O^{(1)}_{1V}&=&\bar \psi \gamma^\mu \psi \langle u_\mu  X^3_+\rangle\ ,\label{eq:o1_1V}\\
 O^{(1)}_{1A}&=&\bar \psi \gamma^\mu \gamma_5 \psi \langle u_\mu X^3_- \rangle \ . \label{eq:o1_1A}
\end{eqnarray}
At order $Q^2$, there are many possibilities; however, as discussed in Appendix~\ref{app:lq2_1},
we consider the following combinations:
\bgroup
\arraycolsep=1.0pt
\begin{eqnarray}
O^{(1)}_1&=&\bar \psi \psi \langle \hat \chi_+  X^3_-\rangle \ ,\label{eq:o1_1}\\
O^{(1)}_2&=&\bar \psi X^3_- \psi \langle \chi_+ \rangle \ ,\label{eq:o1_2}\\
O^{(1)}_3&=&\bar \psi [ \hat \chi_-, X^3_+]  \psi\ ,\label{eq:o1_3}\\
O^{(1)}_4&=&\bar \psi [ X^3_+, [u_\mu, u_\nu]]
\sigma_{\alpha\beta} \epsilon^{\mu\nu\alpha\beta} \psi\ ,\label{eq:o1_4}\\
O^{(1)}_5&=&\bar \psi X^3_- \psi \langle u_\mu u^\mu \rangle \ ,\label{eq:o1_5}\\
O^{(1)}_6&=&\bar \psi \sigma^{\mu\nu} \psi \langle X^3_- i[u_\mu,u_\nu] \rangle \ ,\label{eq:o1_6}\\
O^{(1)}_7&=&\bar \psi u_\mu X^3_- u^\mu \psi \ ,\label{eq:o1_7}\\
O^{(1)}_8&=&{1\over M}\bar \psi X^3_+ \gamma^\mu \gamma^5 D^\nu \psi \langle
u_\mu u_\nu \rangle + {\mathrm{h.c.}}\ ,\label{eq:o1_8}\\
O^{(1)}_{9}&=&{1\over M}\bar \psi ( u_\mu X^3_+ u_\nu\! +\! u_\nu X^3_+ u_\mu
)\gamma^\mu \gamma^5 D^\nu \psi\! +\! {\mathrm{h.c.}} \ ,\label{eq:o1_{9}}\\ 
O^{(1)}_{10}&=&{1\over M}\bar \psi X^3_- \gamma^\mu D^\nu \psi \langle
u_\mu u_\nu \rangle + {\mathrm{h.c.}}\ ,\label{eq:o1_{10}}\\
O^{(1)}_{11}&=&{1\over M}\bar \psi ( u_\mu X^3_-  u_\nu + u_\nu X^3_-  u_\mu )
\gamma^\mu D^\nu \psi + {\mathrm{h.c.}} \ ,\label{eq:o1_{11}}\\ 
O^{(1)}_{12}&=&{1\over M}\bar \psi [ h_{\mu\nu}, X^3_+] \gamma^\mu D^\nu \psi +
     {\mathrm{h.c.}} \ ,\label{eq:o1_{12}}\\
O^{(1)}_{13}&=&{1\over M}\bar \psi [ h_{\mu\nu}, X^3_-] \gamma^\mu \gamma^5 D^\nu \psi +
     {\mathrm{h.c.}} \ ,\label{eq:o1_{13}}\\
O^{(1)}_{14}&=&\bar \psi i[\hat  F_+^{\mu\nu}, X^3_+] \sigma^{\alpha\beta} \epsilon_{\mu\nu\alpha\beta} \psi
\ ,\label{eq:o1_{14}} \\ 
O^{(1)}_{15}&=&\bar \psi \sigma_{\mu\nu} \psi \langle \hat F_+^{\mu\nu} X^3_-\rangle \ ,\label{eq:o1_{15}}\\
O^{(1)}_{16}&=&\bar \psi  X^3_- \sigma_{\mu\nu} \psi \langle F_+^{\mu\nu}\rangle \ .\label{eq:o1_{16}}\\
O^{(1)}_{17}&=&\bar \psi \sigma_{\mu\nu} \psi \langle F_-^{\mu\nu}  X^3_+\rangle \ ,\label{eq:o1_{17}}\\
O^{(1)}_{18}&=&\bar \psi i[ F_-^{\mu\nu}, X^3_- ] \sigma^{\alpha\beta}\epsilon_{\mu\nu\alpha\beta}
\psi\ . \label{eq:o1_{18}}
\end{eqnarray}
\egroup
For operators $O^{(1)}_9$, $O^{(1)}_{11}$, and $O^{(1)}_{12}$,
we have not explicitly written down all possible isospin combinations, but 
have reported only the simplest ones.  When using these operators to construct
interaction vertices by expanding in powers of the pion field, all
allowed possibilities should be considered (see Appendix~\ref{app:lq2_1}).

\subsection{The $\Delta I=2$ sector}
\label{sec:delta2}
The $\Delta I=2$ operators have to be constructed as combinations
of ${\cal I}^{ab}(X^a_R \,O \, X^b_R\pm X^a_L\, O \, X^b_L)$, with $O$ 
an operator transforming as $O\to h\, O h^\dag$.  Since
${\cal I}^{ab}$ is diagonal and traceless, we have ${\cal I}^{ab}X^a_R\,
X^b_R= {\cal I}^{ab}X^a_L \,X^b_L=0$, therefore $O$
cannot be the identity.  Moreover, combinations like
${\cal I}^{ab}(X^a_R \, O\, X^b_L\pm X^a_R \, O\, X^b_L)$ are excluded since they do
not appear in the Standard Model weak Lagrangian~\cite{KS93}.
At order $Q$ there are two possible operators~\cite{KS93},
\begin{eqnarray}
O^{(2)}_{2V}&=&{\cal I}^{ab} \bar \psi ( X^a_R u_\mu X^b_R + X^a_L u_\mu X^b_L) \gamma^\mu
\psi\ , \label{eq:o2_2v}\\
O^{(2)}_{2A}&=&{\cal I}^{ab} \bar \psi ( X^a_R u_\mu X^b_R - X^a_L u_\mu X^b_L) \gamma^\mu \gamma^5
\psi\ . \label{eq:o2_2a}
\end{eqnarray}
At order $Q^2$, as discussed in Appendix~\ref{app:lq2_2},
we find the following eight possible operators:
\bgroup
\arraycolsep=0.5pt
\begin{eqnarray}
O^{(2)}_1&=&{\cal I}^{ab} \bar \psi X^a_R  \hat  \chi_+  X^b_R  \psi
    - (R\leftrightarrow L) \ , \label{eq:o2_1}\\
O^{(2)}_2&=& {\cal I}^{ab} \bar \psi 
    \Bigl[ u_\mu X^a_R u_\nu X^b_R + X^a_R u_\nu X^b_R u_\mu \nonumber \\
    &+& u_\nu X^a_R u_\mu X^b_R\! +\! X^a_R u_\mu X^b_R
   u_\nu\! -\! (R\leftrightarrow L)\Bigr] g^{\mu\nu}\psi
   \ , \label{eq:o2_2}\\
O^{(2)}_3&=&i\, {\cal I}^{ab} \bar \psi 
    \Bigl[ u_\mu X^a_R u_\nu X^b_R - X^a_R u_\nu X^b_R u_\mu \nonumber \\
    && \qquad\quad-u_\nu X^a_R u_\mu X^b_R + X^a_R u_\mu X^b_R
   u_\nu \nonumber \\
    && \qquad\quad - (R\leftrightarrow L)\Bigr]\sigma^{\mu\nu}\psi
   \ , \label{eq:o2_3}\\
O^{(2)}_4&=&i\, {\cal I}^{ab} \bar \psi 
    \Bigl[  X^a_R [u_\mu\, ,\, u_\nu] X^b_R - (R\leftrightarrow L)\Bigr]
    \sigma^{\mu\nu}\psi
   \ , \label{eq:o2_4}\\
O^{(2)}_5&=& \bar \psi 
    \{ D_\alpha \, , \,\overline{Y}^{(1)}_{+,\mu\nu}\}
    g^{\alpha\mu}\gamma^\nu\gamma^5 \psi
   \ , \label{eq:o2_5}\\
O^{(2)}_{6}&=& \bar \psi 
    \{ D_\alpha \, , \,\overline{Y}^{(1)}_{-,\mu\nu}\}
    g^{\alpha\mu}\gamma^\nu \psi
   \ , \label{eq:o2_6}\\
O^{(2)}_{7}&=&{\cal I}^{ab} \bar \psi \Bigl[ X^a_R F^+_{\mu\nu} X^b_R -
    (R\leftrightarrow L)  \Bigr]\sigma^{\mu \nu} \psi \ , \label{eq:o2_7}\\
O^{(2)}_{8}&=&{\cal I}^{ab} \bar \psi \Bigl[ X^a_R  F^-_{\mu\nu} X^b_R +
    (R\leftrightarrow L) \Bigr]\sigma^{\mu \nu} \psi \ , \label{eq:o2_8}
\end{eqnarray}
\egroup
where the quantities $\overline{Y}^{(1)}_{\pm,\, \mu\nu}$
are defined in Eq.~(\ref{eq:barY1}).  In this case too
for some of the operators we have not explicitly written down all possible isospin combinations,
but reported only the simplest ones (see Appendix~\ref{app:lq2_2}).

\subsection{Terms with only pionic degrees of freedom}
\label{sec:pidof}

Possible $\Delta I=0$ PV terms constructed with only pionic
degrees of freedom, namely terms involving $\nabla_\mu U$
together with $\epsilon^{\mu\nu\alpha\beta}$ factors,
turn out to vanish. Considering also terms involving the quantities
$F_{\mu\nu}^\pm$, we find at order $Q^4$ the following $P$- and $C$-odd terms:
\begin{eqnarray}
 &&\langle \nabla^\mu U^\dag\nabla^\nu U F_{\mu\nu}^R
         -\nabla^\mu U\nabla^\nu U^\dag F_{\mu\nu}^L\rangle
         \ ,\label{eq:pvpi1}\\
   && \langle UF_{\mu\nu}^R U^{\dagger}F^{R\mu\nu} -
         UF_{\mu\nu}^LU^{\dagger}F^{L\mu\nu} \rangle\ .\label{eq:pvpi1b}
\end{eqnarray}
Possible PV terms involving $\chi$, such as
$\langle  U\chi^\dag - \chi U^\dag\rangle$, are even under
$C$.

Terms with $\Delta I=1$ and  $\Delta I=2$ can be constructed
using the quantities $X_{L,R}^a$.  We find at order $Q^2$ 
the following two $P$- and $C$-odd operators
\begin{equation}
 O^{(1)}_{\pi\pi\pi}=\langle u_\mu X^3_- u^\mu\rangle\ ,\,\,
 O^{(2)}_{\pi\pi\pi}={\cal I}^{ab}\langle X^a_R u_\mu X^b_R
 u^\mu-(R\rightarrow L)\rangle\ . \label{eq:pvpi2}
\end{equation}
At lowest order, these terms give two three-pion vertices.  However,
their contribution to the PV potential is at least of order $Q^2$.  Therefore,
in the rest of the present work, we disregard the
contributions of these PV terms.

\subsection{Summary}
\label{sec:ham}
The $\chi$EFT PV Lagrangian up to order $Q^2$ includes all terms determined
above, each multiplied by a different low-energy constant (LEC), that is
\begin{eqnarray}
 \mathcal{L}^{PV} &=& \frac{h^1_{\pi}}{2\sqrt{2}}f_{\pi}\overline{\psi}
     X^3_-\psi +{h^0_V\over 2} \overline{\psi}\gamma^\mu u_\mu \psi\nonumber \\
     & +&  \frac{h_V^1}{4}\overline{\psi}\gamma^{\mu}\psi \langle u_{\mu} X^3_+ \rangle
   + \frac{h_A^1}{4}\overline{\psi}\gamma^{\mu}\gamma^5 \psi
   \langle u_{\mu} X^3_-\rangle \nonumber\\ 
    &-&{1\over 3}\, {\cal I}_{ab}\biggl[\frac{h_V^2}{2}\overline{\psi}
    \left(X^a_R u_\mu X^b_R+X^a_L u_\mu X^b_L\right)\gamma^\mu
      \psi\nonumber \\
    && \quad +\frac{h_A^2}{4}\overline{\psi} \left(X^a_R u_\mu X^b_R-X^a_L u_\mu
      X^b_L\right)\gamma^\mu \gamma^5 \psi \biggr]\nonumber\\
    && -\sum_{p=1,2} {h^{0}_p\over f_\pi} O^{(0)}_{p}
       -\sum_{p=1,18} {h^{1}_p\over f_\pi} O^{(1)}_{p}\nonumber \\
    &&   -\sum_{p=1,8} {h^{2}_p\over f_\pi} O^{(2)}_{p}
       + \mathcal{L}^{PV}_{CT}\ , \label{eq:lagKS} 
\end{eqnarray}
where for the LEC's multiplying the terms up to order $Q$ we have adopted the
notation of Ref.~\cite{Zhu00} (the different signs and numerical
factors account for our different definition of $u_\mu$, $X^3_-$ and
${\cal I}_{ab}$).  We have included in the last two lines the 28 terms of order $Q^2$
discussed in Subsec.~\ref{sec:delta0}, \ref{sec:delta1}, and \ref{sec:delta2}. The Lagrangian contains
also four-nucleon contact terms (included in $\mathcal{L}^{PV}_{CT}$),
representing interactions originating from excitation of
$\Delta$-resonances and exchange of heavy mesons. 
At lowest order, $\mathcal{L}^{PV}_{CT}$ contains five independent four-nucleon
contact terms with a single gradient, as discussed in Ref.~\cite{Girlanda08}.

The Lagrangian describing pion-nucleon interactions up to one derivative
was already given by Kaplan and Savage~\cite{KS93}.
The LEC $h^1_\pi$ is the long-sought pion-nucleon PV coupling constant.
By construction, it and all other LEC's are adimensional (the factor
$1/f_\pi$ has been introduced for convenience). 
In principle, the LEC's can be determined by fitting experimental data
or from lattice calculations (or from a combination of both methods).
The order of magnitude of the various constants is
\begin{equation}
  h\sim G_F f_\pi^2 \approx 10^{-7}\ ,\label{eq:hog}
\end{equation}
which is also the order of magnitude of PV effects in few-nucleon
systems.  

In the following, we also need the PC
Lagrangian up to order $Q^2$:
\bgroup
\arraycolsep=1.0pt
\begin{eqnarray}
  {\cal L}^{PC}&=& {\cal L}_{\pi\pi}^{(2)}+{\cal L}_{\pi\pi}^{(4)}+\ldots\nonumber\\
   &+&  {\cal L}_{N\pi}^{(1)}+{\cal L}_{N\pi}^{(2)}+{\cal
    L}_{N\pi}^{(3)}+\ldots+ {\cal L}^{PC}_{CT}\ ,\label{eq:Lpc} \\
  {\cal L}_{\pi\pi}^{(2)}&=& {f_\pi^2\over 4} \langle\nabla_\mu U^\dag
  \nabla^\mu U + \chi^\dag U+\chi U^\dag\rangle\ , \label{eq:Lpc_pipi2}\\
  {\cal L}_{\pi\pi}^{(4)} &= & {1\over 16}\ell_3 \Bigl[\langle \chi^\dag
    U+\chi U^\dag\rangle\Bigr]^2\nonumber \\
      &&+ {1\over 8} \ell_4\Bigl[ \langle \nabla_\mu U^\dag \nabla^\mu
       U\rangle \langle \chi^\dag U+ \chi U^\dag\rangle \nonumber\\
    &&\qquad - {1\over 2} (\langle \chi^\dag U+ \chi U^\dag\rangle)^2\Bigr]
       +\ldots \ , \label{eq:Lpc_pipi4}\\
  {\cal L}_{N\pi}^{(1)} &=& \overline{\psi}\Bigl(i\gamma^\mu D_\mu -M
  +{g_A\over 2} \gamma^\mu\gamma^5 u_\mu \Bigr)\psi\
  , \label{eq:Lpc_pin1}\\
   {\cal L}_{N\pi}^{(2)}  &=& c_1 \overline{\psi}\langle \chi_+\rangle
   \psi +\ldots \ ,\label{eq:Lpc_pin2}\\
  {\cal L}_{N\pi}^{(3)}  &=& d_{16} \overline{\psi}{1\over 2}
  \gamma^\mu\gamma^5 u_\mu \langle \chi_+\rangle \psi\nonumber \\
  &+&  d_{18} \overline{\psi}{i\over 2}
  \gamma^\mu\gamma^5 [D_\mu, \chi_-] \psi+\cdots\ ,\label{eq:Lpc_pin3}
\end{eqnarray}
\egroup
where we have omitted terms not relevant in the present work (the complete
${\cal L}_{\pi\pi}^{(4)}$ can be found in
Ref.~\cite{GL84} and the complete $ {\cal L}_{N\pi}^{(2)}$ and ${\cal
  L}_{N\pi}^{(3)}$ in Ref.~\cite{Fettes00}).  Four-nucleon contact terms (see,
for example, Refs.\cite{Epelbaum09,ME11}) are lumped into ${\cal L}^{PC}_{CT}$.
The parameters $\ell_3$, $\ell_4$, $c_1$, $d_{16}$, and $d_{18}$ are
LEC's entering the PC Lagrangian. To this Lagrangian, we add two mass counterterms
\begin{equation}
  {\cal L}_{MCT}=-{1\over 2} \delta m_\pi^2\, \vec \pi^{\, 2} - \delta M\,
  \overline{\psi} \psi\ ,\label{eq:Lct}
\end{equation}
which renormalize  the pion ($m_\pi$) and nucleon ($M$) masses in ${\cal L}^{PC}$.
The determination of $\delta M$ and $\delta
m^2$ is discussed in Appendix~\ref{app:pvnn}.
\section{The PV potential up to order $Q$}
\label{sec:pvnn}
In this section, we discuss the derivation of the PV $NN$ potential at $\nnlo$.
First, we provide, order by order in the
power counting, formal expressions for it
in terms of time-ordered perturbation theory (TOPT) amplitudes,
and next discuss the various
diagrams associated with these amplitudes (additional details
are given in Appendix~\ref{app:pvnn}).

\subsection{From amplitudes to potentials}
\label{sec:topt}
We begin by considering the conventional perturbative expansion for
the $NN$ scattering amplitude
\begin{eqnarray}
 \!\!\!\!&&\langle N^\prime N^\prime \mid T\mid NN\rangle \nonumber\\ 
\!\!\!\!&&= \langle N^\prime N^\prime \mid H_I \sum_{n=1}^\infty \left( 
 \frac{1}{E_i -H_0 +i\, \eta } H_I \right)^{n-1} \mid NN \rangle \ ,
\label{eq:pt}
\end{eqnarray}
where $\mid NN \rangle$ and $\mid N^\prime N^\prime \rangle$ represent the initial and final
two-nucleon states of energy $E_i$, $H_0$ is the Hamiltonian
describing free pions and nucleons, and $H_I$ is the Hamiltonian
describing interactions among these particles.  The evaluation of this amplitude
is carried out in practice by inserting complete sets of $H_0$ eigenstates
between successive $H_I$ terms.   Power counting is then used
to organize the expansion in powers of $Q/\Lambda_\chi \ll 1 $, where
$\Lambda_\chi \simeq 1$ GeV is the typical hadronic mass scale,
\begin{equation}
 \langle N^\prime N^\prime |T|NN\rangle=\sum_n T^{(n)} ,\label{eq:teft}
\end{equation}
where $T^{(n)}\sim Q^n$.  We note that in Eq.~(\ref{eq:pt}) 
the interaction Hamiltonian $H_I$ is in the Schr\"odinger picture and 
that, at the order of interest here, it follows
simply from $H_I=-\int {\rm d} \bmx\; {\cal L}_I (t=0,\bmx)$.  Vertices from $H_I$
are listed in Appendix~\ref{app:vertex}.

We obtain the $NN$ potential $V$ by requiring that iterations of it
in the Lippmann-Schwinger (LS) equation
\begin{equation}
V+V\, G_0\, V+V\, G_0 \, V\, G_0 \, V+\dots \ ,
\label{eq:lse}
\end{equation}
lead to the $T$-matrix in Eq.~(\ref{eq:teft}), order by order in the power counting.
In practice, this requirement can only be satisfied up to a given order $n^*$, and
the resulting potential, when inserted into the LS equation, will
generate contributions of order $n > n^*$, which do not match $T^{(n)}$.
In Eq.~(\ref{eq:lse}), $G_0$ denotes the free two-nucleon propagator, $G_0=1/(E_i-H_0+i\, \eta)$,
and we assume that
\begin{equation}
\langle N^\prime N^\prime |V|NN\rangle=\sum_n V^{(n)}\ ,
\end{equation}
where the yet to be determined $V^{(n)}$ is of order $Q^n$.  We also note that, generally, a term
like $\left[ V^{(m)}\, G_0 \, V^{(n)}\right]$ is of order $Q^{m+n+1}$, since $G_0$ is of order $Q^{-2}$
and the implicit loop integration brings in a factor $Q^3$ (for a more detailed discussion
see Ref.~\cite{Pastore11}).

We now consider the case of interest here, in which the two nucleons interact via
a PC potential plus a very small PV component.  The $\chi$EFT Hamiltonian
implies the following expansion in powers of $Q$ for $T= T_{PC} + T_{PV}$:
\begin{eqnarray}
 T_{PC}&=&T^{(0)}_{PC}+T^{(1)}_{PC}+T^{(2)}_{PC}+\ldots ,\label{tpc}\\
 T_{PV}&=&T^{(-1)}_{PV}+T^{(0)}_{PV}+T^{(1)}_{PV}+\ldots .\label{tpv}
\end{eqnarray}
We assume that $V= V_{PC}+ V_{PV}$ have a similar expansion,
\begin{eqnarray}
 V_{PC} & = & V_{PC}^{(0)}+V_{PC}^{(1)}+V_{PC}^{(2)}+\dots \\
 V_{PV} & = & V_{PV}^{(-1)}+V_{PV}^{(0)}+V_{PV}^{(1)}+\dots ,
\end{eqnarray}
and to linear terms in $V_{PV}$ we find
\begin{eqnarray}
T & = & V+VG_0V+VG_0 VG_0 V+\cdots  \nonumber\\
  & = & V_{PC}+V_{PV}\nonumber\\
  & + & V_{PC} G_0 V_{PC}\!+\!V_{PV} G_0 V_{PC}\!+\!V_{PC} G_0 V_{PV}\nonumber\\
  & + & V_{PC} G_0 V_{PC} G_0 V_{PC}\!+\! V_{PV} G_0 V_{PC} G_0 V_{PC}\nonumber\\
  & + & V_{PC} G_0 V_{PV} G_0  V_{PC}\!+\!V_{PC} G_0  V_{PC} G_0 V_{PV}\!+\!\cdots\ .\label{eq:tv1}
\end{eqnarray}
By matching up to the order $Q^2$ for $V_{PC}$ and $Q^1$ for
$V_{PV}$, we obtain for the PC potential
\bgroup
\arraycolsep=1.0pt
\begin{eqnarray}
 V_{PC}^{(0)} & = & T_{PC}^{(0)}\ ,\label{eq:vpc0}\\
 V_{PC}^{(1)} & = &
 T_{PC}^{(1)}-\left[V_{PC}^{(0)}G_0 V_{PC}^{(0)}\right]\ , \label{eq:vpc1}\\
 V_{PC}^{(2)}  & = &  T_{PC}^{(2)}-\left[V_{PC}^{(0)}G_0 V_{PC}^{(1)}\right]
               -\left[V_{PC}^{(1)}G_0 V_{PC}^{(0)}\right]\nonumber\\
          & - & \left[V_{PC}^{(0)}G_0 V_{PC}^{(0)}G_0 V_{PC}^{(0)}\right]
             \ , \label{eq:vpc2}
\end{eqnarray}
and for the PV one
\begin{eqnarray}
 V_{PV}^{(-1)} & = & T_{PV}^{(-1)}\ ,\label{eq:vpvml}\\
 V_{PV}^{(0)}  & = &  T_{PV}^{(0)}-\left[V_{PV}^{(-1)}G_0 V_{PC}^{(0)}\right]
                     -\left[V_{PC}^{(0)}G_0 V_{PV}^{(-1)}\right]\ ,\label{eq:vpv0}\\
 V_{PV}^{(1)}  & = &  T_{PV}^{(1)}-\left[V_{PV}^{(0)}G_0 V_{PC}^{(0)}\right]
               -\left[V_{PC}^{(0)}G_0 V_{PV}^{(0)}\right]\nonumber\\
          & - & \left[V_{PV}^{(-1)}G_0 V_{PC}^{(1)}\right]-
                \left[V_{PC}^{(1)}G_0 V_{PV}^{(-1)}\right]\nonumber\\
          & - & \left[V_{PV}^{(-1)}G_0 V_{PC}^{(0)}G_0 V_{PC}^{(0)}\right]
             - \left[V_{PC}^{(0)}G_0 V_{PV}^{(-1)}G_0 V_{PC}^{(0)}\right]
             \nonumber\\
         & - & \left[V_{PC}^{(0)}G_0 V_{PC}^{(0)}G_0 V_{PV}^{(-1)}\right]
             \ .\label{eq:vpv1}
\end{eqnarray}
\egroup
The expressions above relate $V^{(n)}_{PC}$ and $V^{(n)}_{PV}$  to the $T^{(n)}_{PC}$ and $T^{(n)}_{PV}$ amplitudes.

\subsection{The PV potential}
\label{sec:pvp}

It is convenient to define the following momenta
\begin{eqnarray}
  &&\bmK_j=(\bmp_j'+\bmp_j)/2\ , \quad
  \bmk_j=\bmp_j'-\bmp_j\ ,\label{eq:notjb1}
 \end{eqnarray}
where $\bmp_j$ and $\bmp'_j$ are the initial and final momenta
of nucleon $j$.  Since $\bmk_1=-\bmk_2=\bmk$ from overall momentum
conservation $\bmp_1+\bmp_2=\bmp_1'+\bmp_2'$, the momentum-space
matrix element of the potential $V$ is a function of the momentum
variables $\bmk$, $\bmK_1$, and $\bmK_2$, namely
\begin{equation}
 \langle\a_1'\a_2'| V_{PV}| \a_1\a_2\rangle = {1\over \Omega}
  V_{PV}(\bmk,\bmK_1,\bmK_2) \delta_{\bmp_1+\bmp_2,\bmp_1'+\bmp_2'}\ ,
  \label{eq:widetildev}
\end{equation}
where $\alpha_j\equiv\{\bmp_j,s_j,t_j\}$ denotes the momentum, spin
projection, and isospin projection of nucleon $j$, and 
the various momenta are discretized by assuming periodic boundary
conditions in a box of volume $\Omega$. Moreover, we can write
\begin{equation}
  V_{PV}(\bmk,\bmK_1,\bmK_2)= V_{PV}^{(CM)}(\bmk,\bmK)+ 
   V^{(\bmP)}_{PV}(\bmk,\bmK)\ ,\label{eq:widetildev2}
\end{equation}
where $\bmK=(\bmK_1-\bmK_2)/2$, $\bmP=\bmp_1+\bmp_2=\bmK_1+\bmK_2$, and 
the term $ V^{(\bmP)}_{PV}(\bmk,\bmK)$ represents boost
corrections to $  V^{(CM)}_{PV}(\bmk,\bmK)$~\cite{Girlanda10}, the potential
in the center-of-mass (CM) frame.  Below we ignore these boost corrections
and provide expressions for $ V^{(CM)}_{PV}(\bmk,\bmK)$ only.

\begin{figure}[t]
   \includegraphics[scale=.6,clip]{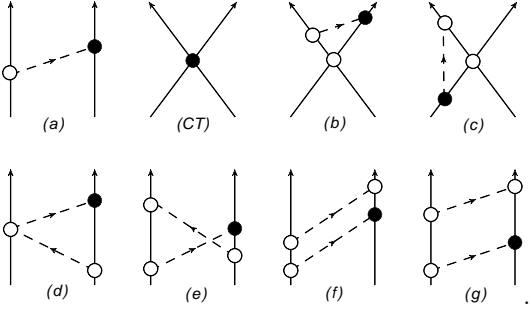}.
   \caption{   \label{fig:pvdiag}
    Time-ordered diagrams contributing to the PV
     potential (only a single time ordering is shown).
    Nucleons and pions are denoted by solid and dashed lines, respectively. The
    open (solid) circle represents a PC (PV) vertex.}
\end{figure}

\begin{figure}[h]
   \includegraphics[scale=.5,clip]{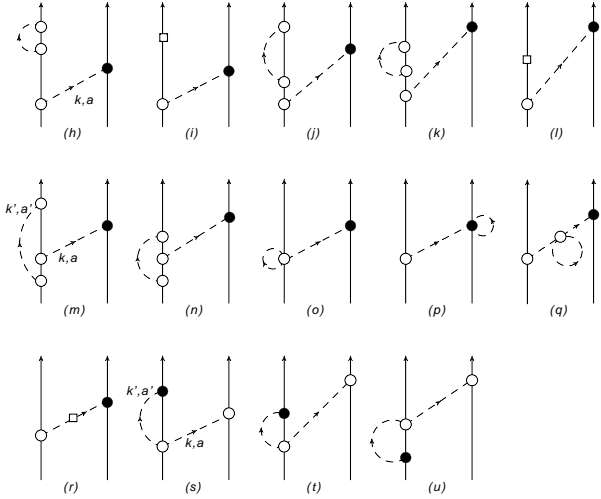}
   \caption{   \label{fig:operen}
     Diagrams contributing to the OPE potential up to order $Q^2$
     (notation as in Fig.~\protect{\ref{fig:pvdiag}}).  Only a subset of the possible
     time-ordered diagrams is shown.}
\end{figure}

Diagrams contributing to the PV potential are shown in
Figs.~\ref{fig:pvdiag} and~\ref{fig:operen}---the diagrams reported in this latter
figure contribute to the renormalization of the LEC
$h^1_\pi$.  An analysis of the $\pi NN$ vertex corrections 
was already carried out in Ref.~\cite{Zhu01} (in that paper the choice $\alpha=1/6$ in
Eq.~(\ref{eq:uumatrix}) was adopted), and we have verified that we obtain
identical expressions to those reported in that work.
Contributions given by the various diagrams are reported
in Appendix~\ref{app:pvnn}.  In this section, we only list the final expression for the
PV potential $V^{(CM)}_{PV}$ as (the dependence on the momenta $\bmk$ and $\bmK$ is
understood)
\begin{eqnarray}
    V^{(CM)}_{PV}&=&
   V^{ ({\rm OPE})}_{PV} +
   V^{ ({\rm TPE})}_{PV} \nonumber\\
   &+&  V^{ ({\rm RC})}_{PV} +
   V^{ ({\rm CT})}_{PV} \ ,\label{eq:dec}
\end{eqnarray}
namely as a sum of terms due to one-pion exchange (OPE), two-pion exchange
(TPE), relativistic corrections (RC), and contact contributions (CT).  Following
the discussion reported in Appendix~\ref{app:pvnn},
the OPE term collects i) the non-relativistic (NR) LO contribution
$V^{(-1)}({\rm NR})$ of diagram (a) in Fig.~\ref{fig:pvdiag},
namely
\begin{equation}
 V^{(-1)}({\rm NR}) =  
{g_A h^1_\pi\over 2\sqrt{2}f_\pi}\, (\vec\tau_1\times\vec\tau_2)_z\;
   {i\, \bmk\cdot(\bmsi_1+\bmsi_2)\over \omega_k^2}\ ,\label{eq:opelo}
\end{equation}
where $\omega_k^2=k^2+m_\pi^2$, 
ii) part of the contribution due to the order $Q^2$ pion-nucleon 
interactions given in Eq.~(\ref{eq:opeL}) (the term proportional to
$V^{(-1)}({\rm NR})$), 
and iii) the various contributions coming from the
diagrams shown in Fig.~\ref{fig:operen}, explicitly
\begin{eqnarray}
  V^{ ({\rm OPE})}_{PV}
   &=& V^{(-1)}({\rm NR}) \biggl[ 1 +
            {2 \, m_\pi^2 \over g_A} \,(2\, d_{16} - d_{18})\nonumber\\
   && - {8 \sqrt{2}\, m^2_\pi\over h^1_\pi
             f_\pi^2}\, (h^{1}_2-h^{1}_3)
      -{2\over 3} {g^2_A\over f^2_\pi} J_{13} 
     \nonumber \\
   &&  - {20\, \alpha-1\over 4f_\pi^2} J_{01}
       -2\, \ell_4 {m^2_\pi\over f^2_\pi}  \nonumber \\
   &&  -{1-10\, \alpha\over 2 f_\pi^2} J_{01}+{1\over
      4f_\pi^2} J_{01}\biggr]\ ,
   \label{eq:opepvren}
\end{eqnarray}
where the (infinite) constants $J_{nm}$ are
\begin{equation}
  J_{mn}= \int \frac{ {\rm d} \bmk}{(2\pi)^3}\, {k^{2m} \over \omega_k^n}\ .\label{eq:funJ}
\end{equation}
In the expression above there is a term proportional 
to $(m_\pi/M)^2$ coming from 
$V^{(1)}({\rm RC})$ which we ignore for simplicity---see 
Appendix~\ref{app:pvnn} and Eq.~(\ref{eq:opennlo}) for more details
on its origin.
Note the cancellation of the terms proportional to $J_{01}$, which removes
the dependence on $\alpha$.  The renormalized OPE potential reads
\begin{equation}
  V^{ ({\rm OPE})}_{PV}
   =   {\overline{g}_A \overline{h}^1_\pi\over 2\sqrt{2}\;\overline{f}_\pi}
        (\vec\tau_1\times\vec\tau_2)_z\; {i\, \bmk\cdot(\bmsi_1+\bmsi_2)\over \omega_k^2} 
   \label{eq:opepvren2}
\end{equation}
where 
\begin{eqnarray}
{\overline{g}_A \overline{h}^1_\pi\over 2\sqrt{2}\;\overline{f}_\pi}&=&
 {g_A h^1_\pi\over 2\sqrt{2}f_\pi } \biggl[ 1 +
            {2 \, m_\pi^2 \over g_A} (2\, d_{16} - d_{18})\nonumber\\
   && - {8 \sqrt{2}\, m^2_\pi\over h^1_\pi
             f_\pi^2}(h^{1}_2-h^{1}_3)
    -{2\over 3} {g^2_A\over f^2_\pi} J_{13}  \nonumber\\
   &&  -2\, \ell_4 {m^2_\pi\over      f^2_\pi}\biggr] \ .\label{eq:hgfren}
\end{eqnarray}
Here the overlined quantities are the renormalized coupling constants
(up to corrections of order $Q^2$).
Performing a similar analysis for the PC OPE potential, 
we find (see also Ref.~\cite{Epel03}):
\bgroup
\arraycolsep=1.0pt
\begin{eqnarray}
  V^{ ({\rm OPE})}_{PC}&=& 
   V^{({\rm OPE},0)}_{PC} 
     \biggl[ 1 -{2\over 3} {g^2_A\over f^2_\pi} J_{13}   \nonumber \\
   && + {4 \,m_\pi^2 \over g_A} (2\, d_{16} - d_{18})
      -2\, \ell_4 {m_\pi^2\over f_\pi^2}\biggr]\ ,
\end{eqnarray}
where
\begin{equation}
    V^{({\rm  OPE},0)}_{PC}=
    -{g^2_A\over 4 f^2_\pi}  \vec\tau_1\cdot\vec\tau_2\, {
   \bmsi_1\cdot\bmk \,\, \bmsi_2\cdot\bmk \over \omega_k^2 }
    \ ,\label{eq:opepc}
\end{equation}
is the PC OPE potential at order $Q^0$. As usual~\cite{Epel03}, the singular 
part coming from $J_{13}$ is absorbed by the LEC's $d_{16}$ and $\ell_4$. The LEC
$d_{18}$ is assumed to have only  a finite part fixed by the
Goldberger-Treiman anomaly.  In summary, the PC OPE potential up to order $Q^2$
is written as
\begin{eqnarray}
    V^{ ({\rm OPE})}_{PC}&=& 
   -{ \overline{g}_A^2\over 4\, \overline{f}^2_\pi}\left(1-{4\, m_\pi^2\, d_{18}\over
      g_A}\right)\nonumber\\
   && \times\,  \vec\tau_1\cdot\vec\tau_2 \, {
   \bmsi_1\cdot\bmk\ \,\, \bmsi_2\cdot\bmk  \over \omega_k^2}
    \ ,\label{eq:pcope2}
\end{eqnarray}
where the renormalized ratio $\overline{g}_A/\overline{f}_\pi$ is given by
\begin{eqnarray}
 { \overline{g}_A\over \overline{f}_\pi }&=&  {g_A\over f_\pi} \biggl( 1
  -{1\over 3} {g^2_A\over f^2_\pi}  J_{13}
    + {4 \, m_\pi^2 \over g_A} d_{16} 
    -\ell_4 {m_\pi^2\over f_\pi^2}\biggr)\ .\label{eq:gfren}
\end{eqnarray}
In the right-hand side of Eqs.~(\ref{eq:hgfren}), (\ref{eq:pcope2}) and~(\ref{eq:gfren}),
$g_A$ and $f_\pi$ can be
replaced by the renormalized (physical) values $\overline{g}_A$ and
$\overline{f}_\pi$, which is correct at this order. 
\egroup
The constant $J_{13}$ is given in dimensional regularization in Eq.~(B38) of
Ref.~\cite{Pastore09},
\begin{equation}
  J_{13}= {3 m^2_\pi\over 8 \pi^2}(d_\epsilon+1)\ ,\quad
  d_\epsilon =  -{2\over \epsilon}+\gamma-\ln
  4\pi +\ln\left({m^2_\pi\over\mu^2}\right)-{4\over
      3} \ ,
\end{equation}
where $\gamma\approx0.5772\cdots$, $\epsilon=3-d$, $d$ being the number of dimensions ($d\rightarrow
3$), and $\mu$ a renormalization scale.  Absorbing $d_\epsilon$ in both
$d_{16}$ and $\ell_4$, we have
\begin{equation}
   { \overline{g}_A\over \overline{f}_\pi }=  {g_A\over f_\pi} \biggl( 1
  - {g_A^2\over f^2_\pi}\, { m_\pi^2\over 8\, \pi^2} 
  + {4\, m_\pi^2 \over g_A} \, \overline{d}_{16}  -\overline{\ell}_4\, {m_\pi^2\over
    f_\pi^2}\biggr)\ ,\label{eq:gfren2} 
\end{equation}
where $\overline{d}_{16}$ and $\overline{\ell}_4$ are the finite
parts of the two corresponding LEC's.  This expression coincides with Eq.~(2.44) of
Ref.~\cite{Epel03}, but for a factor 2 in the second term on the r.h.s.~of
the above equation.  However, this difference is of no import, since the final result depends on
quantities absorbed in the LEC's $d_{16}$ and $\ell_4$ (namely, on
the definition of $d_\epsilon$). 

Using Eq.~(\ref{eq:gfren}) for the renormalized (at order $Q^2$)
ratio $\overline{g}_A/\overline{f}_\pi$, we can extract from
Eq.~(\ref{eq:hgfren}) the expression of the renormalized coupling
constant $h^1_\pi$:
\begin{eqnarray}
  \overline{h}^1_\pi&=&  h^1_\pi \biggl[ 1
    -{1\over 3} { g^2_A\over f^2_\pi } J_{13}  -{2\,
      m_\pi^2\over g_A} d_{18}  - \ell_4 {m_\pi^2\over
      f_\pi^2}\nonumber \\
      && - {8 \sqrt{2}\, m^2_\pi\over h^1_\pi
             f_\pi^2}(h^{1}_2-h^{1}_3)  \biggr]\ ,\label{eq:hren} 
\end{eqnarray}
and the infinite part coming from $ J_{13}$ can be reabsorbed by $\ell_4$, 
$h^{1}_2$, and $h^{1}_3$.

The potential $V^{({\rm TPE})}_{PV}$ comes from the regular contributions of panels
(d)-(g) of Fig.~\ref{fig:pvdiag} reported in 
Eqs.~(\ref{eq:d2R}) and ~(\ref{eq:box2R}), 
\bgroup
\arraycolsep=1.0pt
\begin{eqnarray}
   V^{({\rm TPE})}_{PV}\!&=&\!
  -{g_A h^1_\pi\over 2\sqrt{2} f_\pi \Lambda_\chi^2} 
     (\vec\tau_1\times\vec\tau_2)_z\,
          i\, \bmk\cdot(\bmsi_1+\bmsi_2)\; L(k) \nonumber \\
  &&-{g_A h^1_\pi\over 2\sqrt{2}f_\pi}\, {g_A^2\over \Lambda_\chi^2}
     \biggl[4\, \left(\tau_{1z}+\tau_{2z}\right)\, i\, \bmk\cdot(\bmsi_1\times\bmsi_2)\;
     L(k)\nonumber \\
  && \qquad\qquad +(\vec\tau_1\times\vec\tau_2)_z\,
     i\, \bmk\cdot(\bmsi_1+\bmsi_2)\nonumber\\
  && \qquad\qquad\times \bigl[H(k)-3\, L(k)\bigr]\biggr]\ ,\label{eq:tpe}
\end{eqnarray}
where the loop functions $L(k)$ and $H(k)$ are defined
in Eqs.~(\ref{eq:sL}) and~(\ref{eq:H}).
The TPE potential reported above is in agreement with the 
expression derived in Refs.~\cite{Zhu05,Liu07,Desp08,Vries13}.
In this and following $\nnlo$ terms of the potential, the coupling constants $g_A$,
$f_\pi$, and $h^1_\pi$ can be replaced by the corresponding
physical (renormalized) values.

The potential $V^{({\rm RC})}_{PV}$ coincides, except for the term
proportional to $k^2$ which is reabsorbed in the OPE (as discussed
above) and CT (see below) parts,
with the quantity $V^{(1)}({\rm RC})$ given in Eq.~(\ref{eq:opennlo}), namely
\begin{eqnarray}
 V^{({\rm RC})}_{PV} &=&
  {g_A h^1_\pi\over 2\sqrt{2}f_\pi }\,  {1\over
    4\, M^2}
  (\vec\tau_1\times\vec\tau_2)_z\;   {1 \over \omega_k^2}\nonumber \\
  && \times\,   \bigl[ -4\,i\, K^2\, \bmk\cdot(\bmsi_1+\bmsi_2) \nonumber \\
  && \quad+\,\bmk\cdot\bmsi_1\; (\bmk\times\bmK)\cdot\bmsi_2  \nonumber \\
  && \quad+\, \bmk\cdot\bmsi_2\;
                  (\bmk\times\bmK)\cdot\bmsi_1\bigr] \ .\label{eq:um} 
\end{eqnarray}

Lastly, the potential $V^{({\rm CT})}_{PV}$, derived from the
$4N$ contact diagrams (CT) of Fig.~\ref{fig:pvdiag}, reads
\begin{eqnarray}
 V^{({\rm CT})}_{PV} &=&
    \frac{1}{\Lambda_{\chi}^2f_\pi} \Bigl[ C_1\,
      i\, (\bmsi_1\times\bmsi_2)\cdot\bmk   \nonumber\\
 &&+  C_2 \, \vec\tau_1\cdot\vec\tau_2\, 
    i\, (\bmsi_1\times\bmsi_2)\cdot\bmk   
    \nonumber\\
 &&+ \overline{C_3}\, 
    (\vec\tau_1\times \vec\tau_2)_z \, i\, (\bmsi_1+\bmsi_2)\cdot\bmk   
    \nonumber\\
 && + \overline{C_4} \, 
  (\tau_{1z} + \tau_{2z})\,  i\, (\bmsi_1\times\bmsi_2)\cdot\bmk
    \nonumber\\
 &&+ C_5\, 
    {\cal I}^{ab}\, \tau_{1a}\, \tau_{2b}\, 
    i\, (\bmsi_1\times\bmsi_2)\cdot\bmk\Bigr] \ ,\label{eq:ctren}
\end{eqnarray}
\egroup
where the $\overline{C_i}$ are the renormalized LEC's, in
which the infinite constants coming from the evaluation (in dimensional
regularization) of the TPE diagrams---the
terms given in Eqs.~(\ref{eq:d2S}) and (\ref{eq:box2S})---as well as
the contribution proportional to $\omega_k^2$ in Eq.~(\ref{eq:opennlo}) 
and the contribution proportional to $h^{1}_{12}$ in Eq.~(\ref{eq:opeL}) 
have been reabsorbed (see Appendix~\ref{sec:sub_pvnn} for more details).

In the applications discussed in Sec.~\ref{sec:res}, the configuration-space
version of the potential is needed.  This formally follows from 
\bgroup
\arraycolsep=1.0pt
\begin{eqnarray}
 \!\!\!\!\!\!\!\!\!\!\!\! \langle \bmr_1'\bmr_2'|V|\bmr_1\bmr_2\rangle&=&
  \delta (\bmR-\bmR') \int {d^3k\over (2\pi)^3} {d^3K\over (2\pi)^3}
  \nonumber\\ 
 &\times& e^{i(\bmK+\bmk/2)\cdot\bmr'} V(\bmk,\bmK)
    e^{-i(\bmK-\bmk/2)\cdot\bmr}\ ,
    \label{eq:vrsp}
\end{eqnarray}
\egroup
where $\bmr=\bmr_1-\bmr_2$ and $\bmR=(\bmr_1+\bmr_2)/2$,
and similarly for the primed variables.
In order to carry out the Fourier transforms above, 
the integrand is regularized by including a cutoff of the form
\begin{equation}
   C_\Lambda(k)={\rm e}^{-(k/\Lambda)^4}\ ,\label{eq:cutoff}
\end{equation}
where the cutoff parameter $\Lambda$ is taken in the range
500--600 MeV.  With such a choice the OPE, TPE, and CT
components of the resulting potential are local, i.e.,
$\langle \bmr_1'\bmr_2'|V|\bmr_1\bmr_2\rangle=
\delta (\bmR-\bmR')\, \delta (\bmr-\bmr') V(\bmr)$, while
the RC component contains mild non-localities
associated with linear and quadratic terms in the
relative momentum operator $-i \bmna$.
Explicit expressions for all these
components are listed in Appendix~\ref{app:pvr}.
\section{Results}
\label{sec:res}

In this section, we report results for PV observables in the $A$=2--4 systems.
The $A$=2 calculations are based on the (weak interaction) PV $NN$ potential
derived in the previous section (and summarized in
Appendix~\ref{app:pvr}) and on the (strong interaction) PC $NN$
potential obtained by Entem and Machleidt~\cite{EM03,ME11}
at next-to-next-to-next-to-leading order (N3LO).  This potential is regularized
with a cutoff function depending on a parameter $\Lambda$---its
functional form, however, is different from that adopted here for $V_{PV}$.  Below
we consider the two versions corresponding to $\Lambda$=500 MeV and $\Lambda$=600 MeV, 
labeled N3LO-500 and N3LO-600, respectively.  The $A$=3 and 4 calculations
also include a (strong interaction) PC three-nucleon ($NNN$) potential
derived in $\chi$EFT at  $\nnlo$~\cite{Eea02}.  It too depends on a cutoff parameter 
$\Lambda$, and here we use values for it which are consistent
with those in the PC and PV $NN$ potentials.  
These three-nucleon potentials, labeled, respectively, N2LO-500 and N2LO-600,  
depend in addition on two unknown LEC's, denoted as $c_D$ and $c_E$. 
In this work, they have been determined by reproducing
the $A$=3 binding energies and the Gamow-Teller matrix element in 
tritium $\beta$-decay~\cite{Gazit,Mea12}.  Their values
are listed in Table~\ref{tab:par}.

\begin{table}[t]
\caption{\label{tab:par}
  The $NN$ and $NNN$ chiral potentials used in this work.  In columns 2--4 the values for
  the cutoff parameter $\Lambda$ (in MeV) and adimensional coefficients $c_D$ and $c_E$ in
  the $NNN$ potential are listed.  In the last column the
binding energies predicted for $\heq$ are reported corresponding to the two sets.}
\begin{center}
\begin{tabular}{lcccccc}
\hline 
PC interactions & $\Lambda$  & $c_D$ & $c_E$ & B$(\heq)$ \\
  & [MeV] & & & [MeV] \\
\hline
N3LO/N2LO-500  & $500$ & $-0.12$ & $-0.196$ & $28.49$ \\
N3LO/N2LO-600  & $600$ & $-0.26$ & $-0.846$ & $28.64$ \\
\hline
\end{tabular}
\end{center}
\end{table}

The final expression of the potential is
given in Eq.~(\ref{eq:pvnnr}).  The component $V^{({\rm OPE})}$
is the LO term (of order $Q^{-1}$), although the coupling
constants contain also contributions from the N2LO (order $Q^1$) vertex
corrections, while the components RC, TPE, and CT are N2LO terms.
In the following, the values $\overline{g}_A=1.267$ and
$\overline{f}_\pi=92.4$ MeV are adopted. 

This section is organized as follows.  In Sec.~\ref{sec:ddh}, we provide estimates
for the renormalized LEC's $h^1_\pi$ and $C_i$ (with the overline omitted for brevity)
entering the PV potential,
using a resonance saturation model, in practice by exploiting what is known about
the DDH parameters~\cite{DDH}.  In Sec.~\ref{sec:pp}, we provide constraints
for some of the LEC's by fitting currently available measurements 
of the $\vec p$-$p$ longitudinal asymmetry.  In Secs.~\ref{sec:spinrot}
and~\ref{sec:n3he} we present a study of, respectively, spin-rotation effects
in $n$-$p$ and $n$-$d$ scattering, and of the
longitudinal asymmetry in the $^3$He($\vec{n},p$)$^3$H reaction.

\subsection{Estimates of the LEC's} 
\label{sec:ddh}

In Ref.~\cite{DDH} the pion-nucleon PV coupling constant $h^1_\pi$
was estimated to vary in the ``reasonable range'' $0\le h_\pi^1\le 11.4\times
10^{-7}$ with the ``best value'' $h_\pi^1=4.56\times
10^{-7}$.  More recently, a lattice calculation
has estimated $h_\pi^1\approx1\times10^{-7}$~\cite{Wasem2012} .
Therefore, in the following we perform calculations for three
values of $h^1_\pi$:
\begin{enumerate}
\item $h_\pi^1=1\times 10^{-7}$  (lattice estimate);
\item $h_\pi^1=4.56\times 10^{-7}$ (DDH ``best value'');
\item $h_\pi^1=11.4\times 10^{-7}$ (maximum value allowed in the DDH ``reasonable
  range'').
\end{enumerate}
In order to estimate the LEC's $C_i$
in $V^{({\rm CT})}_{PV}$, we match the components of the DDH potential
mediated by $\rho$ and $\omega$ exchanges to those
of $V^{({\rm CT})}_{PV}$, and obtain in the limit $k\ll m_\rho$, $m_\omega$:
\begin{eqnarray}
 C^{({\rm DDH})}_1 &=& -{3\over 2}g_\rho h^0_\rho D_\rho
         - {3\over 2} g_\omega h^0_\omega D_\omega\ ,
         \label{eq:c1ddh} \\
 C^{({\rm DDH})}_2 &=& - g_\rho h^0_\rho
   \left({1\over2}+\kappa_\rho\right) D_\rho 
        -{1\over 2}g_\omega h^0_\omega
         D_\omega\ ,
         \label{eq:c2ddh} \\
 C^{({\rm DDH})}_3 &=& -{1\over 2}g_\rho (h^{1'}_\rho-h^1_\rho)
        D_\rho-{1\over 2}g_\omega  h^1_\omega 
          D_\omega
         \label{eq:c3ddh} \\
 C^{({\rm DDH})}_4 &=& -{1\over 2}g_\rho h^1_\rho (2+\kappa_\rho)
         D_\rho
         -g_\omega h^1_\omega D_\omega \ ,
         \label{eq:c4ddh} \\
 C^{({\rm DDH})}_5 &=& -{1\over2\sqrt{6}} g_\rho h^2_\rho
 (2+\kappa_\rho) D_\rho\ ,
         \label{eq:c5ddh} 
\end{eqnarray}
where $g_\rho$, $\kappa_\rho$, $g_\omega$, $h^{0,1,2,1'}_\rho$, and $h^{0,1}_\omega$ are 
the vector-meson coupling constants in the DDH potential, and
\begin{eqnarray}
 D_\rho & = &   {\Lambda^2_\chi\over m^2_\rho}\, {f_\pi\over  M}
    \left( 1 -{m_\rho^2 \over \Lambda_\rho^2}\right)^2\ ,
    \label{eq:drho} \\
 D_\omega & = & {\Lambda^2_\chi\over m^2_\omega} \, {f_\pi\over  M}
    \left( 1-{m_\omega^2 \over \Lambda_\omega^2}\right)^2\ .
    \label{eq:domega}
\end{eqnarray}
The cutoff parameters $\Lambda_\rho$ and $\Lambda_\omega$ enter
the vector-meson hadronic form factors used to regularize the
behavior of the associated components of the DDH potential at large momenta
(the values adopted here $\Lambda_\rho$=1.31 GeV and $\Lambda_\omega$=1.50 GeV
are from the  one-boson-exchange charge-dependent Bonn potential~\cite{Machleidt01}).

In the original work~\cite{DDH} ``best values'' and ``reasonable ranges'', derived
from a quark model and symmetry arguments, were proposed also for the DDH
vector-meson PV coupling constants.   In particular, the analysis of Ref.~\cite{H81}
suggested that the coupling constant $h^{1'}_\rho$ is quite small.  In subsequent years, there
were several studies attempting to estimate the values of these coupling constants
either from theoretical models~\cite{DZ86,Feldman1991} or from comparisons between
predictions for PV observables and available experimental data.  For example, in
Ref.~\cite{Carlson02} constraints on the DDH PV vector-meson coupling constants
were obtained by fitting data on the $\vec p\,$-$p$ longitudinal asymmetry (see below).
In a later paper~\cite{Schiavilla04} these constraints and ``best value" estimates---in
particular, the value $h^1_\pi=4.56\times10^{-7}$ was assumed---resulted in
a DDH model, denoted as ``DDH-adj''~\cite{Schiavilla04}.  From the DDH-adj set
of vector-meson PV coupling constants we find via Eqs.~(\ref{eq:c1ddh})--(\ref{eq:c5ddh}),
in units of $10^{-7}$,
\begin{eqnarray}
  &&C^{({\rm DDH})}_1 \approx 1\ , \quad
  C^{({\rm DDH})}_2 \approx +30\ , \quad
  C^{({\rm DDH})}_3 \approx -2\ , \nonumber\\
  &&C^{({\rm DDH})}_4 \approx 0\ , \quad
  C^{({\rm DDH})}_5 \approx +7\ . \label{eq:ciddhadj}
\end{eqnarray}
The large value of $C^{({\rm DDH})}_2$ is due to the tensor
coupling constant $\kappa_\rho$=6.1 of the $\rho$-meson to the
nucleon in Ref.~\cite{Machleidt01}.  Clearly, these values should be taken only as 
indicative, since terms in the DDH vector-meson potential  implicitly also account
for TPE components, which in the $\chi$EFT PV potential are included explicitly.

\subsection{The $\vec p\, $-$p$ longitudinal asymmetry} 
\label{sec:pp}

There exist three accurate measurements of the angle-averaged $\vec p\, $-$p$      
longitudinal asymmetry $\overline{A}^{\, pp}_z(E)$, obtained at different
laboratory energies $E$~\cite{Eversheim91,Kistryn87,Berdoz03}:
\begin{eqnarray}
&& \overline A_z^{\, pp}(13.6 {\rm MeV}) = (-0.97\pm 0.20)\times 10^{-7}
\ ,\nonumber\\
&& \overline A_z^{\, pp}(45 {\rm MeV}) =(-1.53\pm 0.21)\times
10^{-7}\ ,\label{eq:lll}\\
&& \overline A_z^{\, pp}(221 {\rm MeV}) =(+0.84\pm 0.34)\times 10^{-7}\ .\nonumber
\end{eqnarray}
These data points have been obtained by combining results from various
measurements, as discussed in Sec.~IV of Ref.~\cite{Carlson02}.  The
errors reported above include both statistical and systematic errors
added in quadrature.  These experiments measure the asymmetry averaged
over a range $(\theta_1,\theta_2)$ of (laboratory) scattering angles.

The calculation of this observable was carried out with the methods
of Ref.~\cite{Carlson02}.  We have
explicitly verified that the angular distribution of the longitudinal
asymmetry $A_z(E,\theta)$ is approximately constant except at small
angles $\lesssim15$ deg, where Coulomb scattering dominates.
In the following, we have computed the average asymmetry
$\overline{A}^{\, pp}_z(E)$ using for $E=13.6$ MeV $(\theta_1,\theta_2)=(20,78)$
deg, for $E=45$ MeV $(\theta_1,\theta_2)=(23,52)$ deg, and for
$E=221$ MeV $(\theta_1,\theta_2)=(5,90)$ deg. 

For $pp$ scattering it is easily seen that the
longitudinal asymmetry can be expressed as
\begin{equation}
  \overline{A}_z^{\, pp}(E) = a^{(pp)}_0(E) \, h^1_\pi + a^{(pp)}_1(E)\, C
  \ , \label{eq:aaa} 
\end{equation}
where 
\begin{equation}
  C=C_1+C_2 + 2\, (C_4+C_5)\ ,\label{eq:lecc}
\end{equation}
and $a^{(pp)}_0(E)$ and $a^{(pp)}_1(E)$ are
numerical coefficients independent of the LEC values
(however, they do depend on the cutoff $\Lambda$
in the PC and PV chiral potentials).  The dependence of
$\overline{A}^{\, pp}_z(E)$ on $h^1_\pi$ is due to
TPE components in the PV potential.
The coefficients calculated using the PC and PV $NN$ potentials
for the two choices of cutoff parameters are reported
in Table~\ref{tab:pp-eft-coef}.  As is well known, the values of
$a^{(pp)}_i$ at low energy scale as $\sqrt{E}$, since the energy
dependence of the longitudinal
asymmetry in this energy range is driven by that of the S-wave (strong interaction)
phase shift~\cite{Carlson02}.  Because
of this scaling, the experimental points at $E$=13.6 MeV and 45 MeV  
do not provide independent constraints on the LEC's $h^1_\pi$ and $C$.

\begin{table}[t]
\begin{center}
\begin{tabular}{lcc}
\hline
$E$ [MeV]   &   $a^{(pp)}_0$         & $a^{(pp)}_1$         \\
\hline 
\multicolumn{3}{c}{$\Lambda=500$ MeV}\\
\hline
$13.6$      &   $\m 0.26992$      & $  -0.04159$    \\
$45$        &   $\m 0.55528$      & $  -0.07994$    \\
$221$       &   $  -0.24337$      & $\m 0.03134$    \\
\hline
\multicolumn{3}{c}{$\Lambda=600$ MeV}\\
\hline
$13.6$      &   $\m 0.25441$      & $  -0.03990$      \\   
$45$        &   $\m 0.53438$      & $  -0.07841$      \\
$221$       &   $  -0.19341$      & $\m 0.02743$      \\
\hline
\end{tabular}
\caption{ \label{tab:pp-eft-coef}
   Values of the coefficients $a^{(pp)}_i(E)$ at the three energies
   corresponding to the experimental data points for the two choices of
   cutoff parameters $\Lambda$.  The calculations include
   contributions up to $J_{\rm max}=6$ in the partial-wave
   expansion of the $pp$ scattering state.}
\end{center}
\end{table}

\begin{figure}[t]
   \includegraphics[width=8cm,clip]{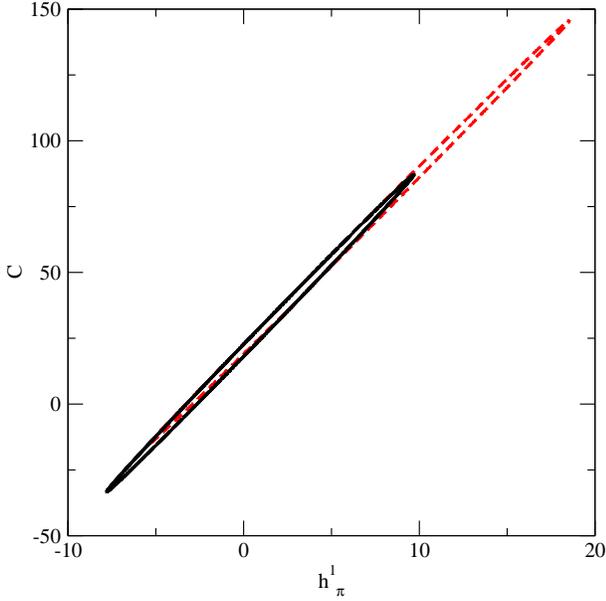}
   \caption{   \label{fig:ellipse}
    (color online). Contours of $h^1_\pi$ and $C$ values
     (in units of $10^{-7}$) corresponding to which $\chi^2=2$ for the $\vec p\,$-$p$
     longitudinal asymmetry. The black solid (red dashed)
     contour is relative to $\Lambda$=500 (600) MeV.
     }
\end{figure}

If we assume $h^1_\pi$=4.56 $\times 10^{-7}$ and $\Lambda$=500
MeV, we obtain $C \approx 50 \times 10^{-7}$ by fitting the
experimental value at $E$=13.6 MeV, in agreement with the result of 
Ref.~\cite{Vries13} (note that the operator proportional to the
LEC $C$ used in that work has a minus sign relative to that
defined here).  In order to take into account
experimental uncertainties, we have performed a $\chi^2$
analysis, and in Fig.~\ref{fig:ellipse} we report the
$h^1_\pi$ and $C$ values for which $\chi^2=2$ when the cutoff
in the PC and PV chiral potentials is fixed at either $\Lambda$=500 or $\Lambda$=600 MeV.
The resulting two elliptic regions almost coincide, and there appears to be
a strong correlation between $h^1_\pi$ and $C$.
The range of allowed $h^1_\pi$ values is rather large
$-1\times 10^{-6}<h^1_\pi< 2\times 10^{-6}$, containing the whole DDH ``reasonable
range''.  Note that the two ellipses are rather narrow and
almost coincident with a straight line.  These conclusions are the
same as those derived in a similar analysis by the authors of Ref.~\cite{Vries13}.

In Table~\ref{tab:lecs}, we report representative values of $C$,
determined from Fig.~\ref{fig:ellipse}, corresponding to the three
choices of $h^1_\pi$ discussed in the previous section.  These values
are used in the following sections to provide estimates for PV observables
in $A$=2--4 systems.

\begin{table}[t]
\begin{center}
\begin{tabular}{l|cc}
\hline 
$\Lambda$ [MeV] &  $\qquad h_\pi^1\quad$  &  $\quad C\qquad$ \\
\hline
     & $1\phantom{.00}$ & $26.7$ \\
500  & $4.56$ & $51.5$ \\  
     & $11.4$ & $99.2$ \\
\hline
     & $1\phantom{.00}$ & $27.6$ \\
600  & $4.56$ & $51.4$ \\  
     & $11.4$ & $97.2$ \\
\hline
\end{tabular}
\end{center}
\caption{\label{tab:lecs}
  Values for $C$ corresponding to different values of $h^1_\pi$
  (both in units of $10^{-7}$) and $\Lambda$, as determined from 
  Fig.~\protect\ref{fig:ellipse}.  The $C$ values are those
   lying on the major axis of the elliptical contours.} 
   \end{table}

\subsection{$\vec n$-$p$ and $\vec n$-$d$ spin rotations}
\label{sec:spinrot}

The rotation of the neutron spin in a plane transverse to the beam
direction induced by PV components in the $NN$ potential is given by
\bgroup
\arraycolsep=1.0pt
\begin{eqnarray}
  \frac{{\rm d}\phi^{(nX)}}{{\rm d}z} &=& \frac{2\pi\rho}{(2S_X+1)\,
    v_{\rm rel}} {\rm Re} 
  \! \sum_{m_n m_X} \epsilon_{m_n}
    \nonumber\\
   & \times& \!^{(-)}\langle p\hat{\bf z}; m_n , m_X | V_{PV} | 
  p\hat{\bf z};m_n , m_X\rangle^{(+)}\ ,
  \label{eq:nphi}
\end{eqnarray}
\egroup
where $\rho$ is the density of hydrogen or deuterium nuclei for $X$=$p$ or $d$,
$V_{PV}$ denotes the PV potential, $\mid\! p \hat{\bf z};m_n,m_X\rangle^{(\pm)}$
are the $n$-$X$ scattering states with outgoing-wave $(+)$ and
incoming-wave $(-)$ boundary conditions and relative momentum
${\bf p}=p\, \hat{\bf z}$ taken along the spin-quantization axis
(the $\hat{\bf z}$-axis), $S_X$ is the $X$ spin, and $v_{\rm rel}=p/\mu$ is
the magnitude of the relative velocity, $\mu$ being the $n$-$X$
reduced mass.  The expression above is averaged over the spin
projections $m_X$; however, the phase
factor $\epsilon_{m_n}= (-)^{1/2-m_n}$ is $\pm 1$ depending
on whether the neutron has $m_n=\pm 1/2$. 

We consider the $\vec n$-$p$ and $\vec n$-$d$ spin
rotations for vanishing incident neutron energy (measurements
of this observable are performed using ultra-cold neutron beams).
In the following, we assume $\rho=0.4\times 10^{23}$ cm$^{-3}$.
The $n$-$d$ wave functions have been obtained with
the hyperspherical-harmonics (HH) method~\cite{Kea08,Mea09} from
the Hamiltonians N3LO/N2LO-500
and N3LO/N2LO-600 of Sec.~\ref{sec:res}.  Details of the calculation of ${\rm d}\phi/{\rm d}z$ for
 $\vec n$-$p$ and $\vec n$-$d$ can be found in Refs.~\cite{Schiavilla04} and~\cite{Schiavilla08,Song11},
 respectively.  In general, the rotation angle depends linearly on the PV LEC's:
\begin{eqnarray}
  \frac{{\rm d}\phi^{(nX)}}{{\rm d}z}&=& h^1_\pi \, b_0^{(nX)}+ C_1 \, b_1^{(nX)}
   + C_2\,  b_2^{(nX)}\nonumber\\
   && + \, C_3\,  b_3^{(nX)}+ C_4 \, b_4^{(nX)}+ C_5\,  b_5^{(nX)}
   \ , \label{eq:nphii}
\end{eqnarray}
where the $b_i^{(nX)}$ for $i=0,\ldots,5$ are numerical coefficients.
Their calculated values for the two choices of cutoff $\Lambda$ are 
listed in Table~\ref{tab:nrot-eft-coef}.

\begin{table*}[t]
\begin{center}
\begin{tabular}{lcccccc}
\hline
\multicolumn{7}{c}{$\vec n$-$p$ scattering}\\
\hline
$\Lambda$ [MeV]  &  $b^{(np)}_0$ & $b^{(np)}_1$ & $b^{(np)}_2$ &
                    $b^{(np)}_3$ & $b^{(np)}_4$ & $b^{(np)}_5$ \\
\hline 
500  &   $1.35997$ &  $0.24399$ & $ 0.17390$ &
         $0.10537$ &  $0.00000$ & $-0.90585$ \\
600  &   $1.26392$ &  $0.23540$ & $ 0.15839$ &
         $0.08483$ &  $0.00000$ & $-0.86459$ \\
\hline
\multicolumn{7}{c}{$\vec n$-$d$ scattering}\\
\hline
$\Lambda$ [MeV]  &  $b^{(nd)}_0$ & $b^{(nd)}_1$ & $b^{(nd)}_2$ &
                    $b^{(nd)}_3$ & $b^{(nd)}_4$ & $b^{(nd)}_5$ \\
\hline 
500  &   $2.17913$   & $-0.01046$ &  $-0.15997$ &
         $0.19117$   & $ 0.06367$ &  $ 0.00032$   \\
600  &   $2.22165$   & $-0.00684$ &  $-0.18226$ &
         $0.18232$   & $ 0.06750$ &  $ 0.00030$  \\
\hline
\end{tabular}
\caption{ \label{tab:nrot-eft-coef}
   Values of the coefficients $b^{(nX)}_i$ in units of Rad m${}^{-1}$
   for the  $\vec n$-$p$ and  $\vec n$-$d$ spin rotations calculated for the two choices of
 cutoff $\Lambda$ at vanishing neutron beam energy.}
\end{center}
\end{table*}

The coefficient $b^{(nX)}_0$ receives contributions from 
the OPE, TPE, and RC components of the PV potential, see Eq.~(\ref{eq:dec}).
For example, for $\vec n$-$p$ scattering and the N3LO-500 PC potential
we find $b^{(np)}_0= 1.23759({\rm OPE})+0.13441({\rm
  TPE})-0.01202({\rm RC})=1.35997$ Rad m$^{-1}$, 
so the TPE (RC) contribution is about 10 (1)\% of the OPE.
Inspection of Table~\ref{tab:nrot-eft-coef} also shows that 
the $\vec n$-$p$ spin rotation is sensitive to all
LEC's except $C_4$ (proportional to $\tau_{1z}+\tau_{2z}$);
in particular, there is a large sensitivity to $C_5$, which 
multiplies the isotensor term of the PV
potential. 

The $\vec n$-$p$ spin rotation was already studied in Ref.~\cite{Liu07} using
the same PV potential as in the present work but the Argonne
$v_{18}$ (AV18) PC potential~\cite{AV18}.  The results obtained
in this work cannot be directly compared with those reported in Ref.~\cite{Liu07}
because of differences in the definition of the LEC's, in the value of the cutoff, and in the presentation
of the results themselves.  In Ref.~\cite{Liu07}
the CT component of the PV potential has a redundant parametrization
in terms of 10 LEC's.  By expressing the 5 CT operators depending on
$\bmK$ in terms of those which only depend on $\bmk$ via Eqs.~(25), (26) and (29) of
Ref.~\cite{Liu07}, we obtain (in units of Rad m$^{-1}$)
\begin{eqnarray}
 \frac{{\rm d}\phi^{(np)}}{{\rm d}z}|_{\rm Liu}&\approx& \Bigl[1.15 \, (h^1_\pi)_{\rm OPE} -0.13\, 
 (h^1_\pi)_{\rm TPE} 
    \nonumber\\
  && +0.15\, C_1 + 0.12\, C_2 \nonumber\\
  && + 0.08\, C_3 -0.56\, C_5 \Bigr]    \ .\label{eq:liu07}
\end{eqnarray}
The coefficients multiplying the various LEC's can be compared
with the $b^{(np)}_i$ reported in Table~\ref{tab:nrot-eft-coef}.   There is qualitative agreement,
given that the coefficients above correspond to the
AV18 model as well as to a larger cutoff $\Lambda$ in the PV potential than adopted here.
In Eq.~(\ref{eq:liu07}), the factors multiplying $(h^1_\pi)_{\rm OPE}$
and $(h^1_\pi)_{\rm TPE}$ are the contributions to $b_0^{(np)}$ from the OPE and TPE components (of
the PV potential), respectively.  These should be compared to
 the OPE and TPE results, obtained here (see above): 1.23759 Rad m$^{-1}$
 and 0.13441 Rad m$^{-1}$.  In the present case, the TPE contribution is
positive and increases $b^{(np)}_0$.  We have verified that this
contribution is very sensitive to the cutoff parameter
$\Lambda$.  For example, using $\Lambda=1.3$ GeV as in Ref.~\cite{Liu07}, we find that the TPE contribution
becomes negative. 

\begin{table}[t]
\begin{center}
\begin{tabular}{l|ccc| ccc}
\hline
 & \multicolumn{3}{c}{$\Lambda=500$ MeV} &
\multicolumn{3}{c}{$\Lambda=600$ MeV}\\
\hline 
LEC  & I & II & III & I  & II  & III \\
\hline
 $h_\pi^1$   & $\n 1.0$ & $\n 4.56$ & $11.4$ & $\n 1.0$ & $\n 4.56$ & $11.4$ \\
 $C_2$ & $13.7$   & $28.5\n $ & $56.2$ & $14.6$   & $28.4\n $ & $54.2$ \\
 $C_5$ & $\n 5.0$ & $10.0\n$  & $20.0$ & $\n 5.0$ & $10.0\n$  & $20.0$ \\
\hline
\end{tabular}
\caption{\label{tab:caselec}
 Values for the LEC's $h_\pi^1$, $C_2$, and $C_5$ correspoding
 to sets I, II, and III (see text).   The remaining LEC's
 are taken to have the values $C_1=1$, $C_3=-1$,
  and $C_4=1$ in each set.  All LEC's are in units of $10^{-7}$.}
\end{center}
\end{table}

In reference to the $\vec n$-$d$ spin rotation, we note the large sensitivity to
$h^1_\pi$ (this fact is well known~\cite{Schiavilla08,Song11}), and to
the LEC's $C_2$ and $C_3$.  A measurement of this observable
could be very useful in constraining their values.

In order to provide an estimate for the magnitude of the $\vec n$-$p$ and
$\vec n$-$d$ spin rotations, we proceed as follows.  We set $C_1=1$,
$C_3=-1$, and $C_4=1$, since estimates from Eq.~(\ref{eq:ciddhadj})
indicate that they are much smaller (in magnitude) than $C_2$ and $C_5$.
For each set of $h^1_\pi$ and $\Lambda$ values, the corresponding value
of the LEC $C_5$ is as reported in Table~\ref{tab:caselec}.  Finally,
the value of the LEC $C_2$ is fixed so that the parameter $C$
in Eq.~(\ref{eq:lecc}) is as given in Table~\ref{tab:lecs}.  All the
LEC values are reported in Table~\ref{tab:caselec}.  The three sets
are denoted hereafter as I, II, and III.

The cumulative contributions to
${\rm d}\phi^{(nX)}/{\rm d}z$ using the LEC's in Table~\ref{tab:caselec} are
reported in Table~\ref{tab:nrot}.  There is a large sensitivity to $h^1_\pi$.   However,
the dependence of the results on $\Lambda$ is also significant, especially
in the $\vec n-p$ case, presumably
due to the fact that the LEC's $C_{1,3,4,5}$ have been taken to have the
same values for the two choices of $\Lambda$.
In general, we expect the values of these LEC's
to vary as $\Lambda$ changes.  The effect of the RC term in the PV
potential is tiny.  For the $\vec n$-$d$ spin rotation angle, we note that the 
 TPE contribution is at few \% level for $\Lambda$=500 MeV, but negligible
for $\Lambda$=600 MeV.

\begin{table}[t]
\begin{center}
\begin{tabular}{l|ccc| ccc}
\hline
 & \multicolumn{3}{c}{$\Lambda=500$ MeV} &
  \multicolumn{3}{c}{$\Lambda=600$ MeV}\\
\hline
\multicolumn{7}{c}{$\vec n$-$p$ spin rotation}\\
\hline
 & I & II & III & I  & II & III \\
\hline 
 OPE  & $\m 1.24$  &   $5.64$  &  $14.11$  & $\m 1.19$  &  $ 5.42$  & $13.56$ \\
 TPE & $\m 1.37$  &   $6.26$  &  $15.64$  & $\m 1.27$  &  $ 5.80$  & $14.50$ \\
 RC & $\m 1.36$  &   $6.20$  &  $15.50$  & $\m 1.26$  &  $ 5.76$  & $14.41$ \\
 CT  & $  -0.65$  &   $2.24$  &  $\n7.30$ & $  -0.60$  &  $ 1.77$  & $\n 5.85$ \\
\hline
\multicolumn{7}{c}{$\vec n$-$d$ spin rotation}\\
\hline
 OPE  & $\m 2.25$ & $10.27$   & $25.67$ & $\m 2.25$ & $10.25$ & $25.62$ \\
 TPE & $\m 2.21$ & $10.06$   & $25.15$ & $\m 2.25$ & $10.26$ & $25.64$ \\
 RC & $\m 2.18$ & $\n 9.94$ & $24.84$ & $\m 2.22$ & $10.13$ & $25.33$ \\
 CT  & $-0.15$   & $\n 5.24$ & $15.72$ & $-0.56$ & $\n 4.84$ & $15.33$ \\
\hline
\end{tabular}
\caption{\label{tab:nrot}
  Cumulative contributions to the $\vec n$-$p$ and $\vec n$-$d$
  spin rotations in units of $10^{-7}$ Rad m${}^{-1}$, corresponding
  to sets I, II, and III of LEC's as specified in Table~\protect\ref{tab:caselec}.}
\end{center}
\end{table}

\subsection{The $^3$He($\vec{n},p$)$^3$H longitudinal
  asymmetry}
\label{sec:n3he}
For ultracold neutrons the longitudinal asymmetry $A_z$ for
the reaction $^3$He($\vec{n},p$)$^3$H  is given by $A_z= a_z \, {\rm cos}\,
\theta$~\cite{Viviani10}, where $\theta$ is the angle between the
outgoing proton momentum and
the neutron beam direction. The coefficient $a_z$ can be expressed in
terms of products of $T$-matrix elements involving 
three PC and three PV transitions (see Ref.~\cite{Viviani10} for
details).  These $T$-matrix elements are calculated by means of the
HH method~\cite{Kea08,Mea09}, using the PC and PV chiral potentials
of the previous sections.   
\begin{table*}[t]
\begin{center}
\begin{tabular}{lcccccc}
\hline
$\Lambda$ [MeV]  &  $a_0$ & $a_1$ & $a_2$ &
                $a_3$ & $a_4$ & $a_5$ \\
\hline 
500  &   $-0.1444$  & $ 0.0061$ &  $ 0.0226$ &
         $-0.0199$  & $-0.0174$ &  $-0.0005$   \\
600  &   $-0.1293$  & $ 0.0081$ &  $ 0.0320$ &
         $-0.0161$  & $-0.0156$ &  $-0.0001$   \\
\hline
\end{tabular}
\caption{ \label{tab:nh-eft-coef}
   Values of the coefficients $a_i$ entering the $^3$He($\vec{n},p$)$^3$H
   longitudinal asymmetry calculated at vanishing neutron beam energy 
   and for two choices of the cutoff parameter $\Lambda$. 
   The calculations are performed summing four-body states up to $J_{\rm max}=1$,
   namely including S--waves in the incident channel.}
\end{center}
\end{table*}

The coefficient $a_z$ is expressed as 
\begin{eqnarray}
  a_z  &=& a_0 \,h^1_\pi + a_1\,C_1 + a_2\, C_2\nonumber\\
   &&+ a_3\, C_3 + a_4\, C_4 + a_5\,  C_5
  \ , \label{eq:azphet} 
\end{eqnarray}
and the calculated coefficients $a_i$ are listed in Table~\ref{tab:nh-eft-coef}.
The largest coefficient is $a_0$, while $a_2$, $a_3$ and $a_4$
are of similar magnitude and about a factor 5 smaller than $a_0$.
The coefficients $a_1$ and $a_5$ are more than an order of magnitude
smaller than the leading $a_0$.  Naively, one would have expected the 
dominant contributions to come from the isoscalar terms (proportional to the LEC's
$C_1$ and $C_2$), since at the low energies of interest here the reaction proceeds 
mainly through the (close) $0^+$ and $0^-$ resonances in the $\heq$ spectrum, having total
isospin $T$=0~\cite{TWH92}.  However, the Coulomb interaction in the final state
induces significant isospin mixing, and the isovector terms in the PV
potential end up giving unexpectedly large contributions.  In any case,
the ``isoscalar'' coefficient $a_2$ when multiplied by $C_2$, which is
expected to be large, leads to a contribution of similar magnitude
as that of $a_0\, h^1_\pi$, but of opposite sign.  The
destructive interference between these two contributions
makes the longitudinal asymmetry rather sensitive to short-range
physics and, in particular, to the cutoff $\Lambda$
(see Table~\ref{tab:nhet} below).

\begin{table}[t]
\begin{center}
\begin{tabular}{l|ccc|ccc}
\hline
 & \multicolumn{3}{c}{$\Lambda=500$ MeV} &
  \multicolumn{3}{c}{$\Lambda=600$ MeV}\\
\hline
 & I & II & III & I  & II & III \\
\hline 
 OPE  & $-0.118$   & $-0.537$ & $-1.34$ & $-0.099$   & $-0.453$   & $-1.13$ \\
TPE & $-0.147$   & $-0.669$ & $-1.67$ & $-0.131$   & $-0.597$   & $-1.49$ \\
RC & $-0.144$   & $-0.658$ & $-1.65$ & $-0.129$   & $-0.589$   & $-1.47$ \\
CT  & $\n 0.171$ & $-0.012$ & $-0.38$ & $\n 0.346$ & $\n 0.326$ & $\n 0.27$ \\
\hline
\end{tabular}
\caption{\label{tab:nhet}
  Cumulative contributions to the coefficient $a_z$ (in units of $10^{-7}$), describing the 
  $^3$He($\vec{n},p$)$^3$H longitudinal asymmetry at vanishing neutron beam energy, corresponding
  to sets I, II, and III of LEC's as specified in Table~\protect\ref{tab:caselec}.}
\end{center}
\end{table}

In Table~\ref{tab:nhet} we report cumulatively the contributions
of the OPE, TPE, RC, CT components of the PV potential
to the parameter $a_z$.   These predictions for $a_z$ correspond to sets I,
II, and III of LEC's, as specified in Table~\protect\ref{tab:caselec}.
The RC contribution is tiny (at the 1\% level), while the
TPE contribution is about 30\% of the OPE one.  Upon
including the CT contribution, we observe a large cancellation,
particularly effective for small values of $h^1_\pi$.
As already noted, this cancelation comes mostly from the term proportional to
the LEC $C_2$.

\section{Conclusions}
\label{sec:conc}

In this paper we have studied the $\chi$EFT Lagrangian describing PV
interactions of nucleons and pions up to order $Q^2$.  This Lagrangian
has been used to derive the PV $NN$ potential at $\nnlo$.  We have also
discussed subleading contributions to the OPE component, and
carried out the renormalization of the pion-nucleon PV coupling constant
$h^1_\pi$.  Finally, we have investigated PV effects in a number
of reactions involving few-nucleon systems.  We find that (i)
the  $\vec p\,$-$p$ longitudinal asymmetry is sensitive to
$h^1_\pi$ (via the TPE component of the PV potential) and to the combination
of LEC's $C=C_1+C_2+2\, C_4+2\, C_5$; (ii) the $\vec n$-$p$
spin rotation is sensitive to all LEC's but $C_4$, while
the $\vec n$-$d$ spin rotation is sensitive to $h^1_\pi$, $C_2$, and $C_3$; and (iii)
the $n$-$\het$ longitudinal asymmetry is sensitive to $h^1_\pi$, $C_2$, $C_3$, 
and $C_4$. 

At the SNS facility at Oak Ridge National Laboratory a measurement of the PV asymmetry
$a_\gamma$ in the $^1$H$(\vec n,\gamma)^2$H radiative capture is in progress
(the NPDGAMMA experiment).  This 
observable is mainly sensitive to $h^1_\pi$~\cite{Desp08}.  It is 
expected that this measurement will provide tight constraints on its value.
Once $h^1_\pi$ is known, measurements of the $\vec n$-$p$ and
$\vec n$-$d$ spin rotations, and of the $\vec n-$$\het$ longitudinal asymmetry,
would provide constraints on the LEC's $C_i$.
It would also be very valuable to have new and more
precise measurements of the $\vec p\,$-$p$ longitudinal asymmetry
at different proton energies. 

An experiment to measure the $\vec n$-$\het$ $A_z$ asymmetry
has already been approved at the SNS facility, and the experimental apparatus 
is in an advanced stage of construction.   The experiment should
start taking data after the conclusion of NPDGAMMA.  An accurate
measurement of  $A_z$ could lead to a precise determination
of $C_2$.  At the present time, experiments to measure $\vec n$-$p$ and $\vec n$-$d$
spin rotation angles are not planned, but could provide useful information on
$C_5$ and $C_3$, respectively.  The only other experiment in progress
we are aware of is a measurement of the $\vec n$-$\heq$
spin rotation at NIST~\cite{Bass09}. 

In the future, we plan to use the $\chi$EFT Lagrangians of order $Q^2$
we have derived here to study PV couplings to the electromagnetic field
and, in particular, to provide estimates for PV observables in the $np$ and
$nd$ radiative captures.
\section*{Acknowledgments}
The work of R.S.~was supported by the U.S.~Department of Energy, Office of
Nuclear Physics, under contract DE-AC05-06OR23177.  The calculations were
made possible by grants of computing time from the National Energy Research
Scientific Computing Center.

\appendix
\section{Chiral transformation properties of $D_\mu X^a_{L/R}$}
\label{app:note1}

\bgroup
\arraycolsep=1.0pt

The transformation properties of the  quantities $X^a_{L/R}$,
as already given in Eq.~(\ref{eq:bbt}), are
\begin{equation}
  X^a_{R}\to h \; u^\dag\;  R^\dag \tau_a R\;  u\; h^\dag\ , \quad
  X^a_{L}\to h \; u \; L^\dag \tau_a L \; u^\dag \; h^\dag\ ,
\end{equation}
i.e., $X^a_{R}\to h
\bigl(X^a_{R}\bigr)^{({\rm constructed\ with\ }R^\dag\tau_a R )}h^\dag$
and similarly for $X^a_{L}$.  For local transformations, the $SU(2)$ matrices
$L$ and $R$ depend on the space-time coordinate $x$, and care must be taken when considering
four-gradients of $X^a_{L/R}$, since under chiral transformations
they contain terms like $h\, u^\dag R^\dag \tau_a (\partial_\mu R)\,
u\, h^\dag$, which transform differently from
$X^a_{L/R}$.  It is convenient to define the following quantities
\begin{eqnarray}
  \left(X^a_R\right)_\mu &=& \left[D_\mu + i \, u^\dag r_\mu u\, ,\,
    X^a_R\right] \ ,\label{eq:xxRL1app} \\
  \left(X^a_L\right)_\mu &=& \left[D_\mu + i\,  u\, \ell_\mu u^\dag\, ,\, 
    X^a_L\right] \ ,\label{eq:xxRL2app}
\end{eqnarray}
which explicitly read
\begin{eqnarray}
\label{eq:a4}
  \!\!\!\!\!\!\!\!\!\!\!\! \left(X^a_R\right)_\mu &=& \partial_\mu(u^\dag \tau_a u) +
   \left[\Gamma_\mu\, ,\, u^\dag \tau_a u\right] + i
   u^\dag\left[r_\mu\, ,\, \tau_a\right] u\ , \\
   \label{eq:a5}
   \!\!\!\!\!\!\!\!\!\!\!\!\left(X^a_L\right)_\mu &=& \partial_\mu(u\, \tau_a u^\dag) +
   \left[\Gamma_\mu\, ,\, u\, \tau_a u^\dag\right] + i
   u \, \left[l_\mu\, ,\, \tau_a\right] u^\dag\ .
\end{eqnarray}
Then, under chiral transformation it is easily seen that
\begin{eqnarray}
  \left(X^a_R\right)_\mu &\to& h\Bigl[ (\partial_\mu u^\dag) R^\dag
  \tau_a R \,u +  u^\dag R^\dag \tau_a R (\partial_\mu u)\Bigr]h^\dag
  \nonumber \\
   && +h \left[ \Gamma_\mu\, ,\, u^\dag R^\dag \tau_a R\,  u\right] h^\dag 
      + i h \,u^\dag [r_\mu\, ,\, R^\dag \tau_a R] u\, h^\dag\ , \nonumber\\
   &=& h \left(X^a_R\right)_\mu^{({\rm constructed\ with\ }R^\dag
        \tau_a R )} h^\dag\ ,
\end{eqnarray}
and similarly
\begin{equation}
 \left(X^a_L\right)_\mu \to h \left(X^a_L\right)_\mu^{({\rm
     constructed\ with\ }L^\dag
        \tau_a L )} h^\dag\ .
\end{equation}
Therefore, the quantities $(X^a_{L/R})_\mu$ transform
consistently as $X^a_{L/R}$.  After inserting the definition
of $\Gamma_\mu$ into Eqs.~(\ref{eq:a4})--(\ref{eq:a5}), straightforward manipulations
allow one to express $(X^a_{L/R})_\mu$ in a more compact form as 
\begin{equation}
   \left(X^a_R\right)_\mu = {i\over 2} \left[ u_\mu\, ,\, X^a_R\right]\ ,\qquad
   \left(X^a_L\right)_\mu = -{i\over 2} \left[ u_\mu\, , \, X^a_L\right]\ .
   \label{eq:xxidapp}
\end{equation}
These identities are used in Appendix~\ref{app:lq2} in order to reduce the number of
terms entering the PV Lagrangian.
\section{Transformation properties of the various fields under $P$ and $C$}
\label{app:trasf}
We list here the transformation properties of various fields and field
combinations under hermitian
conjugation ($H$), parity ($P$), and charge conjugation ($C$).  The nucleon field
$\psi$ transforms as
\begin{eqnarray}
&&\psi \stackrel{P}{\longrightarrow} \gamma^0 \psi\ ,\label{eq:Ppsi}\\
&& \psi \stackrel{C}{\longrightarrow} -i \,\gamma^0 \gamma^2 (\overline\psi)^T
\ ,\label{eq:Cpsi}
\end{eqnarray}
where $A^T$ denotes hereafter the transpose of a given quantity $A$.

For a generic combination $O$ of fields, one has
\begin{eqnarray}
  O^\dag&=& s_H\,  O\ , \nonumber \\
  O_{\mu_1\,\mu_2\, \ldots}  &\stackrel{P}{\longrightarrow}& s_P \,\sigma_{\mu_1}\,
       \sigma_{\mu_2}\, \cdots O_{\mu_1\, \mu_2\, \ldots} \ , \label{eq:hhpc}\\
  O&\stackrel{C}{\longrightarrow}&  s_C \, O^T\ , \nonumber
\end{eqnarray}
where $s_H$, $s_P$, and $s_C$ are $\pm 1$ phase factors,
$\sigma_\mu$ is +1 when $\mu=0$ (time-like) and --1 when $\mu=1,2,3$ (space-like),
and no summation is implied here over the repeated indices $\mu_i$.  The
phase factors $s_H$, $s_P$, and $s_C$ in the case of bilinears $O=\overline{\psi}\, \Gamma \psi$,
where $\Gamma$ is one of the elements of the Clifford algebra, are listed in
Table~\ref{tab:sign1}.  When an operator also includes the Levi-Civita tensor 
$\epsilon^{\mu \nu \rho \sigma}$ as in $\epsilon^{\mu\nu\rho\sigma}
O_{\mu\nu\rho\sigma}$, then 
$\epsilon^{\mu\nu\rho\sigma} O_{\mu\nu\rho\sigma} \stackrel{P}{\longrightarrow} -s_P\,
\epsilon^{\mu\nu\rho\sigma} O_{\mu\nu\rho\sigma}$ since the Lorentz indices
$\mu$, $\nu$, $\rho$, and $\sigma$ must
be all different,
and hence $\epsilon_{\mu \nu \rho \sigma}$ may be considered odd under parity.

\begin{table}[bth]
\begin{tabular}{l|c|c|c|c|c}
\hline
\hline
 & 1& $i\, \gamma_5$ & $ \gamma_\mu$ & $\gamma_\mu \gamma_5$ & $
 \sigma_{\mu \nu}$  \\
\hline
$s_H$ & + & + & + & + & +    \\
\hline
$s_P$ & + & -- & + & -- & +   \\
\hline
$s_C$ & + & + & -- & + & --  \\
\hline
\hline
\end{tabular}
\caption{ \label{tab:sign1}
Transformation properties of fermion bilinears
with different elements of the Clifford algebra
under hermitian conjugation ($H$), parity ($P$), and
charge conjugation ($C$). }
\end{table}

In reference to combinations of pion fields, one has under parity 
\[
u \stackrel{P}{\longrightarrow} u^\dagger \ ,\qquad
u_{\mu} \stackrel{P}{\longrightarrow} - \sigma_{\mu} u_{\mu}\ ,
\]
and
\[
F_{\mu\nu}^R \stackrel{P}{\longleftrightarrow} F_{\mu\nu}^L\ ,\qquad
X^a_L \stackrel{P}{\longleftrightarrow} X^a_R \ ,
\qquad \chi\stackrel{P}{\longrightarrow}\chi^\dag\ ,
\]
and under charge conjugation
\begin{eqnarray}
&&u \stackrel{C}{\longrightarrow} u^T\ , 
    \qquad u_\mu \stackrel{C}{\longrightarrow}  u_\mu^T\ ,\label{eq:Cumu}\\
&&X^2_L \stackrel{C}{\longrightarrow} - (X^2_R)^T , \quad X^2_R
  \stackrel{C}{\longrightarrow} - (X^2_L)^T\ ,\label{eq:CX2}\\
&&X^{1,3}_L \stackrel{C}{\longrightarrow} (X^{1,3}_R)^T , \quad X^{1,3}_R
  \stackrel{C}{\longrightarrow} (X^{1,3}_L)^T \ ,\label{eq:CX13}
\end{eqnarray}
since $(\tau_2)^T=-\tau_2$. 
The transformation properties of other
pion related quantities are summarized in Table~\ref{tab:sign2}.
When considering terms involving $O$ and the covariant derivative $D_\mu$,
it is convenient to introduce the combinations
\begin{equation}
  \{ D_\mu\, ,\, O \}= D_\mu O+O  D_\mu \ , \qquad [D_\mu\, ,\,O]=
   D_\mu O-O  D_\mu \ ,
\end{equation}
and determine how $ \{ D_\mu\, ,\, \ldots \}$
and $[D_\mu\, ,\, \ldots]$ transform under hermitian conjugation,
$P$, and $C$ independently of $O$, as in Table~\ref{tab:sign2}.  In particular,
$C$-even terms must have an even (odd)
number of nested anticommutator terms like
$\{ D_{\mu_1}\, ,\, \{ D_{\mu_2}\, ,\, \ldots \{ D_{\mu_n}\, , O\} \ldots \}\}$,
when the field combination $O$ is $C$-even ($C$-odd). 

\begin{table}[h]
\begin{tabular}{l|c|c|c|c|c|c}
\hline
\hline
 & $u_\mu$ & $\Gamma_\mu$ & $\{ D_\mu\, ,\, \ldots \}$ & $[ D_\mu\, ,\, \ldots ]$ & $
 X^a_{+}$ & $  X^a_{-}$ \\
\hline
$s_H$  & + & -- & -- & + & + & + \\
\hline
$s_P$  & -- & + & + & + & + &  -- \\
\hline
$s_C$  & + & -- & -- & + & $(-)^{a+1}$ & $-(-)^{a+1}$\\
\hline
\hline
\end{tabular}
\caption{ \label{tab:sign2}
 Transformation properties of the quantities $u_\mu$,
 $\Gamma_\mu$,  $\{ D_\mu\, ,\,\ldots \}$,  $[ D_\mu\, ,\, \ldots ]$, 
and $X^a_{\pm}=X^a_{L}\pm X^a_{R}$ under hermitian conjugation ($H$),
parity ($P$), and charge conjugation ($C$). }
\end{table}

Finally, the transformation properties of quantities related to
external fields are reported in Table~\ref{tab:sign3}. 
Note that the ``external'' quantities $r_\mu$, $\ell_\mu$, and $\chi$ are 
considered to transform under $C$ as
\begin{equation}
  r_\mu\stackrel{C}{\longrightarrow}-\ell_\mu^T\ , \qquad
  \ell_\mu\stackrel{C}{\longrightarrow}-r_\mu^T\ ,\qquad 
  \chi\stackrel{C}{\longrightarrow}\chi^T\ ,
\label{eq:Clr}
\end{equation}

\begin{table}[h]
\begin{tabular}{l|c|c|c|c}
\hline
\hline
 & $\ \chi_+\ $ & $\ \chi_-\ $
 & $\ F^{\mu\nu}_+\ $ & $\ F^{\mu\nu}_-\ $
 \\
\hline
$s_H$  & + & -- & + & + \\
\hline
$s_P$  & + & -- & + & --\\
\hline
$s_C$  & + & + & --& +\\
\hline
\hline
\end{tabular}
\caption{ \label{tab:sign3}
 Transformation properties of the quantities $\chi_\pm$ and 
$F^{\mu\nu}_\pm$ under hermitian conjugation ($H$), parity ($P$), and
charge conjugation ($C$). }
\end{table}

\section{Independent PV interaction terms of order $Q^2$ }
\label{app:lq2}

In this Appendix we discuss in detail the selection of 
independent PV $\pi$--$N$ interaction terms of order $Q^2$. 
The transformation properties of various quantities under hermitian
conjugation ($H$), parity ($P$), and charge conjugation ($C$) are
given in Appendix~\ref{app:trasf}.  The following
power counting is assumed
\begin{equation}
u_\mu \sim Q\ , \qquad
F^\pm_{\mu\nu} \sim Q^2\ , \qquad
\chi_\pm \sim Q^2\ .
\end{equation}
The covariant derivative $D_\mu$ is taken as of order $Q$, except
when it acts on a nucleon field, in which case it is of order $Q^0$
due to the presence of the heavy mass scale. 

The (independent) isoscalar ($\Delta I$=0), isovector
($\Delta I$=1), and isotensor ($\Delta I$=2) interaction
terms are constructed in the next three subsections.
In each case, we begin our analysis by considering
quantities constructed first with $\chi_\pm$ or $F_{\mu\nu}^{\pm}$ 
(already of order $Q^2$), then with products $u_\mu u_\nu$ 
(again of order $Q^2$), and lastly with a single $u_\mu$.
For the sake of clarity, a summary of properties of $\gamma$
matrices used below is reported in Appendix~\ref{app:A}.
\subsection{The $\Delta I=0$ sector}
\label{app:lq2_0}
\begin{enumerate}
\item Terms with $\chi_\pm$ or $F^{\mu\nu}_{\pm}$: These
  are already of order $Q^2$, so the simplest $P$-odd and $C$-odd
  quantity is $\bar\psi\, F^-_{\mu\nu}\, \sigma^{\mu\nu}\psi$, listed
  as $O_2^{(0)}$ in Eq.~(\ref{eq:o0_2}).
  Additional terms must involve $\bar \psi\, \{ D_\mu\, ,\, \ldots
  \}\psi$, otherwise the four-gradient acting on a pion field would bring in
  an extra factor of $Q$.  The $P$-odd and $C$-odd quantities are
 \begin{eqnarray}
  && \bar\psi i\{D_\mu\, ,\,\chi_+\} \gamma^\mu\gamma^5\psi\ ,\nonumber\\
  && \bar\psi i\{D_\mu\, ,\,F^+_{\nu\alpha}\} \epsilon^{\mu\nu\alpha\beta}\gamma_\beta\psi\ ,\nonumber\\
  && \bar\psi i\{D_\mu\, ,\,F^-_{\nu\alpha}\}
   \epsilon^{\mu\nu\alpha\beta}\gamma_\beta\gamma^5\psi\ ,\nonumber
  \end{eqnarray}
which, however, turn out to be at least of order $Q^3$.
For example, using the EOM in the first combination, we have
 \begin{equation}
    \bar\psi \{D_\mu\, ,\, \chi_+\} \gamma^\mu\gamma^5\psi =
      \bar\psi\Bigl( D \!\!\!\! \slash \chi_+ \gamma^5 -
            \chi_+ \gamma^5 D \!\!\!\! \slash\Bigr) \psi \sim 
             Q^3\ .
 \end{equation}
 The last two combinations can be reduced similarly via the
 relations given in Appendix~\ref{app:B} (these too are derived from the
 EOM).  Terms with additional
 $D_\mu$'s do not contribute.

\item Terms with $u_\mu$ and $u_\nu$: These are of order
  $Q^2$ too.  The simplest combinations involve $\{u_\mu\, ,\, u_\nu\}$ and
  $[u_\mu\, ,\, u_\nu]$, which are both even under $P$; under $C$, however,
  the first is even and the second is odd. The only possibilities are
  \begin{equation}
   \bar\psi \{u_\mu\, ,\, u_\nu\}
   \sigma_{\alpha\beta}\epsilon^{\mu\nu\alpha\beta}\psi\ , \qquad
   \bar\psi\, [u_\mu\, ,\, u_\nu] g^{\mu\nu}i\gamma^5\psi\ ,
  \end{equation}
 but they vanish identically.  Terms with a $D_\mu$
  can again enter only as $\bar\psi \{D_\mu\, ,\, \ldots\}\psi$, and the only
  possible combinations are
  \begin{eqnarray}
   &&\bar\psi i\{D_\alpha\, ,\, \{u_\mu\, ,\,u_\nu\}\}
   g^{\mu\nu}\gamma^\alpha\gamma^5\psi\ ,\nonumber\\
   && \bar\psi i\{D_\alpha\, ,\, \{u_\mu\, ,\,u_\nu\}\}
   g^{\alpha\mu}\gamma^\nu\gamma^5\psi\ , \label{eq:C3}\\
   &&\bar\psi i\{D_\alpha\, ,\, i[u_\mu\, ,\, u_\nu] \} 
     \epsilon^{\mu\nu\alpha\beta}\gamma_\beta\psi\ .\nonumber
  \end{eqnarray}
However, the first and third terms are of order ${\cal O}(Q^3)$, as can be
seen using the relations~(\ref{eq:Ye1}) and~(\ref{eq:Ye3}).  
In the second term, use of the Cayley-Hamilton 
relation in isospin space ($\{u_\nu\, ,\, u_\alpha\}=\langle u_\nu u_\alpha \rangle$)
allows one to express it as $\bar\psi \, \{D^\mu \, ,\, \langle u_\mu
 u_\nu \rangle \} \gamma^\nu\gamma^5\,\psi\,$, listed
 as $O_1^{(0)}$ in Eq.~(\ref{eq:o0_1}).  It can be shown that 
possible terms with two covariant derivatives would be at least of order $Q^3$.

\item Terms with a single $u_\mu$ plus one or more $D^\mu$'s:
With a single $D^\mu$ we can form
  the combinations $\bar\psi\{D^\mu\, ,\, u_\mu\}\psi$ and
  $\bar\psi[D_\mu\, ,\, u_\nu]\sigma^{\mu\nu}\psi$.  Using the EOM up to
  order $Q$---see Eq.~(\ref{eq:eomn})---the first expression can be reduced to a
  combination of $O^{(0)}_{0V}$ and $O^{(0)}_2$ (defined in Sec.~\ref{sec:delta0})
   by ignoring terms of order $Q^3$.
  The second expression is seen to be identical to $2\, O^{(0)}_2$ via
  Eq.~(\ref{eq:rel2}). 
  Terms with two or more $D^\mu$'s
    can be reduced using the EOM.  In general, each $i D \!\!\!\! \slash \,\psi$ gives a term
    $M\psi$ plus terms of order $Q$ proportional to $u \!\!\! \slash$. Terms with
    the nucleon mass are found to be proportional to those
    without covariant derivatives, which have already been accounted for, while
    terms with the additional $u_\mu$ have been considered above.
    Therefore, at order $Q^2$, no new (independent) terms
    with a single $u_\mu$ and one or more $D^\mu$'s appear.
\end{enumerate}
\subsection{The $\Delta I=1$ sector}
\label{app:lq2_1}

\begin{enumerate}
\item Terms with $\chi_\pm$ or $F_{\mu\nu}^{\pm}$: We can combine
  these quantities with $X^3_\pm$ to form the following $P$-odd and
  $C$-odd combinations
\begin{eqnarray}
&&\bar\psi \{ \chi_+\, ,\,X^3_-\}\psi\ ,\quad
  \bar\psi\, [\chi_-\, ,\, X^3_+] \psi\ , 
   \nonumber\\
&&   \bar\psi\, i\,[ F^{\mu\nu}_+\, ,\, X^3_+
   ]\sigma^{\alpha\beta}\epsilon_{\mu\nu\alpha\beta}\,\psi\ ,\quad  
   \bar\psi \{ F^{\mu\nu}_+\, ,\, X^3_-\}\sigma_{\mu\nu}\psi\ , \label{eq:o1-chiF}\\
&& \bar\psi \{ F^{\mu\nu}_-\, ,\, X^3_+\}\sigma_{\mu\nu}\psi\ ,\quad
  \bar\psi\, i\, [ F^{\mu\nu}_-\, ,\, X^3_- ]\sigma^{\alpha\beta}\epsilon_{\mu\nu\alpha\beta}\psi\ .\nonumber
\end{eqnarray}
As per the isospin structure, for each of these terms one needs
to consider the following possibilities:
\begin{eqnarray}
 &&\bar\psi_{t} A_{tt'} B_{t' t''}\psi_{t''}\ , \quad
 \bar\psi_{t} A_{tt'} B_{t'' t''}\psi_{t'}\ , \quad
 \bar\psi_{t} A_{t't'} B_{t t''}\psi_{t''}\ , \nonumber \\
 &&\bar\psi_{t} A_{t't''} B_{t'' t'}\psi_{t}\ , \quad
 \bar\psi_{t} A_{t't'} B_{t'' t''}\psi_{t}\ ,\nonumber
\end{eqnarray}
where $A$ and $B$ denote schematically the various pairs
of isospin matrices corresponding to $\chi_+X^3_-$ (or $X^3_- \chi_+$)
and so on.
Obviously, if both $A$ and $B$ are traceless, only the first and
the fourth are non vanishing.
Recall that $\langle u_\mu\rangle=\langle X^a_\pm \rangle=
\langle\chi_-\rangle=\langle F^{\mu\nu}_-\rangle=0$. 
The other quantities ($\chi_+$, $F^{\mu\nu}_+$, and $D_\mu$) are 
conveniently written as $A=\hat A +\langle A\rangle\, I/2$ with
$\hat A$ traceless.  A number of manipulations allow one
to express the terms in Eq.~(\ref{eq:o1-chiF}) as the 8 combinations 
$O^{(1)}_{1-3}$ and $O^{(1)}_{14-18}$ listed in subsection~\ref{sec:delta1}. 
Combinations of $\chi_\pm$ or $F_{\mu\nu}^{\pm}$ with one or more
$D_\alpha$'s (in the form $\{D_\alpha\, ,\,\ldots\}$) can be eliminated using
the EOM. For the terms with  $F_{\mu\nu}^{\pm}$ it is necessary to use
the relations reported in Appendix~\ref{app:B}.
\begin{table}[h]
\begin{center}
\begin{tabular}{l|cccccc}
\hline
\hline
 & $ Y^{(1)}_{+,\mu\nu}$ & $ Y^{(2)}_{+,\mu\nu}$ & $ Y^{(3)}_{+,\mu\nu}$ & 
$Y^{(4)}_{+,\mu\nu}$ & $ Y^{(5)}_{+,\mu\nu}$ & $ Y^{(6)}_{+,\mu\nu}$ \\
\hline
$s_H$ & + & + & + & + & + &  +  \\
\hline
$s_P$ & + & + & + & + & + &  +  \\
\hline
$s_C$ & + & -- & -- & + & + & --  \\
\hline
\hline
 & $ Y^{(1)}_{-,\mu\nu}$ & $ Y^{(2)}_{-,\mu\nu}$ & $ Y^{(3)}_{-,\mu\nu}$ & 
$Y^{(4)}_{-,\mu\nu}$ & $ Y^{(5)}_{-,\mu\nu}$ & $ Y^{(6)}_{-,\mu\nu}$  \\
\hline
$s_H$ & + & + & + & + & + &  +  \\
\hline
$s_P$ & -- & -- & -- & -- & -- & -- \\
\hline
$s_C$ &  -- & + & + & -- & -- & + \\
\hline
\hline
\end{tabular}
\caption{ \label{tab:y}
Transformation properties under hermitian conjugation ($H$),
parity $P$, and charge conjugation $C$ of the quantities
  $Y^{(i)}_{\pm,\mu\nu}$ defined in the text.}
\end{center}
\end{table}
\item Terms with $u_\mu$ and $u_\nu$ and no $D_\mu\,$: Combined with a
$X^3_\pm$ we can form the quantities
\begin{equation}
 u_\mu u_\nu X\ , \qquad  u_\mu X u_\nu \ ,  \qquad X u_\mu u_\nu \ , 
\end{equation}
plus exchanges $\mu\longleftrightarrow\nu$, i.e., we can form 6 independent
quantities.  In order to have definite transformations under $H$, $P$, and $C$,
we consider the following combinations:
\begin{eqnarray}
  Y^{(1)}_{\pm,\mu\nu}&=&\{X^3_\pm\, ,\,\{u_\mu\, ,\,u_\nu\}\}\ , \nonumber\\
  Y^{(2)}_{\pm,\mu\nu}&=&\{X^3_\pm\, ,\, i\,[u_\mu\, ,\,u_\nu]\}\ , \nonumber\\
  Y^{(3)}_{\pm,\mu\nu}&=&i\, [X^3_\pm\, ,\, \{u_\mu\, ,\, u_\nu\}]\ ,\nonumber\\
  Y^{(4)}_{\pm,\mu\nu}&=&[X^3_\pm\, ,\, [u_\mu\, ,\, u_\nu]]\ , \label{eq:Y}\\
  Y^{(5)}_{\pm,\mu\nu}&=&u_\mu X^3_\pm u_\nu + u_\nu X^3_\pm u_\mu\ ,\nonumber\\
  Y^{(6)}_{\pm,\mu\nu}&=&i\Bigl(u_\mu X^3_\pm u_\nu - u_\nu X^3_\pm u_\mu\Bigr)\ . \nonumber
\end{eqnarray}
The properties under
$P$ and $C$ of the $Y^{(n)}$'s are summarized in
Table~\ref{tab:y}.  However, $Y^{(3)}_{\pm,\mu\nu}=0$ because
of the Cayley-Hamilton relation $\{u_\mu\, ,\, u_\nu\}=\langle u_\mu u_\nu\rangle$.
Similarly, $Y^{(6)}_{\pm,\mu\nu}$ is proportional to
$Y^{(2)}_{\pm,\mu\nu}$, since $uX=(\{u\, ,\,X\}+[u\, ,\, X])/2$
and hence
\begin{equation}
  u_\mu X u_\nu - u_\nu X u_\mu =
   {1\over 2} \langle X [u_\nu,u_\mu]\rangle\ , \label{eq:rell1}
\end{equation}
where we have used the relation $\{u_\mu\, ,\, u_\nu\} X= \langle u_\mu
u_\nu\rangle X= X \langle u_\mu u_\nu\rangle$, and so on.
Therefore, we will not consider $Y^{(3)}_{\pm,\mu\nu}$
and $Y^{(6)}_{\pm,\mu\nu}$ in the following analysis.
Since the pairs $Y^{(1)}_{\pm,\mu\nu}$,
$Y^{(5)}_{\pm,\mu\nu}$ and $Y^{(2)}_{\pm,\mu\nu}$,
$Y^{(4)}_{\pm,\mu\nu}$ are, respectively, symmetric and antisymmetric under the 
exchange $\mu\longleftrightarrow \nu$, the only allowed combinations are
\begin{eqnarray}
  &&\bar\psi \, Y^{(4)}_{+,\mu\nu}\,
   \epsilon^{\mu\nu\alpha\beta}\sigma_{\alpha\beta}\,\psi\ ,\qquad
  \bar\psi \,Y^{(1)}_{-,\mu\nu}\, g^{\mu\nu}\psi\ , \nonumber\\
  && \bar\psi \, Y^{(2)}_{-,\mu\nu}\, \sigma^{\mu\nu}\psi\ , \qquad
  \bar\psi\, Y^{(5)}_{-,\mu\nu}\, g^{\mu\nu}\psi\ .
\end{eqnarray}
In reference to isospin, we can again form different
structures depending how we contract the isospin indices.  However,
by taking into account that $X^3_\pm$ and $u_\mu$ are traceless, we
obtain the 4 combinations $O^{(1)}_{4-7}$ listed in Sec.~\ref{sec:delta1}.

\item Terms with $u_\mu$ and $u_\nu$, and with a single $D_\mu\,$:
 Since $u_\mu u_\nu$ is already of order $Q^2$,
  we can consider only combinations like
  $\bar\psi\{D_\alpha,Y^{(i)}_{\pm,\mu\nu}\}\psi$.  The three Lorentz indices must
be contracted with 
\begin{equation}
g^{\mu\nu} \gamma^\alpha,\quad
g^{\mu\nu}\gamma^\alpha\gamma^5,\quad
  \epsilon^{\mu\nu\alpha\beta}\gamma_{\beta},\quad
\epsilon^{\mu\nu\alpha\beta}\gamma_{\beta}\gamma^5\ .
\label{eq:3ind}
\end{equation}
The $P$- and $C$-allowed combinations  are:
\begin{itemize}
 \item $\bar\psi\{D_\alpha,Y^{(1)}_{+,\mu\nu}\}
  g^{\mu\nu}\gamma^\alpha\gamma^5\psi$ or $\bar\psi\{D_\alpha,Y^{(1)}_{+,\mu\nu}\}
  g^{\mu\alpha}\gamma^\nu\gamma^5\psi$: The first combination is of
  order $Q^3$ as can be seen from Eq.~(\ref{eq:Ye1}), while the second
  gives $O^{(1)}_{8}$ in Sec.~\ref{sec:delta1}.
%
 \item $\bar\psi\{D_\alpha,Y^{(2)}_{+,\mu\nu}\}
  \epsilon^{\mu\nu\alpha\beta}\gamma_\beta\psi$: From Eq.~(\ref{eq:Ye3})
  we see that it is of order $Q^3$.  This also follows from the fact
  that  $\beta$ must be 0 (otherwise $\bar\psi \,\gamma_i\, \psi\sim Q$) and hence, since $\alpha$ must
 be space-like, $D_i\psi\sim Q$. 

 \item $\bar\psi\{D_\alpha,Y^{(4)}_{+,\mu\nu}\}
    g^{\mu\nu}\gamma^\alpha\gamma^5\psi$ or $\bar\psi\{D_\alpha,Y^{(4)}_{+,\mu\nu}\}
    g^{\mu\alpha}\gamma^\nu\gamma^5\psi$: Since 
    $Y^{(4)}_{+,\nu\mu}=-Y^{(4)}_{+,\mu\nu}$, the first combination
    vanishes, while, using Eq.~(\ref{eq:Ye2}), the second can
    be written as 
    \begin{eqnarray}
    \lefteqn{
      \bar\psi\{D_\alpha,Y^{(4)}_{+,\mu\nu}\}g^{\mu\nu}\gamma^\alpha\gamma^5\psi
     \qquad\qquad\qquad}
      &&\nonumber\\ 
    &=& -{i\over 2}M\bar\psi Y^{(4)}_{+,\mu\nu}
       \epsilon^{\mu\nu\alpha\beta}\sigma_{\alpha\beta}\psi+\dots\ ,
    \end{eqnarray}
   ignoring terms of order $Q^3$, and therefore has already been accounted for above.

 \item $\bar\psi\{D_\alpha,Y^{(5)}_{+,\mu\nu}\}
    g^{\mu\nu}\gamma^\alpha\gamma^5\psi$ or $\bar\psi\{D_\alpha,Y^{(5)}_{+,\mu\nu}\}
    g^{\mu\alpha}\gamma^\nu\gamma^5\psi$: As for 
    $Y^{(1)}_{+,\mu\nu}$, the first combination is of order $Q^3$, while 
    the second gives $O^{(1)}_{9}$ in Sec.~\ref{sec:delta1}. 
    
 \item $\bar\psi\{D_\alpha,Y^{(1)}_{-,\mu\nu}\}
      g^{\mu\nu}\gamma^\alpha\psi$ or $\bar\psi\{D_\alpha,Y^{(1)}_{-,\mu\nu}\}
      g^{\mu\alpha}\gamma^\nu\psi$: The first combination is of order
      $Q^3$, see Eq.~(\ref{eq:Ye4}), while the second gives $O^{(1)}_{10}$
      in Sec.~\ref{sec:delta1}. 

 \item $\bar\psi\{D_\alpha,Y^{(2)}_{-,\mu\nu}\}
      \epsilon^{\mu\nu\alpha\beta}\gamma_\beta\gamma^5\psi$: 
      Using Eq.~(\ref{eq:Ye6}), this combination is proportional to
      $\bar\psi Y^{(2)}_{-,\mu\nu}  \sigma^{\mu\nu}\psi$ ignoring
      terms of order $Q^3$ and hence has already been considered.

\item $\bar\psi\{D_\alpha,Y^{(4)}_{-,\mu\nu}\}
      g^{\mu\nu}\gamma^\alpha\psi$ or $\bar\psi\{D_\alpha,Y^{(4)}_{-,\mu\nu}\}
      g^{\mu\alpha}\gamma^\nu\psi$: Since $Y^{(4)}_{-,\nu\mu}=-Y^{(4)}_{-,\mu\nu}$,
      both terms are of order $Q^3$, see Eqs.(\ref{eq:Ye4}) and~(\ref{eq:Ye5}).

\item $\bar\psi\{D_\alpha,Y^{(5)}_{-,\mu\nu}\}
      g^{\mu\nu}\gamma^\alpha\psi$ or $\bar\psi\{D_\alpha,Y^{(5)}_{-,\mu\nu}\}
      g^{\mu\alpha}\gamma^\nu\psi$: As for $Y^{(1)}_{-,\mu\nu}$, the
      first combination is of order $Q^3$, while the second 
      gives $O^{(1)}_{11}$ in Sec.~\ref{sec:delta1}. 
\end{itemize}
As far as the isospin structure of these combinations is concerned,
they consist of products of 4 $2\times 2$ matrices in the isospin space.
Their isospin indices can
be contracted in many different ways.  However, all matrices (except $D_\mu$)
are traceless and simplifications can be made via the Cayley-Hamilton relation.
For example, the isospin structure of a term like
$\bar\psi D X \{u_\mu, u_\nu\} \psi$ leads to
\begin{equation}
\bar\psi_{t} D_{tt'} X_{t' t''}\psi_{t''} \langle u_\mu u_\nu\rangle\,\,\, {\rm and}\,\,\,
\bar\psi_{t} D_{tt'} \psi_{t'} \langle X \{u_\mu, u_\nu\}\rangle\ , 
\end{equation}
but in the second combination
$\{u_\mu, u_\nu\}$ is proportional to the identity matrix (in isospin space)
and hence $ \langle X \{u_\mu, u_\nu\}\rangle= \langle X\rangle
\langle\{u_\mu, u_\nu\}\rangle=0$.   Combinations which do not include
$\{u_\mu, u_\nu\}$ are more complicated, since there are many ways to contract the
isospin indices.  In the following, we do not explicitly write down all possible
combinations, but just report the simplest one.  When using
these operators to construct interaction vertices by expanding in
powers of the pion field, all allowed possibilities should be considered.
In summary, only 4 combinations, those with $Y^{(1)}_{\pm,\mu\nu}$ and
$Y^{(5)}_{\pm,\mu\nu}$, are found to be independent, and give the interacton
terms $O^{(1)}_{8-11}$ listed in Sec~\ref{sec:delta1}.
Additional terms with two or more covariant derivatives do not
give additional independent terms.

\item Terms with a single $u_\mu$ and one or more $D_\mu$'s:
First consider terms with the anticommutator of the type
  $\bar\psi \{ D_\mu,\{X^3_\pm,u_\nu\}\}\ldots \psi$, which involve a
  $D_\mu$ acting on the nucleon fields $\bar\psi$ or $\psi$.  These terms
  can always be reduced via the EOM to one of the terms of order $Q$ given in
  Eqs.~(\ref{eq:o1_1V})--(\ref{eq:o1_1A}) plus a term $\sim u_\mu u_\nu$,
  already considered at points 2.~and 3.~above.
  Next, we consider terms with the commutator of type 
  $[D_\mu,X^3_{L/R}]$ or $[D_\mu,u_\nu]$.  As discussed in the main text (see also
  Appendix~\ref{app:note1}),
  combinations of $D_\mu$ with $X^a_{L/R}$ must be included via
 $(X^a_{L/R})_\mu$ defined
  in Eqs.~(\ref{eq:xxRL1})--(\ref{eq:xxRL2}).  However, by using the
  identities~(\ref{eq:xxid}), combinations with a single
  $u_\mu$ and a $(X^a_{R/L})_\nu$ reduce to terms $\propto 
  u_\mu u_\nu X^a_{R/L}$, already discussed at points 2.~and 3.~above.
Turning our attention to terms including a commutator
  $[D_\mu,u_\nu]$, we note that, since
  $[D_\mu,u_\nu]-[D_\nu,u_\mu]=F^-_{\mu\nu}$ and we have already
  discussed the operators that can be constructed with
  $F^-_{\mu\nu}$ in point 1.~above, we only need to consider
  operators involving $h_{\mu\nu}$, as defined
  in Eq.~(\ref{eq:hmunu}), which is odd under $P$ and even
  under $C$. In combination with $X^3_\pm$ we can form the 4
  operators listed in Table~\ref{tab:h}, along with their transormation
  properties under $P$ and $C$. Note that $h_{\mu\nu}$ is of order 
  ${\cal O}(Q^2)$.
\begin{table}[h]
\begin{center}
\begin{tabular}{l|cccc}
\hline
\hline
 & $ \{ h_{\mu\nu},X^3_+\}$ & $ i[h_{\mu\nu},X^3_+]$ & 
  $ \{ h_{\mu\nu},X^3_-\}$ & $ i[h_{\mu\nu},X^3_-]$ \\
\hline
$s_H$ & + & + & + & +   \\
\hline
$s_P$ & -- & -- & + & +   \\
\hline
$s_C$ & + & -- & -- & +   \\
\hline
\hline
\end{tabular}
\caption{ \label{tab:h}
Transformation properties under hermitian conjugation ($H$),
parity ($P$), and charge conjugation ($C$) of quantities
constructed in terms of $h_{\mu\nu}$.}
\end{center}
\end{table}
Since
    $h_{\mu\nu}=h_{\nu\mu}$, without any additional covariant
     derivatives we can construct the terms:
   \begin{equation}
     \bar\psi i[h_{\mu\nu},X^3_+] g^{\mu\nu}\psi\ ,\qquad
     \bar\psi \{ h_{\mu\nu},X^3_-\} g^{\mu\nu}\gamma^5\psi\ .
   \end{equation}
   However, using the pion EOM in Eq.~(\ref{eq:eomp}), these terms are
   the same, up to additional terms of order ${\cal O}(Q^4)$, 
    as those constructed with $\chi_-$ (see point 1. above).
   Operators with
   $h_{\mu\nu}$ and an additional covariant derivative enter
      only in combinations with $\{D_\mu,\ldots\}$.  Possible
  ones are:
   \begin{eqnarray}
     &&\bar\psi \{ D_\alpha, i[h_{\mu\nu},X^3_+]\}
     g^{\alpha\mu}\gamma^\nu\psi\ ,\nonumber \\
     &&\bar\psi \{ D_\alpha, i[h_{\mu\nu},X^3_-]\}
     g^{\alpha\mu}\gamma^\nu\gamma^5\psi\ , \nonumber
   \end{eqnarray}
   where we have excluded terms like
   $h_{\mu\nu}g^{\mu\nu}=i\hat{\chi}_-+{\cal O}(Q^4)$. 
   They give the operators $O^{(1)}_{12-13}$ in Sec.~\ref{sec:delta1}.
\end{enumerate}
\subsection{The $\Delta I=2$ sector}
\label{app:lq2_2}
The $\Delta I=2$ operators have to be constructed as combinations
$ {\cal I}^{ab}(X^a_R \,O \,X^b_R\pm X^a_L \,O \,X^b_L)$, with $O$ 
transforming under chiral transformations as $O\longrightarrow  h\, O\, h^\dag$.
At order $Q^2$ we have:
\begin{enumerate}
\item Terms with $\chi_\pm$ or $F_{\mu\nu}^{\pm}$: We can 
 form the following $P$-odd and $C$-odd combinations
\begin{eqnarray}
&& {\cal I}_{ab}\bar\psi \Bigl( X^a_R \chi_+ X^b_R  -  X^a_L \chi_+
     X^b_L\Bigr) \psi\ ,\nonumber \\
&&  {\cal I}_{ab}\bar\psi \Bigl( X^a_R \chi_- X^b_R  -  X^a_L \chi_-
     X^b_L\Bigr)i\gamma^5 \psi\ ,\nonumber \\
&&  {\cal I}_{ab}\bar\psi \Bigl( X^a_R F_+^{\mu\nu} X^b_R  -  X^a_L F_+^{\mu\nu}
     X^b_L\Bigr)\sigma_{\mu\nu} \psi\ ,\nonumber \\
&& {\cal I}_{ab}\bar\psi \Bigl( X^a_R F_-^{\mu\nu} X^b_R  +  X^a_L F_-^{\mu\nu}
     X^b_L\Bigr)\sigma_{\mu\nu} \psi\ .\nonumber
\end{eqnarray}
The second combination is of order $Q^3$, while the remaining three
are the operators $O^{(2)}_{1}$ and  $O^{(2)}_{7-8}$ reported 
in Sec.~\ref{sec:delta2}. 
As  for the $\Delta I=1$ case, combinations of $\chi_\pm$ or
$F_{\mu\nu}^{\pm}$ with one or more operators $D_\alpha$ (in the form
$\{D_\alpha,\ldots\}$) can be eliminated using the EOM.  For the terms
with  $F_{\mu\nu}^{\pm}$ it is necessary to use the relations reported
in Appendix~\ref{app:B}. 
\item  Terms with $u_\mu$ and $u_\nu$ and no $D_\mu$:
We can have the combinations $X u\, u X$ or
  $uXuX\pm Xu Xu$.  We observe that under $P$ and $C$
\begin{eqnarray}
 &&X^a_R u_\mu u_\nu X^a_R \stackrel{P}{\longrightarrow}  X^a_L u_\mu u_\nu
  X^a_L\ ,  \\
 &&X^a_R u_\mu u_\nu X^a_R \stackrel{C}{\longrightarrow}  \Bigl(X^a_L u_\nu u_\mu
    X^a_L\Bigr)^T\ ,
\end{eqnarray}
and
\begin{eqnarray}
 &&u_\mu X^a_R u_\nu X^a_R \stackrel{P}{\longrightarrow}  u_\mu X^a_L u_\nu X^a_L\ , \\
 && u_\mu X^a_R u_\nu X^a_R \stackrel{C}{\longrightarrow}  \Bigl(X^a_L u_\nu X^a_L
 u_\mu \Bigr)^T\ .
\end{eqnarray}
and therefore we consider the combinations
\begin{eqnarray}
 \overline {Y}^{(1)}_{\pm,\mu\nu} &=&  {\cal I}_{ab} \Bigl(
  u_\mu X^a_R u_\nu  X^b_R + X^a_R u_\nu X^b_R u_\mu \nonumber\\
  &&\qquad + u_\nu X^a_R u_\mu X^b_R  + X^a_R u_\mu X^b_R u_\nu
  \Bigr) \nonumber\\
  && \qquad \pm(L\longrightarrow R)\ , \label{eq:barY1}\\
 \overline {Y}^{(2)}_{\pm,\mu\nu} &=&  {\cal I}_{ab} \Bigl(
  u_\mu X^a_R u_\nu  X^b_R + X^a_R u_\nu X^b_R u_\mu \nonumber\\ 
  &&\qquad - u_\nu X^a_R u_\mu X^b_R  - X^a_R u_\mu X^b_R u_\nu
  \Bigr) \nonumber\\
  &&\qquad \pm(L\longrightarrow R)\ ,  \label{eq:barY2}\\
 \overline {Y}^{(3)}_{\pm,\mu\nu} &=& i\, {\cal I}_{ab} \Bigl(
  u_\mu X^a_R u_\nu  X^b_R - X^a_R u_\nu X^b_R u_\mu \nonumber\\
  &&\qquad +u_\nu X^a_R u_\mu X^b_R  - X^a_R u_\mu X^b_R u_\nu 
   \Bigr) \nonumber\\
  && \qquad \pm(L\longrightarrow R)\ , \label{eq:barY3}\\
 \overline {Y}^{(4)}_{\pm,\mu\nu} &=& i\, {\cal I}_{ab} \Bigl(
  u_\mu X^a_R u_\nu  X^b_R - X^a_R u_\nu X^b_R u_\mu \nonumber\\
  &&\qquad -u_\nu X^a_R u_\mu X^b_R  + X^a_R u_\mu X^b_R u_\nu  
 \Bigr)\nonumber\\
  && \qquad \pm(L\longrightarrow R)\ , \label{eq:barY4}\\
 \overline {Y}^{(5)}_{\pm,\mu\nu}&=& {\cal I}_{ab} X^a_R\{u_\mu,u_\nu\}X^b_R
  \pm(L\longrightarrow R)\ , \label{eq:barY5}\\
 \overline {Y}^{(6)}_{\pm,\mu\nu}&=&i\, {\cal I}_{ab} X^a_R[u_\mu,u_\nu]X^b_R
  \pm(L\longrightarrow R)\ , \label{eq:barY6}
\end{eqnarray}
with the transformation properties under $P$ and $C$
summarized in Table~\ref{tab:bary}. 
\begin{table}[h]
\begin{center}
\begin{tabular}{l|cccccc}
\hline
\hline
 & $ \overline {Y}^{(1)}_{+,\mu\nu}$ & $ \overline {Y}^{(2)}_{+,\mu\nu}$ & $ \overline {Y}^{(3)}_{+,\mu\nu}$ & 
$\overline {Y}^{(4)}_{+,\mu\nu}$ & $ \overline {Y}^{(5)}_{+,\mu\nu}$ & $ \overline {Y}^{(6)}_{+,\mu\nu}$ \\
\hline
$s_H$ & + & + & + & + & + &  +  \\
\hline
$s_P$ & + & + & + & + & + &  +  \\
\hline
$s_C$ & + & + & -- & -- & + & --  \\
\hline
\hline
 & $ \overline {Y}^{(1)}_{-,\mu\nu}$ & $ \overline {Y}^{(2)}_{-,\mu\nu}$ & $ \overline {Y}^{(3)}_{-,\mu\nu}$ & 
$\overline {Y}^{(4)}_{-,\mu\nu}$ & $ \overline {Y}^{(5)}_{-,\mu\nu}$ & $ \overline {Y}^{(6)}_{-,\mu\nu}$  \\
\hline
$s_H$ & + & + & + & + & + &  +  \\
\hline
$s_P$ & -- & -- & -- & -- & -- & -- \\
\hline
$s_C$ &  -- & -- & + & + & -- & + \\
\hline
\hline
\end{tabular}
\caption{ \label{tab:bary}
Transformation properties under hermitian conjugation ($H$), parity ($P$),
and charge conjugation ($C$) of the quantities
  $\overline {Y}^{(i)}_{\pm,\mu\nu}$ defined in the text.}
\end{center}
\end{table}
We further note that
\begin{eqnarray}
   \lefteqn{  u_\mu X u_\nu X + X u_\nu X u_\mu \qquad\qquad\qquad}
   &&\nonumber\\
   &=& {1\over 2} \{u_\mu,X\} \{u_\nu,X\}
      +{1\over 4} [u_\mu,X] [u_\nu,X] \nonumber \\
   &&   +{1\over 4} [u_\nu,X] [u_\mu,X]\ , \nonumber
\end{eqnarray}
where use has been made of the identity $\{u_\mu,X\}[u_\nu,X]+[X,u_\nu]\{X,u_\mu\}=0$
which follows from the Cayley-Hamilton relation $\{u_\mu,X\}=\langle u_\mu X\rangle$
and the fact that $\langle u_\mu X\rangle$ commutes with $[u_\nu,X]$.  The final expression is
symmetric under the exchange $\mu \longleftrightarrow\nu$ and hence
$\overline {Y}^{(2)}_{\pm,\mu\nu}=0$.  Similarly,
\begin{eqnarray}
  \overline {Y}^{(5)}_{\pm,\mu\nu}&=&
 {\cal I}_{ab} \Bigl[X^a_R \langle u_\mu u_\nu\rangle X^b_R\pm
 (L\longrightarrow R)\Bigr] \ ,\nonumber \\
   &=&   {\cal I}_{ab} \Bigl[X^a_R X^b_R\pm
 (L\longrightarrow R)\Bigr] \langle u_\mu u_\nu \rangle=0 \ ,
  \nonumber
\end{eqnarray}
and the combinations $\overline {Y}^{(2)}_{\pm,\mu\nu}$ and
$\overline {Y}^{(5)}_{\pm,\mu\nu}$ can be disregarded in the
analysis that follows.
By taking into account that the pairs $\overline {Y}^{(1)}_{\pm,\mu\nu}$,
$\overline {Y}^{(3)}_{\pm,\mu\nu}$ and $\overline {Y}^{(4)}_{\pm,\mu\nu}$,
$\overline {Y}^{(6)}_{\pm,\mu\nu}$ are, respectively, symmetric and antisymmetric under the
exchange $\mu\longleftrightarrow \nu$, the possible combinations are
\begin{equation}
  \bar\psi \overline {Y}^{(1)}_{-,\mu\nu} g^{\mu\nu}\psi\ , \qquad
  \bar\psi \overline {Y}^{(4)}_{-,\mu\nu} \sigma^{\mu\nu}\psi\ ,\qquad
  \bar\psi \overline {Y}^{(6)}_{-,\mu\nu} \sigma^{\mu\nu}\psi\ .\nonumber
\end{equation}
Note that allowed terms such as $\bar\psi Y^{(3)}_{+,\mu\nu}
g^{\mu\nu}i\gamma^5\psi$ are of order $Q^3$.
As per isospin, we can again form different
structures depending on how the isospin indices are contracted.  However, since
the $X_{L/R}$ and $u_\mu$ are traceless, we arrive at the 3 operators
$O^{(2)}_{2-4}$ listed in Sec.~\ref{sec:delta2}.

\item Terms with $u_\mu$ and $u_\nu$ and a single $D_\mu$:
Since $u_\mu u_\nu$ is already of order $Q^2$,
  we need consider only combinations like
  $\bar\psi\{D_\mu,\overline{Y}^{(i)}_{\pm,\nu\alpha}\}\psi$.  These quantities
  have three Lorentz indices, which must be contracted with the
  operators given in Eq.~(\ref{eq:3ind}).  Typical combinations
  are $\bar\psi\{D_\alpha,\overline {Y}_{\mu\nu}\}
  g^{\mu\nu}\gamma^\alpha\gamma^5\psi$ or
  $\bar\psi\{D_\alpha,\overline{Y}_{\mu\nu}\}
  g^{\alpha\mu}\gamma^\nu\gamma^5\psi$, where 
  $\overline {Y}_{\mu\nu}$ may be one of the operators defined in
  Eqs.~(\ref{eq:barY1})--(\ref{eq:barY6}).  The various terms can then be
  reduced by using the relations given in
  Eqs.~(\ref{eq:Ye1})--(\ref{eq:Ye6}).  We can
 construct the following $P$-odd and $C$-odd quantities:
\begin{itemize}
\item $\bar\psi\{D_\alpha,\overline {Y}^{(1)}_{+,\mu\nu}\}
  g^{\mu\nu}\gamma^\alpha\gamma^5\psi$ or $\bar\psi\{D_\alpha,\overline {Y}^{(1)}_{+,\mu\nu}\}
  g^{\alpha\mu}\gamma^\nu\gamma^5\psi$: The first combination is seen to be of
  order $Q^3$, while the second gives the operator $O^{(2)}_5$ 
listed in Sec.~\ref{sec:delta2}.
\item $\bar\psi\{D_\alpha,\overline {Y}^{(3)}_{+,\mu\nu}\}
  \epsilon^{\mu\nu\alpha\beta}\gamma_\beta\psi$ or  $\bar\psi\{D_\alpha,\overline {Y}^{(3)}_{-,\mu\nu}\}
  \epsilon^{\mu\nu\alpha\beta}\gamma_\beta\gamma^5\psi$: These combinations
  vanish since  $\overline{Y}^{(3)}_{\pm,\mu\nu}$ is symmetric in the
  indices $\mu\nu$.
 \item $\bar\psi\{D_\alpha,\overline {Y}^{(4)}_{+,\mu\nu}\}
  \epsilon^{\mu\nu\alpha\beta}\gamma_\beta\psi$ or
   $\bar\psi\{D_\alpha,\overline {Y}^{(6)}_{+,\mu\nu}\}
  \epsilon^{\mu\nu\alpha\beta}\gamma_\beta\psi$: These combinations
  are at least of order $Q^3$.
\item  $\bar\psi\{D_\alpha,\overline {Y}^{(1)}_{-,\mu\nu}\}
  g^{\mu\nu}\gamma^\alpha\psi$ or  $\bar\psi\{D_\alpha,\overline {Y}^{(1)}_{-,\mu\nu}\}
  g^{\alpha\mu}\gamma^\nu\psi$: Using Eq.~(\ref{eq:Ye4}) and ignoring terms
  of order $Q^3$, the first combination reduces to
  $-2iM O^{(2)}_2$, while the
  second gives the operator $O^{(2)}_{6}$
  of Sec.~\ref{sec:delta2}.
\item
 $\bar\psi\{D_\alpha,\overline {Y}^{(4)}_{-,\mu\nu}\}
  \epsilon^{\mu\nu\alpha\beta}\gamma_\beta\gamma^5\psi$:
  Using Eq.~(\ref{eq:Ye6}) and ignoring terms of order $Q^3$, this combination reduces to
  $2iM \bar\psi  \, \overline{Y}^{(4)}_{-,\mu\nu}\sigma^{\mu\nu}
   = 2iM O^{(2)}_3$.
\item
 $\bar\psi\{D_\alpha,\overline {Y}^{(6)}_{-,\mu\nu}\}
  \epsilon^{\mu\nu\alpha\beta}\gamma_\beta\gamma^5\psi$:
  As for $ \overline{Y}^{(4)}_{-,\mu\nu}$, this combination can be disregarded.
\end{itemize}
\item Terms with a single $u_\mu$ and one or more $D_\mu$'s:
First, we consider combinations with the anticommutator like
  $\bar\psi\, {\cal I}_{ab}\{ D_\mu,X^a_R u_\nu X^b_R \pm
  (L\longrightarrow R) \}\cdots \psi$, namely with
  $D_\mu$ acting on the nucleon fields $\bar\psi$ or $\psi$.  Using the EOM,
these can always be reduced to i) one of the order $Q$ terms given in
  Eqs.~(\ref{eq:o2_2v})--(\ref{eq:o2_2a}), ii) terms involving $u_\mu u_\nu$
  which have already been considered at points 2.~and 3.~above, and iii)
  terms with the commutator of $D_\mu$, as shown below. For example, 
  for
  \begin{eqnarray}
  O_\nu&=& {\cal I}_{ab}(X^a_R u_\nu X^b_R + X^a_L u_\nu X^b_L) \ ,\\
 \overline{O}_\nu&=& {\cal I}_{ab}(X^a_R u_\nu X^b_R - X^a_L u_\nu X^b_L) \ ,
 \end{eqnarray}
  we have, respectively
  \begin{eqnarray}
    \lefteqn{\bar\psi\{ D_\mu \, ,\, O_\nu\}
      g^{\mu\nu}\psi\qquad\qquad\qquad}&&
       \nonumber \\
    &=& -2iM O^{(2)}_{2V} + \bar\psi\Bigl[ D_\mu\, ,\, O_\nu\Bigr]
     i\sigma^{\mu\nu}\psi \nonumber\\
    && + ({\rm terms\ with\ }u_\mu,u_\nu)\ ,\nonumber \\
    \lefteqn{\bar\psi\{ D_\mu \, ,\, \overline{O}_\nu\} \epsilon^{\mu\nu\alpha\beta}
    \sigma_{\alpha\beta}\psi\qquad\qquad\qquad}&&
       \nonumber \\
    &=& 4iM O^{(2)}_{2A} -2 \bar\psi\Bigl[ D_\mu\, ,\, \overline{O}^\mu\Bigr]\psi
    \nonumber \\
    && + ({\rm
       terms\ with\ }u_\mu,u_\nu)\ ,\nonumber
  \end{eqnarray}
  where the operators $O^{(2)}_{2V}$ and $O^{(2)}_{2A}$ are 
   given in Eqs.~(\ref{eq:o2_2v}) and~(\ref{eq:o2_2a}), respectively.

  Next, we consider the terms with the commutator of type 
  $[D_\mu,X^3_{L/R}]$.  As discussed in Appendix~\ref{app:note1},
  combinations of $D_\mu$ with $X^a_{L/R}$ must be included via
  $(X^a_{L/R})_\mu$.  However, by using the
  identities~(\ref{eq:xxRL1app}) and~(\ref{eq:xxRL2app}), terms with a single
  $u_\mu$ and a $(X^a_{R/L})_\nu$ are $\propto 
  u_\mu u_\nu X^a_{R/L}$, already discussed at points 2.~and 3.~above.
Turning our attention to terms including a commutator
  $[D_\mu,u_\nu]$, we note that, since $[D_\mu,u^\mu]=(i/2) 
   \hat{\chi}_-+{\cal O}(Q^4)$ and 
  $[D_\mu,u_\nu]-[D_\nu,u_\mu]=F^-_{\mu\nu}$ and we have already
  discussed the operators that can be constructed with $\chi_-$ and 
  $F^-_{\mu\nu}$ in point 1.~above, we need only consider
  operators involving $h_{\mu\nu}$. We can form two combinations:
   \begin{eqnarray}
   &&  \bar\psi \,{\cal I}_{ab} \Bigl(X^a_R h_{\mu\nu} X^b_R + X^a_L
     h_{\mu\nu} X^b_L\Bigr) \sigma^{\mu\nu}\psi\ ,\\
  &&  \bar\psi \, {\cal I}_{ab} \Bigl(X^a_R h_{\mu\nu} X^b_R - X^a_L
     h_{\mu\nu} X^b_L\Bigr) g^{\mu\nu}i\gamma^5\psi\ .
   \end{eqnarray}
   They both can be disregarded: the first vanishes
     because of the symmetry of $h_{\mu\nu}$ and
   the second is of order $Q^3$ because of the $\gamma^5$ and
   the fact that $h_{\mu\nu}$ is already of order $Q^2$. 
   Combinations with additional $D_\alpha$'s can only be of the form
   $\{D_\alpha,\ldots h_{\mu\nu}\ldots \}$. However, the only $P$-odd
   and $C$-odd combinations that can be formed,
      \begin{eqnarray}
     &&{\cal I}_{ab} \, \bar\psi\Bigl\{ D_\alpha\, ,\, X^a_R h_{\mu\nu} X^b_R + X^a_L
     h_{\mu\nu} X^b_L\Bigr\}\nonumber\\
     &&\qquad\quad\times \epsilon^{\mu\nu\alpha\beta}\gamma_\beta\gamma^5\psi\ ,\\
     &&{\cal I}_{ab} \,  \bar\psi\Bigl\{ D_\alpha\, ,\, X^a_R h_{\mu\nu} X^b_R - X^a_L
     h_{\mu\nu} X^b_L\Bigr\}\nonumber\\
     &&\qquad\quad\times \epsilon^{\mu\nu\alpha\beta}\gamma_\beta\psi\ ,
   \end{eqnarray}
    vanish since $h_{\mu\nu}=h_{\nu\mu}$, and therefore there are no $\Delta I= 2$ terms with $h_{\mu\nu}$.
\end{enumerate}
\section{Properties of the $\gamma$ matrices}
\label{app:A}
The $\gamma^\mu$ matrices are in the standard form as given, for example, in
Ref.~\cite{Gross93}.  They satisfy the following identities:
\begin{eqnarray}
  \sigma^{\mu\nu}&=&{i\over 2}\,[\gamma^\mu,\gamma^\nu]\ ,\label{eq:sigma}\\
  \gamma^5\sigma^{\mu\nu}&=& {i\over 2}\, \epsilon^{\mu\nu\alpha\beta}
  \sigma_{\alpha\beta} \ , \label{eq:sigmagamma5}\\
  i\, \sigma^{\mu\nu} &=& g^{\mu\nu} -\gamma^\mu\gamma^\nu\ ,\label{eq:sigmamunu}\\
  {1\over 2} \{ \sigma^{\mu\nu},\gamma^\alpha \} &=&
  \epsilon^{\mu\nu\alpha\beta}\gamma^5\gamma_\beta\ , \label{eq:sigmagamma1}\\
  {1\over 2}\,  [\sigma^{\mu\nu},\gamma^\alpha] &=& -i\,
  g^{\mu\alpha}\gamma^\nu + i\, g^{\nu\alpha}\gamma^\mu \ ,\label{eq:sigmagamma2}\\
  \sigma^{\mu\nu} \gamma^\alpha  &=&
  \epsilon^{\mu\nu\alpha\beta}\gamma^5\gamma_\beta
 -i \, g^{\mu\alpha}\gamma^\nu + i \, g^{\nu\alpha}\gamma^\mu \ ,\label{eq:sigmagamma3}\\
  \gamma^\alpha \sigma^{\mu\nu}  &=&
  \epsilon^{\mu\nu\alpha\beta}\gamma^5\gamma_\beta
 +i \, g^{\mu\alpha}\gamma^\nu - i\, g^{\nu\alpha}\gamma^\mu \ .\label{eq:sigmagamma4}
\end{eqnarray}
\section{Useful relations}
\label{app:B}
Let ${Y}_{\mu\nu}$ be a (pion field dependent) quantity of order $Q^2$.
Using the EOM and the properties of the $\gamma$ matrices given in 
Eqs.~(\ref{eq:sigma})--(\ref{eq:sigmagamma4}), it follows that
\begin{eqnarray}
     \lefteqn{ \bar\psi\{D_\alpha\, ,\,{Y}_{\mu\nu}\}
      g^{\mu\nu}\gamma^\alpha\gamma^5\psi\qquad\qquad\qquad}
     && \nonumber \\
     &=&  -iM \bar\psi \Bigl({Y}_{\mu\nu} g^{\mu\nu}\gamma^5
              -{Y}_{\mu\nu} g^{\mu\nu}\gamma^5\Bigr)\psi + {\cal
                O}(Q^3) \nonumber \\
     &=& {\cal  O}(Q^3)\ ,\label{eq:Ye1}
\end{eqnarray}
and
\begin{eqnarray}
   \lefteqn{ \bar\psi\{D_\alpha\, ,\,{Y}_{\mu\nu}\}
      g^{\alpha\mu}\gamma^\nu\gamma^5\psi\qquad\qquad\qquad}
     && \nonumber \\
     &=&  \bar\psi \Bigl[ D_\alpha {Y}_{\mu\nu}
        (\gamma^\alpha\gamma^\mu+i\, \sigma^{\alpha\mu})\gamma^\nu\gamma^5\nonumber \\
     &&\qquad   + {Y}_{\mu\nu}\gamma^\nu\gamma^5 
           (\gamma^\mu\gamma^\alpha+i\, \sigma^{\mu\alpha})D_\alpha \Bigr]\psi 
        \nonumber \\
     &=& -iM\bar\psi\Bigl(  {Y}_{\mu\nu} \gamma^\mu\gamma^\nu\gamma^5
         + {Y}_{\mu\nu}\gamma^\nu\gamma^5\gamma^\mu\Bigr)\psi +  {\cal
           O}(Q^3)\nonumber\\
      && \qquad + i\bar\psi \Bigl(D_\alpha {Y}_{\mu\nu}
         \sigma^{\alpha\mu}\gamma^\nu\gamma^5
          +  {Y}_{\mu\nu}D_\alpha \gamma^\nu\gamma^5
           \sigma^{\mu\alpha}\Bigr)\psi \nonumber\\
     &=& -iM\bar\psi {Y}_{\mu\nu}\epsilon^{\mu\nu\alpha\beta}\sigma_{\alpha\beta}
           \psi +{\cal O}(Q^3) \nonumber \\
     &&  \qquad -i \bar\psi \Bigl[D_\alpha\, ,\,{Y}_{\mu\nu}\Bigr] \epsilon^{\mu\nu\alpha\beta}\gamma_\beta\psi
        \nonumber\\
     &&  \qquad + \bar\psi \{D_\alpha\, ,\,{Y}_{\nu\mu}\}
      g^{\alpha\mu}\gamma^\nu\gamma^5\psi\ . \label{eq:Ye2}
\end{eqnarray}
Note that in the last row we obtain the same starting expression, but
with ${Y}_{\nu\mu}$ instead of ${Y}_{\mu\nu}$ and that
$[D_\alpha\, ,\,Y_{\mu\nu}]\sim {\cal O}(Q^3)$, since the four-gradient now
acts on the pion field.  Similarly, one shows that
\begin{eqnarray}
    \lefteqn{ \bar\psi\{D_\alpha\, ,\, {Y}_{\mu\nu}\}
      g^{\mu\nu}\gamma^\alpha\psi=}\qquad\qquad
     &&\nonumber\\
     &=&  -2iM \bar\psi  {Y}_{\mu\nu}  g^{\mu\nu}\psi +
     {\cal  O}(Q^3)\ ,\label{eq:Ye4}\\
    \lefteqn{\bar\psi\{D_\alpha\, , \, {Y}_{\mu\nu}\}
      g^{\alpha\mu}\gamma^\nu\psi=}\qquad\qquad
  &&\nonumber\\
     &=&  \bar\psi\{D_\alpha, {Y}_{\nu\mu}\}
      g^{\alpha\mu}\gamma^\nu\psi+
     {\cal  O}(Q^3)\ ,\label{eq:Ye5}\\
    \lefteqn{\bar\psi\{D_\alpha\, ,\, {Y}_{\mu\nu}\} \epsilon^{\mu\nu\alpha\beta}
      \gamma_\beta\psi=}\qquad\qquad
  &&\nonumber\\
     &=&   {\cal  O}(Q^3)\ ,\label{eq:Ye3}\\
     \lefteqn{\bar\psi\{D_\alpha\, ,\, {Y}_{\mu\nu}\} \epsilon^{\mu\nu\alpha\beta}
      \gamma_\beta\gamma^5\psi=}\qquad\qquad
  &&\nonumber\\
    &=&  2iM \bar\psi  {Y}_{\mu\nu}  \sigma^{\mu\nu}\psi+
     {\cal  O}(Q^3)\ .\label{eq:Ye6}
\end{eqnarray}
\egroup
\section{Interaction vertices}
\label{app:vertex}

The various building blocks, $u_\mu$, $\Gamma_\mu$, and so on, are
expanded in powers of the pion field.  It is convenient to decompose
the interaction Hamiltonian $H_I$ as follows
\bgroup
\arraycolsep=0.5pt
\begin{eqnarray}
 H_I\!=\!H^{00}\!+\!H^{01}\!+\!H^{10}\!+\!H^{02}\!+\!H^{11}\!+\!H^{20}\!+\!\cdots\ ,
\end{eqnarray}
where $H^{nm}$ has $n$ creation and $m$ annihilation 
operators for the pion, and
\begin{eqnarray}
 H^{00} & = & {1\over \Omega}\sum_{\a_1' \a_1\a_2' \a_2} 
 b^{\dag}_{\a_1'}b_{\a_1}  b^{\dag}_{\a_2'} b_{\a_2} \nonumber\\
    &&\qquad\times M^{00}_{\a_1'\a_1\a_2'\a_2}
 \delta_{\bmp_1'+\bmp_2' ,\bmp_1+\bmp_2}\ , \label{eq:m00} \\
 H^{01}& = & \frac{1}{\sqrt{\Omega}} \sum_{\a' \a}\sum_{\bmq \, a} 
   b^{\dag}_{\a'}b_{\a}a_{\bmq\, a}  M^{01}_{\a'\a,\bmq\, a}
   \delta_{\bmq+\bmp,\bmp'}\ ,\label{eq:m01}\\
 H^{10}& = & \frac{1}{\sqrt{\Omega}} \sum_{\a' \a}\sum_{\bmq\, a} 
   b^{\dag}_{\a'}b_{\a}a^\dag_{\bmq\, a}  M^{10}_{\a'\a,\bmq\, a}
   \delta_{\bmq+\bmp',\bmp}\ ,\label{eq:m10}\\
 H^{02} &= & \frac{1}{\Omega}\sum_{\a' \a}\sum_{\bmq' a'\,\bmq\, a}
  b_{\a'}^{\dag}b_{\a}a_{\bmq' a'}a_{\bmq\, a}  \nonumber\\
    &&\qquad\times
  M^{02}_{\a'\a,\bmq'  a'\, \bmq \, a}
 \delta_{\bmq+\bmq'+\bmp,\bmp'}\ ,\label{eq:m02}\\
 H^{11} &= & \frac{1}{\Omega}\sum_{\a' \a}\sum_{\bmq' a'\, \bmq\, a}
  b_{\a'}^{\dag}b_{\a}a^\dag_{\bmq' a'}a_{\bmq\,a } \nonumber\\
    &&\qquad\times
  M^{11}_{\a'\a,\bmq'a'\bmq \, a}
 \delta_{\bmq+\bmp,\bmq'+\bmp'}\ ,\label{eq:m11}\\
 H^{20} &= & \frac{1}{\Omega}\sum_{\a' \a}\sum_{\bmq'a' \,\bmq\,a}
  b_{\a'}^{\dag}b_{\a}a^\dag_{\bmq' a'}a^\dag_{\bmq \, a} \nonumber\\
    &&\qquad\times
  M^{20}_{\a'\a,\bmq' a'\, \bmq \,a}
 \delta_{\bmp,\bmq+\bmq'+\bmp'}\ .\label{eq:m20}
\end{eqnarray}
\egroup
Here $\alpha_j\equiv \bmp_j,s_j,t_j$ denotes the
momentum, spin projection, isospin projection of nucleon $j$ with energy
$E_j=\sqrt{p_j^2+M^2}$, $\bmq$ and $a$ denote the momentum and isospin
projection of a pion with energy $\omega_q=\sqrt{\bmq^2+m_\pi^2}$, and
$M^{nm}$ are the vertex functions listed below.
The various momenta are discretized by assuming periodic boundary
conditions in a box of volume $\Omega$.  We note that in the expansion
of the nucleon field $\psi$ we have only retained the nucleon degrees of freedom,
since anti-nucleon contributions do not enter the PV $NN$ potential at the order
$Q$ of interest here.  We also note that in general the creation and annihilation
operators are not normal-ordered.  Of course, after normal-ordering them,
tadpole-type contributions result, which can contribute
to transition amplitudes and turn out to be relevant when discussing
renormalization issues in Appendix~\ref{app:pvnn}.

The vertex functions $M^{nm}$ involve bilinears
\begin{equation}
   \label{eq:bil}
  {\overline u(\bmp',s') \over
   \sqrt{2E_{\bmp'}}} \Gamma { u(\bmp,s) \over
   \sqrt{2E_{\bmp}}}= 
    \chi_{s'}^\dag B({\Gamma})_{\a' \a}\chi_s\ ,
\end{equation}
where $\Gamma$ denotes generically an element of the Clifford
algebra and $\chi_s$, $\chi_{s'}$ are spin states.  
These bilinears are expanded non-relativistically in powers
of momenta, and terms up to order $Q^3$ are included.  We obtain
(subscripts are suppressed for brevity):
\bgroup
\arraycolsep=1.0pt
\begin{eqnarray}   
   B({1})&=& 1-{F^{(2)}(1)\over 4M^2} \ , 
    \label{eq:bil1}\\
   B({i\gamma^5})&=& -{i\bmsi\cdot\bmk\over 2M} 
     +{F^{(3)}({\gamma^5})\over 16M^3}\ ,
    \label{eq:bil5}\\
   B({\gamma^0})&=& 1+{F^{(2)}({\gamma^0})\over 4M^2} \ ,
    \label{eq:bilm0}\\
   {\bm B}({{\bm \gamma}})&=& {2 \,\bmK -i \,\bmk\times\bmsi 
     \over 2M}- {\bmG^{(3)}({\bm \gamma})\over 16M^3}\ ,    \label{eq:bilmi}\\
   B({\gamma^0\gamma^5})&=& {\bmK\cdot\bmsi\over M} 
     -{F^{(3)}({\gamma^0 \gamma^5})\over 8M^3} \ ,
    \label{eq:bil5m0}\\
   {\bm B}({{\bm \gamma}\gamma^5})&=& \bmsi+ {\bmG^{(2)}({{\bm \gamma}\gamma^5})\over
     4M^2}\ ,    \label{eq:bil5mi}\\
   {\bm B}({\sigma^{0i}})&=&-{ i\, \bmk+2 \bmK\times\bmsi
     \over 2M}+{{\bm G}^{(3)}({\sigma}^{0i}) \over 16M^3}\ ,    \label{eq:bilmn0i}\\
   B({\sigma^{ij}})&=& \epsilon_{ijl}\Bigl[ \sigma_l-
    {G^{(2)}_l(\sigma^{ij})\over 4M^2}\Bigr] ,   \label{eq:bilmnij}
\end{eqnarray}
where $F^{(n)}(\Gamma)$ and $\bm G^{(n)}(\Gamma)$ are, respectively,
scalar and vector quantities of order $Q^n$, explicitly given by
\begin{eqnarray}   
   F^{(2)}({1})&=& 2 K^2 + i (\bmk\times\bmK)\cdot\bmsi\ , 
    \label{eq:xbil1}\\
   F^{(3)}({\gamma^5})&=& i\bigl[ \bmsi\cdot\bmk \, (4K^2+k^2) \nonumber\\
    &&\qquad\times + 4\, \bmsi\cdot\bmK\, \bmk\cdot\bmK\bigr]\ ,
    \label{eq:xbil5}\\
   F^{(2)}({\gamma^0})&=&   - k^2/2 + i
   (\bmk\times\bmK)\cdot\bmsi\ ,
    \label{eq:xbilm0}\\
   \bmG^{(3)}({\bm \gamma})&=& \left(2\bmK-i \bmk\times\bmsi \right)
    \left(4K^2+k^2\right)\nonumber\\
   &&  + 2\left(\bmk-2i \bmK\times\bmsi \right)\,
    \bmK\cdot\bmk \ ,    \label{eq:xbilmi}\\
   F^{(3)}({\gamma^0\gamma^5})&=& \bmk\cdot\bmsi\,  \bmk\cdot\bmK\,
     +  \bmK\cdot\bmsi (4K^2+k^2)\ ,
    \label{eq:xbil5m0}\\
   \bmG^{(2)}({\bm \gamma}\gamma^5)&=& 
       2\, \left(\bmK\cdot\bmsi\right)\, \bmK-
    \left( \bmk\cdot\bmsi\right)\, \bmk/2 \nonumber\\
   &&- 2 K^2 \bmsi - i \,\bmk\times\bmK
     \ ,    \label{eq:xbil5mi}\\
   {\bm G}^{(3)}({\sigma}^{0i})&=& \left(i\bmk+2 \bmK\times\bmsi \right)
    \left(4K^2+k^2\right)\nonumber\\
   &&  + 2\left(2i\bmK+ \bmk\times\bmsi \right)\,
    \bmK\cdot\bmk \ ,  \label{eq:xbilmn0i}\\
    \bmG^{(2)}(\sigma^{ij})&=& 2\,(\bmK\cdot\bmsi)\,\bmK-
     (\bmk\cdot\bmsi)\,\bmk/2
     + \bmsi\, k^2/2 \nonumber\\
    && - i\, \bmk\times\bmK
    \ ,   \label{eq:xbilmnij}
\end{eqnarray}
with the momenta $\bmK=(\bmp'+\bmp)/2$ and $\bmk=\bmp'-\bmp$.
We also expand $K^0$ and $ K_\mu K^\mu$ as
\begin{eqnarray}
   K^0 &=& \frac{E+E'}{2} \rightarrow
   M\left(1+{2K^2+k^2/2\over 4M^2}\right)\ ,\\
   K_\mu K^\mu &=& (K^0)^2-K^2 \rightarrow
   M^2\left(1+{k^2\over 4M^2}\right)\ .
\end{eqnarray}
\begin{figure}[t]
   \includegraphics[width=8cm]{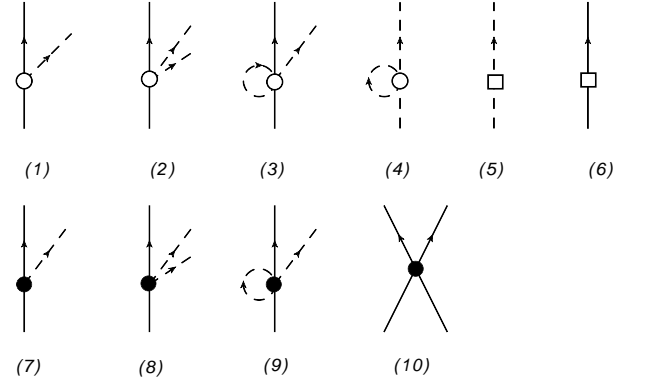}
   \caption{   \label{fig:vert}
     Vertices entering the PV potential at $\nnlo$. The
     solid (dash) lines represent nucleons (pions). The open (solid) symbols
     denote PC (PV) vertices.  In diagrams
     (3) and (4) the vertices are tadpole contributions from the $3\pi NN$ and $4\pi$
     interaction Hamiltonians, respectively.  The open square on a pion line as in 
     diagram (5) represents a $\pi\pi$ vertex coming from the $\ell_3$ and
     $\ell_4$ terms in ${\cal L}_{\pi\pi}^{(4)}$ and the $\delta m_\pi^2$
     term. The open square on a nucleon line as in 
     diagram (6) represents insertions
     from  the $c_1$ term in ${\cal L}_{\pi 
       N}^{(2)}$ and the $\delta M$ term. } 
\end{figure}The interaction vertices needed for the construction of the PV potential
are summarized in Fig.~\ref{fig:vert}.  Note that in the power counting
of these vertices below, we do not include the $1/\sqrt{\omega_k}$ normalization
factors in the pion fields.  We obtain:
\begin{enumerate}
\item $\pi NN$ vertices: The LO PC interaction term (of order $Q$) is
\begin{equation}
   H_I^{\pi NN} = \int {\rm d}^3 x \; {g_A\over 2 f_\pi } \overline{\psi}
   \gamma^\mu\gamma^5 \vec\tau\cdot\partial_\mu\vec\pi \psi   \ .
\end{equation}
giving the following vertex functions, see Eqs.~(\ref{eq:m01}) and (\ref{eq:m10}),
\begin{eqnarray}
 {}^{PC}M^{01}_{\alpha' \alpha, \bmq\, a}  &=&  -i {g_A\over 2
   f_\pi}{ \xi_{t'}^\dag\tau_a\xi_t\over \sqrt{2\omega_k}}
 {\overline{u}_{\alpha'}\over \sqrt{2E'}} \, q \!\!\!/\,\gamma^5
 {u_\alpha\over\sqrt{2E}}\ ,\label{eq:m01gen}\\
 {}^{PC}M^{10}_{\alpha' \alpha, \bmq\, a}  &=&  +i {g_A\over 2
   f_\pi}{\xi_{t'}^\dag\tau_a\xi_t\over \sqrt{2\omega_k}}
 {\overline{u}_{\alpha'}\over \sqrt{2E'}} \, q \!\!\!/\,\gamma^5
 {u_\alpha\over\sqrt{2E}}\ ,\label{eq:m10gen}
\end{eqnarray}
where $u_\alpha\equiv u(\bmp,s)$, etc., and $\xi_t$, $\xi_{t'}$ are isospin states.
The non-relativistic (NR) expansion of these amplitudes is needed up to order $Q^2$.
Other PC $\pi NN$ vertices follow from the interactions terms in ${\cal L}_{N\pi}^{(3)}$ 
proportional to the LEC's $d_{16}$ and $d_{18}$. Thus we find up to order $Q^3$
(spin-isospin states are suppressed for brevity):
\begin{eqnarray}
 {}^{PC}M^{\pi NN,01}_{\alpha' \alpha,
     \bmq\,a}  &=&  {g_A \over 2 f_\pi}
 {\tau_a\over \sqrt{2\omega_q}}\Bigl[ i\, \bmq\cdot\bmsi-{i\over
     M}\omega_q\;\bmK\cdot\bmsi \nonumber \\
   &+& {i\over  4M^2}\Bigl(2\bmK\cdot\bmq\;\bmK\cdot\bmsi-
      2K^2\;\bmq\cdot\bmsi\nonumber\\
   && \qquad -{1\over 2}\bmk\cdot\bmsi\; \bmq\cdot\bmk\Bigr)\Bigr]
    \nonumber\\
     &+& {m_\pi^2 \over f_\pi}(2d_{16}-d_{18})
    {\tau_a\over \sqrt{2\omega_q}}\; i\bmq\cdot\bmsi\ ,\label{eq:MpiNN01b}\\
   {}^{PC}M^{\pi NN, 10}_{\alpha'\alpha,
      \bmq\,a}   &= &  -{}^{PC}M^{\pi NN,01}_{\alpha' \alpha,                                                                     \bmq\,a}  .\label{eq:MpiNN10b}
\end{eqnarray}
In diagrams, these PC vertex functions are represented as open circles.
The PV $\pi NN$ vertices are due to interaction terms
proportional to the LEC's $h^1_\pi$, $h^0_V$,
$h^1_V$, $h^2_V$, $h^{1}_2$, $h^{1}_3$, and $h^{1}_{12}$,
and read up to order $Q^2$
\begin{eqnarray}
 {}^{PV}M^{\pi NN, 01}_{\alpha' \alpha,
     \bmq\,a} & = &  -{h^1_\pi
   \over \sqrt{2} }
 {\epsilon_{3ab}\tau_b\over \sqrt{2\omega_q}}\Bigl[1-{
     2K^2+i(\bmk\times\bmK)\cdot\bmsi \over 4 M^2}\Bigr]
    \nonumber\\
     && -{i\over f_\pi}\Bigl({h^0_V  \over 2 } \tau_a + h^1_V\,
    \delta_{a,3}+{2\over 3}\,h^2_V\,
        {\cal I}^{ab}\,\tau_b\Bigr)\nonumber\\
      &&\times  {1\over \sqrt{2\omega_q}}
        \Bigl(\omega_q-{\bmq\cdot\bmK\over M}\Bigr)
        + {8\over f_\pi^2}   {\epsilon_{3ab} \tau_b\over
          \sqrt{2\omega_q} }
           \nonumber \\
     &&\times \Bigl[ (h^{1}_2-h^{1}_3)\,
          m_\pi^2-2h^{1}_{12}\, \omega_q^2\Bigr]         
   \ ,\label{eq:MpiNN01c}\\
{}^{PV}M^{\pi NN, 10}_{\alpha' \alpha,
     \bmq\,a} & = &  -{h^1_\pi
   \over \sqrt{2} }
 {\epsilon_{3ab}\tau_b\over \sqrt{2\omega_q}}\Bigl[1-{
     2K^2+i(\bmk\times\bmK)\cdot\bmsi \over 4 M^2}\Bigr]
    \nonumber\\
     && +{i\over f_\pi}\Bigl({h^0_V  \over 2 } \tau_a + h^1_V \,
    \delta_{a,3}+{2\over 3}\, h^2_V
        \, {\cal I}^{ab}\,\tau_b\Bigr)\nonumber\\
     &&\times   {1\over \sqrt{2\omega_q}}
        \Bigl(\omega_q-{\bmq\cdot\bmK\over M}\Bigr)
        + {8\over f_\pi^2}   {\epsilon_{3ab} \tau_b\over
          \sqrt{2\omega_q} }
     \nonumber \\
     && \times\Bigl[ (h^{1}_2-h^{1}_3)\,
          m_\pi^2-2h^{1}_{12}\, \omega_k^2\Bigr]
   \ .\label{eq:MpiNN10c}
\end{eqnarray}
\item $\pi\pi NN$ vertices: The PC
interaction is due to the Weinberg-Tomozawa term
  \begin{equation}
   H_I^{\pi\pi NN} = \int {\rm d}^3 x \; {1\over 4f_\pi^2}\overline{\psi}\,
\gamma^\mu
    \vec\tau\cdot(\vec\pi\times\partial_\mu\vec\pi)\, 
\psi   \ ,
\end{equation}
and in the following only the LO in the NR expansion is needed at the order
we are interested in.  The
corresponding vertex functions read
\begin{eqnarray}
   {}^{PC}M^{\pi\pi NN,02}_{\alpha' \alpha,\bmq' a'\,\bmq\, a}&=&
    {i\over 8f_\pi^2}  \epsilon_{aa'b}\tau_b
    {\omega_q-\omega_{q'}\over\sqrt{2\omega_q}\sqrt{2\omega_{q'}}}
    \ , \\ 
  {}^{PC}M^{\pi\pi NN,11}_{\alpha' \alpha,\bmq' a'\,\bmq \,a}&=&
    {i\over 4 f_\pi^2} \epsilon_{aa'b}\tau_b
    {\omega_q+\omega_{q'}\over\sqrt{2\omega_q}\sqrt{2\omega_{q'}}}
    \ , \\ 
   {}^{PC}M^{\pi\pi NN,20}_{\alpha'\alpha,\bmq' a'\,\bmq \,a}&=&
    {i\over 8f_\pi^2} \epsilon_{aa'b}\tau_b
    {\omega_{q'}-\omega_{q}\over\sqrt{2\omega_q}\sqrt{2\omega_{q'}}}
    \ ,
\end{eqnarray}
and tadpole contributions vanish because of the isospin factor
$\epsilon_{aa'b}$.  The PV vertices follow from the interaction
terms proportional to $h^1_A$ and $h^2_A$, and are given by
\begin{eqnarray}
   {}^{PV}M^{\pi\pi NN,02}_{\alpha' \alpha,\bmq' a'\, \bmq \,a}&=&
    {i\over 2f_\pi^2} {1\over\sqrt{2\omega_q}\sqrt{2\omega_{q'}}}
    \Bigl[-h^1_A\, \epsilon_{3aa'} (\bmq-\bmq')\cdot\bmsi \nonumber \\ 
  && -  {1\over3}h^2_A \, \epsilon_{aa'b}\, I^{b}\, \tau_b\,
              (\bmq-\bmq')\cdot\bmsi \nonumber\\
  &&  + {1\over3}h^2_A \, \epsilon_{aa'b}\, \tau_b \,(
              I^a\, \bmq-I^{a'}\, \bmq')\cdot\bmsi\Bigr]\ , \\ 
   {}^{PV}M^{\pi\pi NN,11}_{\alpha' \alpha,\bmq' a'\,\bmq \,a}&=&
    {i\over f_\pi^2} {1\over\sqrt{2\omega_q}\sqrt{2\omega_{q'}}}
    \Bigl[-h^1_A \, \epsilon_{3aa'} (\bmq+\bmq')\cdot\bmsi \nonumber \\ 
  &&  -  {1\over3}h^2_A \,\epsilon_{aa'b}\, I^{b}\,\tau_b\,
              (\bmq+\bmq')\cdot\bmsi \nonumber\\
  &&   + {1\over3}h^2_A\, \epsilon_{aa'b}\,\tau_b\, (
              I^a\,\bmq+I^{a'}\, \bmq')\cdot\bmsi\Bigr]\ , \\ 
   {}^{PV}M^{\pi\pi NN,20}_{\alpha' \alpha,\bmq' a'\,\bmq \,a}&=&
    {i\over 2f_\pi^2} {1\over\sqrt{2\omega_q}\sqrt{2\omega_{q'}}}
    \Bigl[-h^1_A \, \epsilon_{3aa'}\ (\bmq'-\bmq)\cdot\bmsi \nonumber \\ 
  && -  {1\over3}h^2_A \,\epsilon_{aa'b}\, I^{b}\tau_b\,
              (\bmq'-\bmq)\cdot\bmsi \nonumber\\
  &&  + {1\over3}h^2_A \, \epsilon_{aa'b}\, \tau_b \, (
              I^{a'}\, \bmq'-I^{a}\, \bmq)\cdot\bmsi\Bigr]\ , 
\end{eqnarray}
where the factor $I^a$ has been defined as $I^a=(-1,-1,2)$.
\item $3\pi NN$ vertices: We need consider the 
  $3\pi NN$ interactions deriving from the expansion
   of the $U$ matrix given in Eq.~(\ref{eq:uumatrix}) in both
  the PC and PV LO Lagrangians. The corresponding Hamiltonians read
\begin{eqnarray}
   H_I^{3\pi NN}&=&-{g_A\over 2}\int {\rm d}^3x\; \overline{\psi}
   \Bigl[ {4\alpha-1\over 2f_\pi^3}  \gamma^\mu\gamma^5(\vec\tau\cdot\vec\pi) (\vec \pi
          \cdot\partial_\mu \vec\pi)\nonumber\\
    &&\qquad\quad +{\alpha\over f_\pi^3} \gamma^\mu\gamma^5\, \vec\pi^2\,  (\vec \tau
          \cdot\partial_\mu \vec\pi)\Bigr]\psi \\
  && + {h^1_\pi\over\sqrt{2}} {\alpha\over f_\pi^2}  \int {\rm d}^3x\; \overline{\psi}
   \, \pi^2 \, (\vec\pi\times\vec\tau)_3\psi\ .
\end{eqnarray}
Here we are interested only in tadpole contributions associated with these
interactions.  We denote them as $T^{3\pi NN}_I=T^{3\pi NN,01}+T^{3\pi NN,10}$, where
\begin{eqnarray}
 T^{3\pi NN,01} & =& {1\over\sqrt{\Omega}}\sum_{\alpha' \alpha}\sum_{\bmq\, a}
   b^{\dag}_{\alpha'}b_{\alpha}a_{\bmq\,a}\;
  M^{3\pi NN,01}_{\alpha' \alpha, \bmq\, a}\delta_{\bmq+\bmp,\bmp'}\ ,\nonumber\\
 T^{3\pi NN,10} & = &  {1\over\sqrt{\Omega}}\sum_{\alpha' \alpha}\sum_{\bmq\,a}
   b^{\dag}_{\alpha'}b_{\alpha}a^\dag_{\bmq\,a}\;
  M^{3\pi NN,10}_{\alpha' \alpha,\bmq\,a}\delta_{\bmq+\bmp',\bmp}\ .\nonumber
\end{eqnarray}
The PC part is given by
\begin{eqnarray}
 {}^{PC}M^{3\pi NN,01}_{\alpha'\alpha, \bmq\,a} & = &  -i\,{g_A\over 8f_\pi^3} (10\,\alpha-1)
{ \tau_a \over \sqrt{2\omega_q}} \,\bmsi\cdot \bmq \, J_{01} \ , \nonumber\\
 {}^{PC}M^{3\pi NN,10}_{\alpha'\alpha, \bmq\,a} & = &  i\, {g_A\over 8f_\pi^3} (10\, \alpha-1)
  { \tau_a \over  \sqrt{2\omega_q}}\, \bmsi\cdot \bmq \, J_{01}\ ,\nonumber
\end{eqnarray}
where the (infinite) constants $J_{mn}$ have been defined as
\begin{equation}
  J_{mn}= {1\over \Omega} \sum_\bmk {k^{2m} \over \omega_k^n}\ .\label{eq:funJapp}
\end{equation}
The PV part reads
\begin{equation}
 {}^{PV}\!M^{3\pi NN,01}_{\alpha' \alpha, \bmq\,a} =  
 {}^{PV}\!M^{3\pi NN,10}_{\alpha' \alpha, \bmq\,a} = 
   {5\over 2}{h^1_\pi\over\sqrt{2}}  {\alpha\over f_\pi^2}
  {\epsilon_{3ab}\tau_b\over \sqrt{2\omega_q}} J_{01}\ .\label{eq:tp3pi}
\end{equation}
\item $4\pi$ vertices: The relevant interaction Hamiltonian follows from the PC
Lagrangian ${\cal L}^{(2)}_{\pi\pi}$ (there is no a PV contribution in this case), and is
given by
\begin{eqnarray}
 H_I^{4\pi}&=& \int d^3x\; \Bigl[- {1-4\alpha\over 2f_\pi^2} (\vec \pi\cdot\partial_\mu \vec\pi)
  (\vec \pi \cdot\partial^\mu \vec\pi) \nonumber \\
   && +{\alpha\over f_\pi^2} \vec \pi^2
 (\partial^\mu \vec \pi\cdot\partial_\mu \vec\pi)-{8\alpha-1\over 8 f_\pi^2} m_\pi^2
     \vec\pi^4 \Bigr] \ . 
\end{eqnarray}
The associated tadpole contributions read
$T^{4\pi}_I=T^{4\pi,02}+T^{4\pi,11}+T^{4\pi,20}$ are:
\begin{eqnarray}
 T^{4\pi,02} & = &  \sum_{\bmq\, a} a_{\bmq\,a}a_{-\bmq\,a}\;
    M^{4\pi,02}_{\bmq}\ ,\\
 T^{4\pi,11} & = &  \sum_{\bmq \, a} a_{\bmq\,a}^\dag a_{\bmq\,a}\;
    M^{4\pi,11}_{\bmq}\ ,\\
 T^{4\pi,20} & = &  \sum_{\bmq \, a} a_{\bmq\,a}^\dag a_{-\bmq\,a}^\dag\;
    M^{4\pi,20}_{\bmq}\ ,
\end{eqnarray}
with
\begin{eqnarray}
   {}^{PC}M^{4\pi,02}_{\bmq}&=& {1-10\alpha\over 4f_\pi^2} \omega_q J_{01}
   +{16\alpha-3\over 4 f_\pi^2} J_{00}\nonumber\\
   && +{m_\pi^2\over 16 f_\pi^2}{1\over\omega_q} J_{01}\ , \\ 
   {}^{PC}M^{4\pi,11}_{\bmq}&=& {m_\pi^2\over 8 f_\pi^2}{1\over\omega_q} J_{01} \ , \\ 
   {}^{PC}M^{4\pi,20}_{\bmq}&=& {1-10\alpha \over 4f_\pi^2} \omega_q J_{01} 
   -{16\alpha-3\over 4 f_\pi^2} J_{00}\nonumber\\
   &&+{m_\pi^2\over 16 f_\pi^2}{1\over\omega_q} J_{01}\ .
\end{eqnarray}
\item $\pi\pi$ contact interactions: These follow from
the two terms in ${\cal L}^{(4)}_{\pi\pi}$ and the $\delta
m_\pi^2$ counter-term, and are represented by an open square
in diagram 5 of Fig.~\ref{fig:vert}.  They are given by
  $H^{\pi\pi}_I=H^{\pi\pi,02}+H^{\pi\pi,11}+H^{\pi\pi,20}$, where
\begin{eqnarray}
H^{\pi\pi,02} & = &  \sum_{\bmq a} a_{\bmq\,a}a_{-\bmq\,a}\;
    M^{\pi\pi,02}_{\bmq}\ ,\\
H^{\pi\pi,11} & = &  \sum_{\bmq a} a_{\bmq\,a}^\dag a_{\bmq\,a}\;
    M^{\pi\pi,11}_{\bmq}\ ,\\
H^{\pi\pi,20} & = &  \sum_{\bmq a} a_{\bmq\,a}^\dag a_{-\bmq\,a}^\dag\;
    M^{\pi\pi,20}_{\bmq}\ ,
\end{eqnarray}
with
\begin{eqnarray}
   {}^{PC}M^{\pi\pi,02}_{\bmq}&=& \Bigl[\ell_3 {m_\pi^4\over f_\pi^2}+{\delta m_\pi^2
       \over 2}\Bigr]  {1\over 2\omega_q}+\ell_4 {m_\pi^2\over f_\pi^2}
   \omega_q\ , \\
   {}^{PC}M^{\pi\pi,11}_{\bmq}&=& \Bigl[\ell_3 {m_\pi^4\over f_\pi^2}+{\delta m_\pi^2
       \over 2}\Bigr]  {1\over \omega_q}
  \ ,
\end{eqnarray}
and ${}^{PC}M^{\pi\pi,20}_{\bmq}=  {}^{PC}M^{\pi\pi,02}_{\bmq}$.

\item $NN$ contact interactions: These follow from the $c_1$ term in ${\cal L}_{\pi
       N}^{(2)}$ and the $\delta M$ counter-term, and are represented by an open
       square on a nucleon line in diagram 6 of Fig.~\ref{fig:vert}.   They are given
       by
\begin{equation}
  H^{NN}= \sum_\alpha b^\dag_\alpha b_\alpha M^{NN}_\bmp \ , \qquad
  M^{NN}_\bmp=-4c_1 m_\pi^2  \ .
\end{equation}

\item $4N$ contact interaction: The EFT Hamiltonian includes also the term
given in Eq.~(\ref{eq:m00}) derived from a contact Lagrangian.  We only need its PV
part of order $Q$, arising from the five independent interaction terms
discussed in Ref.~\cite{Girlanda08}.
At order $Q$, with a suitable choice of the LEC's, 
the vertex function ${}^{PV}M^{00}$ can be written as
\begin{eqnarray}
 {}^{PV}M^{00}_{\a_1'\a_1\a_2'\a_2} &=&  
         {1 \over 2\Lambda_\chi^2 f_\pi} \Bigl[ 
       C_1 (\bmsi_1\times\bmsi_2)\cdot  \bmk_1 \nonumber\\
       &+& C_2\,  \vec\tau_1\cdot\vec\tau_2\,
       (\bmsi_1\times\bmsi_2)\cdot  \bmk_1\;
        \nonumber\\
       &+&  C_3\,  (\vec\tau_1\times\vec\tau_2)_z\,
        (\bmsi_1+\bmsi_2)\cdot\bmk_1\;
      \nonumber\\
       &+&  C_4\,(\tau_{1z}+\tau_{2z})\,
        (\bmsi_1\times\bmsi_2)\cdot\bmk_1\;
         \nonumber \\
       &+&C_5\, {\cal I}^{ab}\, \tau_{1a} \, \tau_{2b} \,
        (\bmsi_1\times\bmsi_2)\cdot \bmk_1\;     
             \Bigr]\ , \label{eq:m00pv}
\end{eqnarray}
where $\bmk_1=\bmp_1'-\bmp_1=-\bmp_2'+\bmp_2$.

\end{enumerate}
\egroup

\section{The PV potential}
\label{app:pvnn}

In this Appendix we report on the derivation of the PV $NN$
potential.  In Sec.~\ref{sec:np} we discuss the nucleon and
pion propagators, and in Sec.~\ref{sec:sub_pvnn} provide
explicit expressions for the contributions of various diagrams.

\subsection{The nucleon and pion propagators}
\label{sec:np}
\begin{figure}[h]
   \includegraphics[width=8cm]{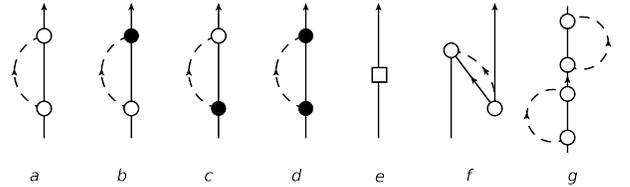}
   \caption{   \label{fig:nprop}
    Diagrams describing the $N\rightarrow N$ transition amplitude.
   The open (solid) circles represent contributions due to PC (PV) $\pi
     NN$ vertices, while the open square represents the contribution from 
     the $c_1$ term in ${\cal L}_{\pi N}^{(2)}$ and the $\delta M$ counterterm.
     The notation is as in Fig.~\protect\ref{fig:vert}.}
\end{figure}

We begin by considering the propagation of an
isolated nucleon.  Diagrams contributing to the
transition amplitude $\langle
\alpha'|T|\alpha\rangle=T\,
\delta_{\alpha'\, \alpha}$ are displayed in Fig.~\ref{fig:nprop}.
Diagrams (a)-(d) are pion-loop contributions due to PC and
PV $\pi NN$ vertices.  The contribution of diagram (a) is of order $Q^3$,
while those of diagrams (b) and (c) vanish, since the integrand is odd with
respect to the loop momentum.  The contribution of diagram (d) is ignored,
since it contains two PV vertices.  Diagram (e) represents the contribution
from the two terms proportional to $\overline{\psi} \psi$ (namely, the $c_1$
term in  ${\cal L}_{\pi N}^{(2)}$ of order $Q^2$ and the
$\delta M$ counter-term), diagram (f) represents contributions associated
with antinucleons, which enter first at order $Q^4$, and lastly diagram (g) is an
example of a reducible two-loop contribution (of
order $Q^4$).  Contributions from reducible diagrams can 
be summed up analytically as shown below.
Explicitly, the contribution of diagram (a) is
\begin{equation}
  T({\rm a})={1\over \Omega} \sum_\bmk  {3\, g^2_A\over 8\, f^2_\pi} \, {k^2\over
    \omega_k} {1\over E_p-E_{|\bmp-\bmk|}-\omega_k}\ ,
\end{equation}
which to leading order $Q^3$ gives
\begin{equation}
    T^{(3)}({\rm a})= -{1\over \Omega} \sum_\bmk 
    {3\, g^2_A\over 8\, f^2_\pi} \, {k^2\over
    \omega_k^2} \ ,
\end{equation}
while the contributions of diagrams (e) and (f) read, respectively, 
\begin{eqnarray}
    T({\rm e})&=& -4 \,m^2_\pi\, c_1 +\delta M\ , \\
    T^{(4)}({\rm f})&=& {1\over \Omega} \sum_\bmk 
    {3\, g^2_A\over 16\,f^2_\pi} \, {\omega_k\over
    M} \ ,
\end{eqnarray}
where only the leading order $Q^4$ has been retained in the
case of diagram (f). We now set the $N\rightarrow N$ amplitude to zero
order by order in the power counting by assuming
\begin{equation}
 \delta M= \delta M^{(2)}+\delta M^{(3)}+\delta M^{(4)}+\cdots \ ,
\end{equation}
where $\delta M^{(n)}$ is of order $Q^n$, and by determining
$\delta M^{(n)}$ so that $T^{(n)}=0$.  Up to order $Q^4$, we obtain
\begin{eqnarray}
 \delta M^{(2)}&=& 4 \, m_\pi^2\, c_1\ , \qquad \delta M^{(3)} =
 -T^{(3)}({\rm a})\ , \label{eq:deltaM23}\\
 \delta M^{(4)} &=& -T^{(4)}({\rm a})-T^{(4)}({\rm f})\ .\label{eq:deltaM4}
\end{eqnarray}
\begin{figure}[h]
   \includegraphics[width=8cm]{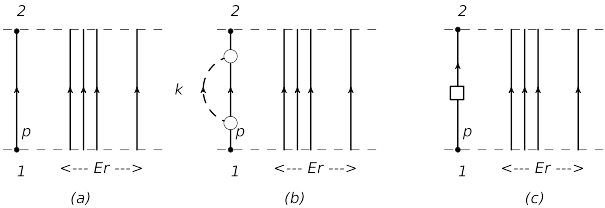}
   \caption{   \label{fig:nprop1}
    Parts of a general diagram with the propagation of nucleons only.}
\end{figure}

Next we consider the ``dressing'' of a nucleon line belonging to a more
complicated diagram, see Fig.~\ref{fig:nprop1}.  Panel (a) on this
figure represents a diagram in which one nucleon of momentum $\bmp$
is created at vertex 1 and annihilated at vertex 2 (shown by the two dots
at the beginning and end of the nucleon line).  The other nucleons
have energies collectively denoted by $E_r$.  Note that there are no
pions in flight in the intermediate state.
The energy denominator of the diagram in panel (a) is
\begin{equation}
   P_0(E)= {1\over E_0-(E_p+E_r)+i\epsilon}= {1\over E+i\epsilon}\ ,
\end{equation}
where $E= E_0-E_p-E_r$ and $E_0$ is the initial energy (which
depends on the particular process under consideration).

Panels (b) and (c) in Fig.~\ref{fig:nprop1} represent, respectively,
the contribution in which nucleon 1 emits and reabsorbs a pion of momentum
$\bmk$ and that in which a contact interaction occurs.  These contributions are given by
\begin{eqnarray}
  S(E)&=& {1\over \Omega} \sum_\bmk {3\, g^2_A\over 8\,f^2_\pi}\, {k^2\over
    \omega_k}\, {1\over E+E_p-E_{|\bmp-\bmk|}-\omega_k}\nonumber\\
   &&-4\,m_\pi^2 \,c_1 +  \delta M + \ldots\ , 
\end{eqnarray}
and $S(0)=0$ follows from the choice of $\delta M$ discussed
previously for a single nucleon (of course, energy denominators
in the diagrams of Figs.~\ref{fig:nprop} and~\ref{fig:nprop1} are different,
and $S(E)$ only vanishes for $E=0$). 

By summing up repeated (b)- and (c)-type insertions, we obtain the
well known result
\begin{eqnarray}
  P_D(E)&=& {1\over E+i\epsilon}+  {1\over E+i\epsilon}S(E)
  {1\over E+i\epsilon}+\cdots \nonumber\\
    &=&  {1\over E-S(E)+i\epsilon} \ .
\end{eqnarray}
By expanding $S(E)$ in powers of $E$ ($E$ is assumed
to be small) and by keeping only linear terms in $E$, we find
\begin{eqnarray}
   P_D(E)&\simeq&
 {1\over 1-  S'(0)} {1\over E + i\epsilon} =
     {Z_N\over E+i\epsilon}\ ,
\end{eqnarray}
where $Z_N=1/[1-  S'(0)]$,
\begin{equation}
  S'(0) =  - {1\over \Omega} \sum_\bmk {3\, g^2_A\over 8\, f^2_\pi} \,{k^2\over
    \omega_k^3} =-{3\, g^2_A\over 8\, f^2_\pi}\, J_{13}\ ,
\end{equation}
and the (infinite) constant $J_{13}$ is defined in Eq.~(\ref{eq:funJ}).
Since $-E=E_p+E_r-E_0$ is the energy of the intermediate
state relative to the initial energy, it is physically sensible
that for $E\rightarrow 0$ the dressed operator should have the
same form as the bare propagator $1/(E+i\epsilon)$ up
to the (nucleon wave function) renormalization factor $Z_N$.  In the following we adopt
the common practice of attaching a $\sqrt{Z_N}$ at each of
the two vertices of an internal nucleon line, and of multiplying by
an extra $\sqrt{Z_N}$ each external nucleon line.
The renormalization of nucleon lines when additional pions are present
must be discussed case by case.

\begin{figure}[h]
   \includegraphics[width=8cm]{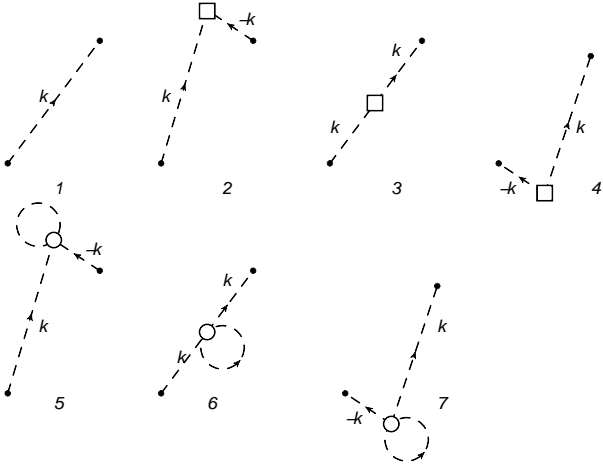}
   \caption{   \label{fig:pprop}
   Propagation of a pion in the intermediate
   state of a generic diagram.  Only the pion line is shown.}
\end{figure}

We now consider a diagram with an intermediate state
involving a single pion of momentum $\bmk$ and energy $\omega_k$,
and denote with $E_0$ the energy of the initial state and with $E_r$
the total energy of particles other than the pion present in this
intermediate state.  We assume that $|E_0-E_r|\ll \omega_k$.
In Fig.~\ref{fig:pprop}, the various panels only show the (internal)
pion line of this generic diagram.    
The energy denominator associated with a single pion
propagation in panel (1) is $(E_0-\omega_k-E_r+i\epsilon)^{-1}\simeq
-\omega_k^{-1}$.  The two vertices at the beginning and end of the pion
line each bring in a $(1/\sqrt{2\omega_k})$.
Furthermore, there is factor $2$ coming from the two possible time orderings
in the propagation of the pion (``left-to-right'' and ``right-to-left'').  Thus the
bare pion propagator  is 
$-1/\omega_k^2$.  By taking into account the remaining contributions from panels (2)-(7)
in Fig.~\ref{fig:pprop}, we find the ``dressed'' propagator up to corrections of
order $Q^2$ included to be given by
\begin{equation}
   S_\pi(\bmk) \simeq -{1\over \omega_k^2}\left(1-{A_\pi\over \omega_k^2}+\delta Z_\pi
    \right) \ ,
\end{equation}
where
\begin{eqnarray}
  A_\pi&=&{2\, \ell_3 \, m^4_\pi\over f_\pi^2}+{m^2_\pi\over 4\,  f_\pi^2} \, J_{01}+\delta m_\pi^2 \ ,
  \label{eq:Api}\\
  \delta Z_\pi&=&-{2\, \ell_4\,  m^2_\pi\over f_\pi^2}-{1-10\, \alpha \over 2\, f_\pi^2}
 \, J_{01} \ .\label{eq:AdZ}
\end{eqnarray}
Note that $A_\pi/\omega_k^2$ and $\delta Z_\pi$ are both of order $Q^2$.
The dressed propagator $S_\pi(\bmk)$ can also be written as 
\begin{equation}
   S_\pi(\bmk) \simeq 
    -{1+\delta Z_\pi\over \omega_k^2+A_\pi} \ .
\end{equation}
Here $Z_\pi=1+\delta Z_\pi$ represents the renormalization of the pion
wave function, and again we attach a factor $\sqrt{Z_\pi}$ at each of the two
ends of the pion line.  Therefore this factor contributes to the
renormalization of the coupling constants.  The term $A_\pi$ represents
the shift in the square of the pion mass, and in order to have $m^2_\pi$ be
the physical pion mass, we choose $\delta m_\pi^2$ so that $A_\pi=0$.
The expressions above for $A_\pi$ and $\delta Z_\pi$ are
the same as those reported in Eq.~(2.39) of Ref.~\cite{Epel03}.
\subsection{The PV $NN$ potential}
\label{sec:sub_pvnn}

Diagrams contributing to the PV $NN$ amplitude up to order $Q$ are shown in
Figs.~\ref{fig:pvdiag} and~\ref{fig:operen}.  The power counting is as follows:
(i) a PC (PV) $\pi NN$ vertex is of order $Q$ ($Q^{0}$); (ii) a PC or PV $\pi\pi NN$
vertex is of order $Q^1$; (iii) a PC (PV) $NN$ contact vertex is of order $Q^0$ ($Q$);
(iv) an energy denominator without (with one or more) pions is of order $Q^{-2}$ ($Q^{-1}$);
(v) factors $Q^{-1}$ and $Q^{3}$ are associated with, respectively, each pion line and each loop
integration.  The momenta are defined as given below Eq.~(\ref{eq:notjb1}), and in what follows
use is made of the fact that $\bmk\cdot\bmK$ vanishes in the CM frame.

The PV $NN$ potential is derived from the amplitudes in Figs.~\ref{fig:pvdiag} and~\ref{fig:operen}
via Eqs.~(\ref{eq:vpvml})--(\ref{eq:vpv1}).  Up to order $Q$ included, we obtain for the OPE
component in panel (a) of Fig.~\ref{fig:pvdiag} :
\begin{equation}
 V({\rm a}) =
 V^{(-1)}({\rm NR})+ V^{(1)}({\rm RC}) + V^{(1)}({\rm LEC})\ , \label{eq:ope2}
 \end{equation}
 where
\bgroup
\arraycolsep=0.5pt
 \begin{eqnarray}
 V^{(-1)}({\rm NR})\!\!&=&  
  \!\!{g_A h^1_\pi\over2 \sqrt{2}f_\pi} (\vec\tau_1\times\vec\tau_2)_z
   {i\bmk\cdot(\bmsi_1\!+\!\bmsi_2)\over \omega_k^2}\ ,\label{eq:opeloapp}\\
 V^{(1)}({\rm RC})\!\! &=&
  \!\!{g_A h^1_\pi\over 2\sqrt{2}f_\pi }\, {1\over 4M^2}\,
  (\vec\tau_1\times\vec\tau_2)_z\;
   {1 \over \omega_k^2}\nonumber \\
  & \times \biggl[& -\frac{i}{2} \left( 8  K^2 +k^2\right) \,\bmk\cdot(\bmsi_1+\bmsi_2) \nonumber\\
  &+&\!  \bmk\cdot\bmsi_1 \,\, (\bmk\times\bmK)\!\cdot\!\bmsi_2\nonumber \\
  &+&\!\bmk\!\cdot\!\bmsi_2\,\, (\bmk\times\bmK)\cdot\bmsi_1\biggr]
                 \ ,\qquad\quad\label{eq:opennlo} 
\end{eqnarray}
\vspace{-1cm}
\begin{eqnarray}
  V^{(1)}({\rm LEC})&=& V^{(-1)}({\rm NR}) 
            \biggl[ {2 \,m_\pi^2 \over g_A}\, (2\,d_{16} - d_{18})\nonumber\\
     &&\qquad - {8 \sqrt{2}\, m^2_\pi\over h^1_\pi \,f_\pi^2}(h^{1}_2-h^{1}_3) \biggr]   \nonumber \\
     &+& {16 \,h^{1}_{12}\over f_\pi^2} {g_A\over 2\, f_\pi} (\vec\tau_1\times\vec\tau_2)_z\;
              i\bmk\cdot(\bmsi_1\!+\!\bmsi_2) \ .\qquad\label{eq:opeL}
\end{eqnarray}
The potential does not depend on the LEC's $h^0_V$, $h^1_V$, 
and $h^2_V$, since the associated contributions cancel out when summing over the
different time orderings.  The factor $k^2=\omega_k^2-m_\pi^2$ in
$V^{(1)}({\rm RC})$ leads to a piece that can be reabsorbed in
the contact term proportional to
$C_3$ in Eq.~(\ref{eq:ctren}) and a piece proportional to $m_\pi^2$ that simply
renormalizes the LEC $h^1_\pi$.  Similarly, the piece in
$V^{(1)}({\rm LEC})$ proportional to $V^{(-1)}({\rm NR})$
leads to renormalization of $h_\pi^1$ and the remaining
term can be reabsorbed in $C_3$.  We are then left with
the components $V_{PV}^{\rm (OPE)}$ and $V_{PV}^{\rm (RC)}$
given in Eqs.~(\ref{eq:opepvren2}) and~(\ref{eq:um}).

The component of the PV potential due to the
contact terms in panel CT of Fig.~\ref{fig:pvdiag}
derives directly from the vertex function
${}^{PV}M^{00}$ given in Eq.~(\ref{eq:m00pv}). The final expression 
has already been given in Eq.~(\ref{eq:ctren}). 

The panels (b) and (c) contain a combination of a contact
interaction with the exchange of a pion. However, it can be shown that 
their contribution is at least of order $Q^3$.

Next we consider the TPE components in panels (d)-(g).  The contribution
from panel (d) reads  
\begin{eqnarray}
 V^{(1)}({\rm d})&=& 
  {g_A h^1_\pi\over 8\sqrt{2} f_\pi^3} \,
     (\vec\tau_1\times\vec\tau_2)_z\,
          i\,\bmk\cdot(\bmsi_1+\bmsi_2)\nonumber\\
  &&  \int {{\rm d}^3q\over (2\pi)^3}
  \; {1\over \omega_+ \omega_- (\omega_+ + \omega_-)}\ , \label{eq:d}
\end{eqnarray}
where $\omega_\pm=\sqrt{(\bmk\pm \bmq)^2 + 4 \, m_\pi^2}$.
Use of dimensional regularization allows one to obtain the
finite part as~\cite{Pastore09}
\begin{equation}
 \overline{V}^{(1)}({\rm d}) =
  -{g_A h^1_\pi\over 2\sqrt{2} f_\pi\Lambda_\chi^2} 
     (\vec\tau_1\times\vec\tau_2)_z\,
          i\bmk\cdot(\bmsi_1+\bmsi_2)\; L(k)\ , \label{eq:d2R}
\end{equation}
where $\Lambda_\chi=4\pi f_\pi$ and the loop function $L(k)$ is defined as
\begin{equation}
  L(k)= {1\over 2} {s\over k} \ln\left({s+k\over s-k}\right)\ ,\quad
  s=\sqrt{k^2+4\, m^2_\pi}\ .\label{eq:sL}
\end{equation}
The singular part is given by
\begin{equation}
V^{(1)}_\infty({\rm d}) =
  -{g_A h^1_\pi\over 4\sqrt{2} f_\pi\Lambda_\chi^2} 
     (\vec\tau_1\times\vec\tau_2)_z
       i\bmk\cdot(\bmsi_1+\bmsi_2)\;(d'_\epsilon-2) \ , \label{eq:d2S}
       \end{equation}
where
\begin{equation}
  d'_\epsilon =  -{2\over \epsilon}+\gamma-\ln
  \pi +\ln\left({m^2_\pi\over\mu^2}\right)
      \ ,
\end{equation}
$\epsilon=3-d$, $d$ being the number of dimensions ($d\rightarrow 
3$), and $\mu$ is a renormalization scale.  This singular contribution
is absorbed in the $V^{({\rm CT})}$ term proportional to $C_3$.

The contributions from panels (e), (f), and (g) in Fig.~\ref{fig:pvdiag} are collectively
denoted as ``box'' below, and the non-iterative pieces in reducible
diagrams of type (g) are identified via Eq.~(\ref{eq:vpv1})---elsewhere~\cite{Pastore09},
they have been referred to as the ``recoil corrections''.  We obtain
\begin{widetext}
\begin{eqnarray}
 V^{(1)}({\rm box})\!&=&\!\! 
  {g_A h^1_\pi\over 2\sqrt{2}f_\pi} {g_A^2\over 4f_\pi^2} 
    \int {{\rm d}^3q\over (2\pi)^3} \; 
   {\omega_+^2+\omega_+ \omega_- + \omega_-^2\over \omega_+^3
     \omega_-^3 (\omega_+ + \omega_-)} 
    \biggl[-2i\,(\tau_{1z}+\tau_{2z})\; 
     \Bigl[ \bmq\cdot \bmsi_1 \, (\bmq\times\bmk)\cdot \bmsi_2 \nonumber\\
     &&\!\! -\bmq\cdot \bmsi_2 \, (\bmq\times\bmk)\cdot \bmsi_1\Bigr]\!
       -\!2i\, (\tau_{1z}-\tau_{2z})\, 
     \Bigl[ \bmq\cdot \bmsi_1\,\, (\bmq\times\bmk)\cdot \bmsi_2
           +\bmq\cdot \bmsi_2\,\, (\bmq\times\bmk)\cdot \bmsi_1\Bigr] \nonumber\\
       &&+i\,(\vec\tau_1\times\vec\tau_2)_z \, (k^2-q^2) \, \, \bmk\cdot(\bmsi_1+\bmsi_2)\biggr]
    \ ,\label{eq:box}
\end{eqnarray}
\end{widetext}
and, after dimensional regularization, the finite part reads
\begin{eqnarray}
  \lefteqn{\overline{V}^{(1)}({\rm box}) =\qquad\qquad\qquad\qquad}
  &&\nonumber\\  
  &&-{g_A^3 \, h^1_\pi\over 2\sqrt{2}f_\pi \Lambda_\chi^2} 
     \Bigl[4(\tau_{1z}+\tau_{2z})\; i\bmk\cdot(\bmsi_1\times\bmsi_2)
     L(k)\nonumber \\
  &&+(\vec\tau_1\times\vec\tau_2)_z \,i\bmk\!\cdot\!(\bmsi_1+\bmsi_2)
     \bigl[ H(k)-3\, L(k)\bigr]\Bigr]\ ,\label{eq:box2R}
\end{eqnarray}
where
\begin{equation}
   H(k)= {4\, m^2_\pi\over s^2} L(k)\ ,\label{eq:H}
\end{equation}
while the singular part is given by
\begin{eqnarray}
  \lefteqn{\overline{V}^{(1)}({\rm box}) =\qquad\qquad\qquad}
  &&\nonumber\\
  &&-{g_A^3\, h^1_\pi\over 2\sqrt{2}f_\pi \Lambda_\chi^2} 
     \biggl[2\, (\tau_{1z}+\tau_{2z})\; i\bmk\cdot(\bmsi_1\times\bmsi_2)\;
     \Bigl(d'_\epsilon-{4\over 3}\Bigr)\nonumber \\
  && -(\vec\tau_1\times\vec\tau_2)_z\, i\bmk\cdot(\bmsi_1+\bmsi_2)\; 
     \Bigl({3\over 2}d'_\epsilon-1\Bigr)\biggr]\ .\label{eq:box2S}
\end{eqnarray}
The latter is absorbed in the $V^{({\rm CT})}$ terms proportional to
$C_3$ and $C_4$.

We now turn our attention to the contributions from panels (h)-(u)
in Fig.~\ref{fig:operen}.  Those from panels (h) and (i) represent the
renormalization of nucleon external lines, discussed in
subsection~\ref{sec:np}, and cancel out due the choice of the
mass counter-term $\delta M$.  However, there is a factor
of $\sqrt{Z_N}$ which needs to be included for each of the
nucleon external lines.  A correction of order $Q$ to the PV OPE potential follows given by
\begin{eqnarray}
 V^{(1)}({\rm h}\!+\!{\rm i})&=& 
   \Bigl[\Bigl(\sqrt{Z_N}\Bigl)^4-1\Bigl]\,\,
   V^{(-1)}({\rm NR}) \nonumber\\
     & =& 
  -{3 \, g^2_A\over 4\, f^2_\pi}\,  J_{13} \,\,
V^{(-1)}({\rm NR}) 
 \ .\label{eq:h+i}
\end{eqnarray}
The contributions from panels (j), (k), and (l), which are of order $Q$,
cancel out; in particular, diagrams of the (j)- and (k)-type, but where the
PV $\pi NN$ vertex occurs in the pion loop, vanish (the integrand
in the loop integration is odd).
The contributions from panels (m) and (n)
represent vertex corrections of order $Q$, given by
\begin{equation}
  V^{(1)}({\rm m}\!+\!{\rm n})= { g^2_A\over 12\, f^2_\pi}\, J_{13}\,\,
  V^{(-1)}({\rm NR})
    \ .\label{eq:m+n}
\end{equation}
Note that hereafter we ignore factors of $\sqrt{Z_N}$, since they
would lead to corrections of order higher than $Q$.
The contributions from panels (o) and (p) are tadpoles originating from the
interaction Hamiltonian $H^{3\pi NN}$,
\begin{equation}
  V^{(1)}({\rm o}\!+\!{\rm p})=
- {20\, \alpha-1\over 4f_\pi^2} \, J_{01}\,\, V^{(-1)}({\rm NR})  \ .
\end{equation}
The contributions from panels (q) and (r) represent renormalizations
of the (internal) pion line (see Sec.~\ref{sec:np}),
\begin{equation}
  V^{(1)}({\rm q}\!+\!{\rm r})=
   \left( -{A_\pi\over \omega_k^2} +\delta Z_\pi \right) \,\, V^{(-1)}({\rm NR})\ ,
\end{equation}
where $A_\pi$ and $\delta Z_\pi$ are the quantities defined in
Eqs.~(\ref{eq:Api}) and~(\ref{eq:AdZ}); in particular, 
 $A_\pi=0$ because of our choice to work with the physical pion
mass, and $\delta Z_\pi$ contributes to the renormalization of the PC $\pi NN$
vertex.  Finally, the contributions from panels (s), (t), and (u)
represent vertex corrections.  Those involving the PC $\pi\pi NN$ vertex
read
\begin{equation}
  V^{(1)}({\rm s}\!+\!{\rm t}\!+\!{\rm u})=  {1\over 4\, f_\pi^2}\, J_{01}\,\,
  V^{(-1)}({\rm NR}) \ ,
\end{equation}
However, contributions from diagrams involving the PV $\pi\pi NN$ vertex
are at least of order $Q^2$.

We conclude by noting that an analysis of $\pi NN$ vertex corrections
was also carried out in Ref.~\cite{Zhu01} (in that paper the choice $\alpha=1/6$ in
Eq.~(\ref{eq:uumatrix}) was adopted).  Our expressions are in agreement
with those reported there.
\egroup

\section{The potential in configuration space}
\label{app:pvr}

In this Appendix, all the LEC's are considered to
be the renormalized ones (with the overlines omitted
for simplicity).
The configuration space expressions follow from Eq.~(\ref{eq:vrsp}) and read
\begin{equation}
 V_{PV}(\bmr,\bmp) =
 V^{({\rm OPE})}(\bmr)+V^{({\rm TPE})}(\bmr)+V^{({\rm RC})}(\bmr,\bmp)+V^{({\rm CT})}(\bmr)\ , \label{eq:pvnnr}
\end{equation}
where $\bmp=-i\bmna$ is the relative momentum operator and
\bgroup
\arraycolsep=1.0pt
\begin{eqnarray}
 V^{({\rm OPE})}(\bmr) &=& 
  {g_A h^1_\pi\over 2\sqrt{2}f_\pi}\, (\vec\tau_1\times\vec\tau_2)_z\;
   (\bmsi_1+\bmsi_2)\cdot\hat\bmr\, g'(r)\ ,\label{eq:opelor}\\
 V^{({\rm TPE})}(\bmr) &=& 
  -{g_A h^1_\pi\over 2\sqrt{2} f_\pi} {m_\pi^2\over \Lambda_\chi^2} 
     (\vec\tau_1\times\vec\tau_2)_z
          (\bmsi_1+\bmsi_2)\cdot\hat\bmr\, L'(r)\nonumber \\
  &&-{g_A h^1_\pi\over 2\sqrt{2}f_\pi} {g_A^2 m_\pi^2\over \Lambda_\chi^2} 
     \biggl[4\, (\tau_{1z}+\tau_{2z})\; (\bmsi_1\times\bmsi_2)\cdot\hat\bmr\, L'(r)\;
     \nonumber \\
  &&+(\vec\tau_1\times\vec\tau_2)_z (\bmsi_1+\bmsi_2)\cdot\hat\bmr
\Bigl[H'(r)-3L'(r)\Bigr]\biggr]\ ,\label{eq:boxr}\\
 V^{({\rm RC})}(\bmr,\bmp) &=&-
{g_A h^1_\pi\over 2\sqrt{2} f_\pi 4M^2}(\vec\tau_1\times\vec\tau_2)_z\nonumber \\
  && \biggl[  \bigl\{p_j\, ,\,\bigl\{ p_j\, ,\, 
     (\bmsi_1+\bmsi_2)\cdot\hat\bmr\, g'(r)\bigr\}\bigr\} \nonumber\\
  &+& {\epsilon_{j\ell m}\over 2} \Bigl( \sigma_{1i}\, \sigma_{2j} 
                                  +\sigma_{1j}\, \sigma_{2i} \Bigr) \,\Bigl\{
                 p_m \, , \, \partial_i\, \partial_\ell g(r) \Bigr\}\biggr]
                  \ ,\label{eq:opennlor}\\
 V^{({\rm CT})}(\bmr) &=&  
         {m_\pi^2 \over \Lambda_\chi^2 f_\pi} \Bigl[ 
       C_1 (\bmsi_1\times\bmsi_2)\cdot  \hat\bmr \nonumber\\
       && \quad +C_2\,  \vec\tau_1\cdot\vec\tau_2\,
       (\bmsi_1\times\bmsi_2)\cdot  \hat\bmr\;
        \nonumber\\
       && \quad +C_3\,  (\vec\tau_1\times\vec\tau_2)_z\,
        (\bmsi_1+\bmsi_2)\cdot\hat \bmr\;
      \nonumber\\
       &&\quad  +C_4\,(\tau_{1z}+\tau_{2z})\,
        (\bmsi_1\times\bmsi_2)\cdot  \hat\bmr\;
         \nonumber \\
       &&\quad +C_5\, {\cal I}^{ab}\, \tau_{1a} \, \tau_{2b} \,
        (\bmsi_1\times\bmsi_2)\cdot \hat\bmr\;     
             \Bigr] Z'(r)\ , \label{eq:ctr}
\end{eqnarray}
with
\begin{eqnarray}
 g(r) &=& \int {{\rm d}^3k\over (2\pi)^3}\; {C_\Lambda(k)\over k^2+m_\pi^2}\,
 e^{i\bmk\cdot\bmr} \ ,\label{eq:g}\\
 L(r) &=& \int {{\rm d}^3k\over (2\pi)^3}\; {C_\Lambda(k)\over m_\pi^2} L(k)\,
 e^{i\bmk\cdot\bmr} \ ,\label{eq:lr}\\
 H(r) &=& \int {{\rm d}^3k\over (2\pi)^3}\; {C_\Lambda(k)\over m_\pi^2} H(k)\,
 e^{i\bmk\cdot\bmr} \ ,\label{eq:hr}\\
 Z(r) &=& \int {{\rm d}^3k\over (2\pi)^3}\; {C_\Lambda(k)\over m_\pi^2}\,
 e^{i\bmk\cdot\bmr} \ .\label{eq:z}
\end{eqnarray}
\egroup   
Note that
\begin{eqnarray} 
 \bigl\{p_j\, ,\,\bigl\{ p_j\, ,\, 
    O\bigr\}\bigr\} &=& -\left(\bmna^2 O\right) \nonumber\\
    &&-
    4\, \left[ \left( \bmna  O \right)\cdot \bmna+4 \, O \bmna^2 \right] \ ,
\end{eqnarray}   
and
\begin{eqnarray} 
 \lefteqn{\bmna^2 (\bmsi_1+\bmsi_2)\cdot\hat\bmr \,g'(r)
 \qquad\qquad\qquad\ }&& \nonumber\\
  &=&  (\bmsi_1+\bmsi_2)\cdot\hat\bmr \,
      \Bigl[g'''(r) +2\, {g''(r)\over r}-2 \,{g'(r)\over
        r^2}\Bigr]\ ,
\end{eqnarray}   
\begin{eqnarray}
 \lefteqn{ \partial_j (\bmsi_1+\bmsi_2)\cdot\hat\bmr\, g'(r)
   =(\sigma_{1j}+\sigma_{2j})\, {g'(r)\over r}
   \qquad\qquad\qquad\ }&& \nonumber\\
  &&+
   (\bmsi_1+\bmsi_2)\cdot\hat\bmr \,
     \Bigl[{g''(r)\over r}- {g'(r)\over r^2}\Bigr]\,{r_j\over r}\ ,
\end{eqnarray}   
It is convenient to define the operators
\begin{eqnarray}
  S_r^\pm&=&(\bmsi_1\pm\bmsi_2)\cdot\hat\bmr\ , \\
  S_p^\pm&=&(\bmsi_1\pm\bmsi_2)\cdot \bmp \ , \\
  S_r^\times &=& (\bmsi_1\times\bmsi_2)\cdot\hat\bmr \ ,\\
  S_{L} &=& \bmsi_1\cdot\hat\bmr\,\,\bmsi_2\cdot\hat{\bmL}+
             \bmsi_2\cdot\hat\bmr\,\, \bmsi_1\cdot\hat \bmL\ ,
\end{eqnarray}
where $\hat\bmL=\hat\bmr\times\bmp$ is the ``reduced'' orbital angular momentum operator.
In terms of these, $V^{({\rm RC})}(\bmr,\bmp)$  can be written as
\begin{eqnarray}
 \lefteqn{V^{({\rm RC})}(\bmr,\bmp)= {g_A h^1_\pi\over 4\sqrt{2} F_\pi M^2}
  (\vec\tau_1\times\vec\tau_2)_z
 \qquad\qquad\qquad}&&\nonumber\\
  &&  \Biggl[ 
     \Bigl[g'''(r) +2{g''(r)\over r}-2 {g'(r)\over r^2}\Bigr]
       S^+_r\nonumber\\
  &&  \ \   +4\,i\biggl[  {g'(r)\over r}\, S^+_p
                     +\Bigl[ g''(r)-{g'(r)\over r}\Bigr]\,S_r^+ \,\hat\bmr\cdot\bmp\biggr]
      \nonumber\\
  &&  \ \ -4\, g'(r) \;  S^+_r\; \bmp^2 \;
   -\Bigl[ g''(r)-{g'(r)\over r}\Bigr] \,S_L\Biggr]
                  \ .\label{eq:opennlo2}
\end{eqnarray}
The functions $g(r)$, $L(r)$, $H(r)$, and $Z(r)$ are calculated numerically by
standard quadrature techniques.  It is easily seen that
$g''(r)-g'(r)/ r$ and $g'''(r) +2\,g''(r)/ r-2\,
g'(r)/r^2$ are well-behaved as $r\rightarrow0$.
\newpage
\end{document}